\documentclass[11pt,a4paper]{article}
\usepackage{jheppub} 
\pdfoutput=1
\usepackage{amsmath}
\usepackage[normalem]{ulem}

\usepackage{subcaption}
\usepackage[T1]{fontenc}
\usepackage[utf8]{inputenc}
\usepackage{braket}
\usepackage{slashed}
\usepackage{placeins}
\usepackage{xcolor}

\DeclareUnicodeCharacter{3B3}{ }
\DeclareUnicodeCharacter{204E}{ }

\DeclareRobustCommand{\eq}[1]{Eq.~\eqref{eq:#1}}
\DeclareRobustCommand{\eqs}[2]{Eqs.~\eqref{eq:#1} and \eqref{eq:#2}}
\DeclareRobustCommand{\eqss}[3]{Eqs.~\eqref{eq:#1}, \eqref{eq:#2} and \eqref{eq:#3}}
\DeclareRobustCommand{\fig}[1]{Fig.~\ref{fig:#1}}
\DeclareRobustCommand{\figs}[2]{Figs.~\ref{fig:#1} and \ref{fig:#2}}

\DeclareRobustCommand{\app}[1]{appendix~\ref{app:#1}}
\DeclareRobustCommand{\sec}[1]{Sec.~\ref{sec:#1}}

\DeclareRobustCommand{\tbl}[1]{Table~\ref{tbl:#1}}
\DeclareRobustCommand{\mycite}[1]{Ref.~\cite{#1}}
\DeclareRobustCommand{\mycites}[1]{Refs.~\cite{#1}}

\DeclareRobustCommand{\Li}[1]{\mathrm{Li}_{#1}}
\newcommand{\MS}{\overline{\mathrm{MS}}}
\newcommand{\df}{\mathrm{d}}
\newcommand{\img}{\mathrm{i}}

\newcommand{\eps}{\epsilon}
\newcommand{\cA}{\mathcal{A}}
\newcommand{\cB}{\mathcal{B}}

\newcommand{\cO}{\mathcal{O}}
\newcommand{\cP}{\mathcal{P}}

\newcommand{\nn}{\nonumber}
\newcommand{\bn}{{\bar{n}}}
\newcommand{\as}{\alpha_s}
\newcommand{\LQCD}{\Lambda_\mathrm{QCD}}

\newcommand{\UV}{\mathrm{UV}}
\newcommand{\IR}{\mathrm{IR}}
\newcommand{\Tr}{\mathrm{Tr}}
\newcommand\TMD{\mathrm{TMD}}

\newcommand{\Lb}[1]{\ln^{#1}\frac{b_T^2 \mu^2}{b_0^2}}
\newcommand{\LPz}[1]{\ln^{#1}\frac{(2 P^z)^2}{\mu^2}}

\newcommand{\bt}{\vec b_T}
\newcommand{\kt}{\vec k_T}

\newcommand{\qt}{\vec q_T}
\newcommand{\sgn}[1]{\mathrm{sgn}(#1)}
\newcommand{\plusinf}{_+^{\pm\infty}}
\newcommand{\GammaC}{\Gamma_{\rm cusp}}
\newcommand{\unsub}{{\rm(unsub)}}
\newcommand{\ns}{{\rm ns}}

\title{Towards Quasi-Transverse Momentum Dependent PDFs Computable on the Lattice}

\author{Markus A.~Ebert,}
\author{Iain W.~Stewart,}
\author{and Yong Zhao}
\affiliation{Center for Theoretical Physics, Massachusetts Institute of Technology, Cambridge, Massachusetts 02139, USA}
\emailAdd{ebert@mit.edu}
\emailAdd{iains@mit.edu}
\emailAdd{yzhaoqcd@mit.edu}

\abstract{
Transverse momentum dependent parton distributions (TMDPDFs) which appear in factorized cross sections involve infinite Wilson lines with edges on or close to the light-cone. Since these TMDPDFs are not directly calculable with a Euclidean path integral in lattice QCD,  we study the construction of quasi-TMDPDFs with finite-length spacelike Wilson lines that are amenable to such calculations.
We define an infrared consistency test to determine which quasi-TMDPDF definitions are related to the TMDPDF, by carrying out a one-loop study of infrared logarithms of transverse
position $b_T\sim \LQCD^{-1}$, which must agree between them. 
This agreement is a necessary condition for the two quantities to be related by perturbative matching.
TMDPDFs necessarily involve combining a hadron matrix element, which nominally depends on a single light-cone direction, with soft matrix elements that necessarily depend on two light-cone directions.
We show at one loop that the simplest definitions of the quasi hadron matrix element, the quasi soft matrix element, and the resulting quasi-TMDPDF all fail the infrared consistency test.
Ratios of impact parameter quasi-TMDPDFs still provide nontrivial information about the TMDPDFs, and are more robust since the soft matrix elements cancel. We show at one loop that such quasi ratios can be matched to ratios of the corresponding TMDPDFs.  We also introduce a modified ``bent'' quasi soft matrix element which yields a quasi-TMDPDF that passes the consistency test with the TMDPDF at one loop, and discuss potential issues at higher orders.
}

\preprint{\vbox{
\hbox{MIT--CTP 5039}
}}

\arxivnumber{}

\begin{document}
\maketitle

\section{Introduction}

Transverse momentum dependent (TMD) distributions are important ingredients for describing high-energy scatterings at low transverse momentum, such as for Drell-Yan production at the Tevatron~\cite{Affolder:1999jh,Abbott:1999yd,Abazov:2007ac,Abazov:2010kn} or LHC~\cite{Aad:2011gj,Chatrchyan:2011wt, Aad:2014xaa, Khachatryan:2015oaa, Aad:2015auj,  Khachatryan:2016nbe} which is a benchmark observable of the Standard Model (SM).
For Higgs bosons produced in proton collisions, the transverse momentum spectrum is also one of the primary observables describing the production kinematics and is of great interest in LHC measurements \cite{Aad:2014lwa,Aad:2014tca,Aad:2016lvc,Aaboud:2017oem,Aaboud:2018xdt,Khachatryan:2015rxa,Khachatryan:2015yvw,Khachatryan:2016vnn,Sirunyan:2018kta}.
TMDs also play an important role in measuring semi-inclusive deep-inelastic scattering (SIDIS) at low energies \cite{Ashman:1991cj,Derrick:1995xg,Adloff:1996dy,Aaron:2008ad,Airapetian:2012ki,Adolph:2013stb,Aghasyan:2017ctw},
and in improving our understanding of hadron structure \cite{Boer:2011fh,Accardi:2012qut}.

A characteristic feature of TMDs is that they depend on both the longitudinal momentum fraction $x$ and transverse momentum $q_T$ carried by the struck parton.  Currently, TMDs are only directly calculable for the perturbative regime $q_T\gg \LQCD$, %
where they can be obtained in terms of collinear parton distribution functions (PDFs). In contrast, nonperturbative TMDs with small $q_T\sim \LQCD$ have only been extracted from measurement by performing global fits to a variety of experimental data sets, see e.g.~\mycites{Landry:1999an,Landry:2002ix,Konychev:2005iy,DAlesio:2014mrz,Bacchetta:2017gcc,Scimemi:2017etj}.
Imprecise knowledge of these TMDs limits the predictive power at small transverse momenta and constitutes a significant theoretical uncertainty. At small $q_T$ the TMDs also provide a crucial window into the structure of the proton. It is therefore desirable to find a method to calculate them from first principles.

For nonperturbative objects, lattice QCD currently provides the only practical means for first-principle calculations, and studies have been performed in \mycites{Musch:2010ka,Musch:2011er,Engelhardt:2015xja,Yoon:2016dyh,Yoon:2017qzo}, where ratios of $x$-moments of quark TMDPDFs were determined.
This gets around the fact that the $x$-dependence of the TMDPDFs is not directly accessible from the Euclidean lattice, since it involves light-cone correlations which depend on the Minkowski time, in direct parallel to the same issue for calculating the $x$-dependence of the longitudinal PDFs. These analyses use Lorentz invariance to connect the TMD nucleon matrix element of interest to spatial matrix elements that are accessible from the lattice.

In \mycites{Ji:2013dva,Ji:2014gla} the large momentum effective theory (LaMET) was proposed as a method to overcome the hurdle of calculating light-cone quantities by relating them to time-independent quasi observables in a large-momentum nucleon state. The latter can be directly calculated on the Euclidean lattice and matched to the corresponding light-cone quantity through a factorization theorem that is based on a systematic expansion in the nucleon momentum.
For example, the unpolarized collinear PDF of a proton moving along the $n$-direction is usually defined in the $\overline{\mathrm{MS}}$ scheme as
\begin{align} \label{eq:pdf}
f_i(x,\mu) &= \int\frac{\df b^+}{4\pi} \, e^{-\img \frac{1}{2} b^+ (x P^-)} \,
\Bigl<p(P) \Big| \Bigl[\bar q_i\Big(b^+ \frac{\bn}{2}\Bigr)
\frac{\gamma^-}{2} W\Bigl(b^+\frac{\bn}{2},0\Bigr)  q_i(0)\Bigr]_{\!\mu} \,\Big| p(P) \Bigr>
\,,\end{align}
where $i$ is the flavor index and the square bracket with subscript $\mu$ denotes that the operator is renormalized at scale $\mu$. For later use we introduce the lightlike reference vectors $n^\mu=(1,0,0,1)$ and $\bn^\mu=(1,0,0,-1)$. The light-cone coordinates are defined as $b^{\pm} = b^0 \mp b^3$,
and $p(P)$ denotes a proton state with momentum $P^\mu = (E,0,0,P^z)$.
The path-ordered light-cone collinear Wilson line is
\begin{align}
W\Big(b^+\frac{\bn}{2},0\Big) &= P \exp\biggl[ -\img g \int^{b^+/2}_0 \df s\, \bn \cdot \cA(s \bn^\mu) \biggr]
\,.\end{align}
To calculate the PDF $f_i(x,\mu)$ using LaMET, one starts from a quasi-PDF, which is defined from equal-time correlation functions,
\begin{align} \label{eq:quasipdf}
\tilde{f}_i(x,P^z,\tilde\mu) &= \int\frac{\df z}{2\pi} \, e^{\img z (x P^z)} \,
\Bigl< p(P) \Big| \Bigl[\bar q_i(z)\frac{\Gamma}{2} W_z (z,0) q_i(0) \Bigr]_{\!\tilde\mu} \,\Big| p(P) \Bigr>
\,,\end{align}
where the operator is renormalized with a lattice friendly renormalization scheme and $\tilde\mu$ denotes a corresponding scale.
The spacelike Wilson line is
\begin{align}
W_z(z,0) &= P \exp\left[ \img g \int^z_0 \df z'\, \cA^z(z') \right]
\,,\end{align}
and $\Gamma=\gamma^0$ or $\Gamma=\gamma^3$ as they both belong to the same universality class of operators that can be related to the PDF through an infinite Lorentz boost along the $z$ direction~\cite{Hatta:2013gta}. Unlike the PDF $f_i(x,\mu)$ that is boost invariant, the quasi-PDF $\tilde{f}_i(x,P^z,\tilde\mu)$ depends nontrivially on the nucleon momentum $P^z$. When $P^z$ is large compared to $\LQCD$ as well as the nucleon mass $M$, the quasi-PDF satisfies the following factorization theorem~\cite{Ji:2013dva,Ji:2014gla,Ma:2014jla,Ma:2017pxb,Izubuchi:2018srq},
\begin{align}\label{eq:fimatch}
 \tilde{f}_i(x,P^z,\tilde\mu) = \int_{-1}^1 \frac{\df y}{|y|}
  C_{ij}\biggl(\frac{x}{y}, \frac{\tilde\mu}{\mu}, \frac{\mu}{|y| P^z} \biggr) f_j(y,\mu)
  + \mathcal{O}\biggl({\LQCD^2\over P_z^2}, {M^2\over P_z^2} \biggr)
\,,\end{align}
where $\mathcal{O}(\LQCD^2/P_z^2,M^2/P_z^2)$ terms are power corrections. Here $f_j(y,\mu)$ for $-1<y<0$ corresponds to the anti-quark PDF.  
The  $C_{ij}$ are perturbative matching coefficients which come from a hard region of momentum space, see~\mycites      {Ma:2014jla,Izubuchi:2018srq} for further details.

Recently, significant progress has been made on various aspects of the LaMET procedure, including the renormalization and matching~\cite{Xiong:2013bka,Ma:2014jla,Ma:2014jga,Ji:2015jwa,Ji:2015qla,Xiong:2015nua,Li:2016amo,Ishikawa:2016znu,Chen:2016fxx,Carlson:2017gpk,Briceno:2017cpo,Xiong:2017jtn,Constantinou:2017sej,Rossi:2017muf,Ji:2017rah,Ji:2017oey,Ishikawa:2017faj,Green:2017xeu,Wang:2017qyg,Chen:2017mie,Stewart:2017tvs,Wang:2017eel,Spanoudes:2018zya,Izubuchi:2018srq,Xu:2018mpf,Rossi:2018zkn,Zhang:2018diq,Li:2018tpe,Liu:2018tox} of the quasi-PDF, the power corrections~\cite{Chen:2016utp,Radyushkin:2017ffo,Braun:2018brg}, as well as the lattice calculation of the $x$-dependence of PDFs and distribution amplitudes~\cite{Lin:2014zya,Alexandrou:2015rja,Chen:2016utp,Alexandrou:2016jqi,Zhang:2017bzy,Alexandrou:2017huk,Chen:2017mzz,Green:2017xeu,Chen:2017lnm,Chen:2017gck,Alexandrou:2018pbm,Chen:2018xof,Chen:2018fwa,Alexandrou:2018eet,Liu:2018uuj,Lin:2018qky,Fan:2018dxu,Liu:2018hxv}. Notably, the most recent lattice results at physical pion mass~\cite{Alexandrou:2018pbm,Chen:2018xof,Alexandrou:2018eet,Lin:2018qky,Liu:2018hxv} and large nucleon momenta~\cite{Chen:2018xof,Lin:2018qky,Liu:2018hxv} have shown encouraging signs that the LaMET approach can lead to a precise determination of the PDFs.

Due to the interest in TMDPDFs it is natural to consider the extension of the LaMET approach to transverse momentum observables. 
Due to the required focus on spatial matrix elements for TMDPDFs, studies based on LaMET are actually related to the lattice methods  developed in \mycites{Engelhardt:2015xja,Yoon:2016dyh,Yoon:2017qzo}.
While applying LaMET to TMDPDFs might seem straightforward,
the richer structure of TMD factorization, which we review in \sec{tmd_review}, actually makes this quite non-trivial.
In contrast to the case for collinear factorization, TMD physics is plagued by so-called
rapidity divergences and the need for combining collinear\footnote{Note that the second use of the word ``collinear'' here is in the context of factorized collinear and soft fields as defined for example in SCET, not to distinguish between collinear and TMD factorization.} proton matrix elements with soft vacuum matrix elements. Such soft matrix elements
retain a minimal amount of information about both incoming protons (their direction and the color charge of the probing parton).
The importance of the soft matrix elements to cancel the analog of rapidity divergences in the spatial matrix elements for TMDPDFs has been discussed in \mycite{Ji:2014hxa}, which was aimed at constructing a new TMD factorization theorem for the Drell-Yan process in terms of the so-called quasi-TMDPDFs in LaMET. More recently, the matching relationship between the quasi-TMDPDF on a finite-volume lattice and standard TMDPDF was studied at one-loop order in~\mycite{Ji:2018hvs}, where it was shown that the finite lattice size regulates these divergences, thus eliminating the need to introduce a dedicated regulator in the lattice calculation. 
However, we argue here that the final result of \mycite{Ji:2018hvs}, which agrees at one loop with one of the results studied here, cannot be used for nonperturbative $b_T$ or $q_T$, for reasons that will be discussed in detail.

In this work we consider the problem of constructing a quasi-TMDPDF that can be used to study the TMDPDF for nonperturbative $q_T\sim \LQCD$. Here the quasi-TMDPDF must be chosen such that it can be calculated with lattice QCD, agrees with the physical TMDPDF in the infrared, and differs only by short distance contributions from the ultraviolet.
This is required in order for a matching equation connecting the quasi-TMDPDF and TMDPDF to exist, analogous to \eq{fimatch}, with a coefficient $C$ that is not sensitive to infrared physics. These requirements still leave some freedom in the construction of a suitable quasi-TMDPDF.
We therefore propose to test which quasi-TMDPDF definitions are feasible by carrying out a perturbative study of the infrared logarithms of $q_T$, or equivalently the transverse position $b_T$, which must agree order by order in $\alpha_s$ with the same logarithms in the TMDPDF.
Collinear infrared divergences related to the collinear momentum fraction $x$ must
of course also agree between TMDPDF and quasi-TMDPDF.
We construct our quasi-TMDPDFs from distinct collinear proton and soft vacuum matrix elements, in direct analogy with the TMDPDF, where the quasi adjustment is made through the form of the operators appearing in these matrix elements. We carry out our study of infrared logarithms separately for the collinear and soft matrix elements, thus enabling us to separately probe the form of the corresponding collinear and soft operators.
We also discuss in detail the role of rapidity divergences,
which play an important role in the construction of TMDPDFs on the light-cone,
which for the quasi-TMDPDF are regulated by having Wilson lines of finite length $L$, as required on lattice. The cancellation of the leftover $L$-dependence constrains how one must combine quasi soft and quasi collinear matrix elements.

This paper is structured as follows.
In \sec{tmd_review}, we review the TMD factorization theorem,
discussing in particular the role of rapidity divergences and regulators, and how the TMDPDF is constructed from combining a proton matrix element and a vacuum soft matrix element.
We then discuss in \sec{towards_qtmd} the expected form of a matching relation between TMDPDF and quasi-TMDPDF, how Wilson lines of finite length naturally regulate the analog of rapidity divergences in lattice calculations, and discuss the quasi-construction of the matrix elements required to define the quasi-TMDPDF.
In \sec{nlo_results}, we explicitly test the suggested quasi-TMDPDF by comparing to the TMDPDF at one loop, showing that the simplest quasi collinear proton matrix element fails the infrared test. The simplest attempt of constructing a quasi soft matrix element also fails this test. Finally, combining these into the simplest quasi-TMDPDF also gives a result that fails this test.
To resolve this issue we introduce a bent quasi soft function, which leads to a quasi-TMDPDF whose infrared divergences properly match those of the TMDPDF at one-loop.
Discussion of the implications of these results are given in \sec{results}, 
including issues that may still spoil the bent quasi soft function construction at higher orders, and how the main issues can be avoided by only studying ratios of TMDPDFs in impact parameter space.
We conclude in \sec{outlook}.  
Further details that are important for our analysis are provided in appendices. In \app{conventions} we summarize our notations and conventions for light-cone coordinates and $\overline{\rm MS}$, and contrast them with another popular convention used in the literature. In \app{overview_tmdpdfs} we discuss different schemes for TMDPDFs that are used in the literature, demonstrating that they all satisfy the general functional form for the TMDPDF that we use for our analysis.
In \app{qbeamfunc} we provide details on the perturbative calculation of the quasi proton matrix element.

\section{Review of TMD Factorization}
\label{sec:tmd_review}

A precise understanding of the TMD factorization theorem and its different formulations in the literature is important to properly connect a lattice determination of the TMDPDF to the phenomenological TMDPDF.
In this section, we provide a detailed review of TMD factorization and set up a general notation for the definition of the TMDPDF that encompasses most of the available definitions in the literature.

\subsection{TMD Factorization and TMDPDFs}
\label{sec:tmd_factorization}

TMD factorization was originally derived by Collins, Soper and Sterman (CSS)
in \mycites{Collins:1981uk,Collins:1981va,Collins:1984kg}.
\mycites{Collins:1985ue,Collins:1988ig,Collins:1989gx,Collins:1350496,Diehl:2015bca} showed the cancellation of potentially factorization-violating Glauber modes,
and the factorization was further elaborated on and extended
in \mycites{Catani:2000vq,deFlorian:2001zd,Catani:2010pd,Collins:1350496}.
It has also been considered in the framework of Soft-Collinear Effective Theory (SCET)
\cite{Bauer:2000ew, Bauer:2000yr, Bauer:2001ct, Bauer:2001yt}
by various authors \cite{Becher:2010tm, Becher:2011xn, Becher:2012yn, GarciaEchevarria:2011rb, Echevarria:2012js, Echevarria:2014rua, Chiu:2012ir}.
For a relation of the different approaches to each other, see e.g.~\mycite{Collins:2017oxh}, and for a historical review on TMDPDFs we refer the reader to~\cite{Collins:1350496}.
In \app{conventions} we give a summary of our notation for light-cone coordinates. 
In \app{overview_tmdpdfs} we give a comparison between different TMDPDF constructions from the literature, and provide explicit relations to the notation we use here.

For simplicity we consider TMD factorization in the context of the production of a color-singlet final state $F$ in the scattering of two unpolarized
energetic protons moving along the
$n^\mu = (1,0,0,1)$ and $\bn^\mu = (1,0,0,-1)$ directions.
Examples of such cross sections include Drell-Yan, $W$, and Higgs production.
Here we only measure the total four momentum of $F$ through its invariant mass $Q$, rapidity $Y$ and transverse momentum $\qt$, so that we only have unpolarized TMDPDFs.
In the limit $q_T\equiv |\qt| \ll Q$ the cross section can be factorized as
\begin{subequations} \label{eq:sigma}
\begin{align}
 \label{eq:sigmaa}
 \frac{\df\sigma}{\df Q \df Y \df^2\qt}
 &= \sum_{i,j} H_{ij}(Q,\mu) \int\! \df^2\bt \, e^{\img \bt \cdot \qt} \,
   B_{i}\Bigl(x_a, \bt, \mu, \frac{\zeta_a}{\nu^2}\Bigr)
   B_{j}\Bigl(x_b, \bt, \mu, \frac{\zeta_b}{\nu^2}\Bigr)
   S_{ij}(b_T,\mu,\nu)
   \nn\\&\hspace{1cm}
   \times \biggl[ 1 + \cO\biggl(\frac{q_T^2}{Q^2}, \frac{\LQCD^2}{Q^2}\biggr) \biggr]
 \\
\label{eq:sigmab}
 &= \sum_{i,j} H_{ij}(Q,\mu) \int\! \df^2\bt \, e^{\img \bt \cdot \qt} \,
   f^\TMD_{i}(x_a, \bt, \mu, \zeta_a) \, f^\TMD_{j}(x_b, \bt, \mu, \zeta_b)
   \nn\\&\hspace{1cm}
   \times \biggl[ 1 + \cO\biggl(\frac{q_T^2}{Q^2}, \frac{\LQCD^2}{Q^2}\biggr) \biggl]
\,.\end{align}
\end{subequations}
The factorized cross section receives power corrections in $q_T^2/Q^2$ and $\LQCD^2/Q^2$,
and first studies of their perturbative and nonperturbative structure have been performed in \mycites{Balitsky:2017flc,Balitsky:2017gis,Ebert:2018gsn}.
Importantly, \eq{sigma} remains valid for nonperturbative $q_T \sim \LQCD$.
The factorization is most elegantly written in impact parameter space,
with $\bt$ being Fourier conjugate to the measured transverse momentum $\qt$.
In \eq{sigma} $i,j$ are parton indices for quark flavors or gluons,%
\footnote{For gluon-induced processes, $H$ and $B$ also carry helicity indices even for unpolarized TMDPDFs}
which we suppress, and $H_{ij}$ is the hard function containing virtual corrections
to the underlying hard process $a b \to F$. The $x_{a,b}$ are the fractions of the proton momenta carried by the partons $i$ and $j$ participating in the hard collision, so $x_a = Qe^Y/E_{\rm cm}$ and $x_b=Q e^{-Y}/E_{\rm cm}$ where $E_{\rm cm}$ is the center of mass energy of the $pp$ collision.
Finally, $\zeta_a$ and $\zeta_b$ are related to the momenta of the struck partons and given by
\begin{align} \label{eq:zeta}
 \zeta_a = (x_a P_a^-)^2 e^{-2y_n}
 \,, \qquad
 \zeta_b = (x_b P_b^+)^2 e^{2y_n}
 \,, \qquad
 \zeta_a \zeta_b = Q^4
\,.\end{align}
Here $P_a^\mu = P_a^- \frac{n^\mu}{2}+ \frac{m_P^2}{P_a^-}\frac{\bn^\mu}{2}$ and $P_b^\mu = P_b^+ \frac{\bn^\mu}{2}+ \frac{m_P^2}{P_b^+}\frac{n^\mu}{2}$ are the momenta of the protons,
such that $P_a^- P_b^+ = E_{\rm cm}^2$, and $y_n$ is a parameter that encodes scheme dependence. Note that whenever possible we neglect target mass corrections from $m_P^2\ll Q^2< E_{\rm cm}^2$.

The small transverse momentum $q_T \ll Q$ of the final state is generated
from soft and collinear radiation off the incoming protons.
In \eq{sigmaa}, the collinear and soft radiation are described separately.
$B_{i}$ and $B_{j}$ are collinear proton matrix elements that measure the transverse momentum
originating from energetic radiation close to the $n$- and $\bn$-collinear protons,
and to distinguish them from the final TMDPDF we will follow the language of \mycite{Stewart:2009yx}
and refer to $B_{i}$ and $B_{j}$ as beam functions.
The soft function $S^i\equiv S_{ij}$ is a vacuum matrix element that encodes soft exchange between the incoming partons.
It only differs for gluons, $S^g \equiv S_{gg}$, and quarks, $S^q \equiv S_{q\bar q}$,
but is independent of the light quark flavor.
$S^i$ tracks the direction of both incoming partons, and hence depends on both light-cone directions $n$ and $\bn$.

The different ingredients in \eq{sigma} depend on both the renormalization scale $\mu$ and an additional \emph{rapidity} renormalization scale $\nu$.%
\footnote{Some regulators and schemes have two distinct scales $\nu_n, \nu_\bn$ where $\nu_n$ appears in the $n$-collinear beam function $B_i$, $\nu_\bn$ appears in the $\bn$-collinear beam function $B_j$, and both scales appear in $S_{ij}$. We suppress this possibility in our review.}
The latter scale arises because the matrix elements not only suffer from UV divergences, but also from so-called rapidity divergences that require a dedicated regulator \cite{Soper:1979fq,Collins:1981uk,Collins:1992tv,Collins:2008ht,Becher:2010tm,GarciaEchevarria:2011rb,Chiu:2011qc,Chiu:2012ir}, and we will discuss their physical origin in more detail in \sec{rapidity_divergences}.
Crucially, the regularization of these divergences requires an explicit rapidity regulator, which we generically denote as $\tau$.
One can distinguish two classes of schemes by whether introducing the rapidity regulator affects the hard function $H_{ij}$ or not.
The hard function describes virtual corrections to the underlying Born process and thus carries the full process dependence (the TMDPDFs are only sensitive to the hard initial state, namely $q\bar q$ or $gg$), so it is simplest to consider regulators that do not affect $H_{ij}$.
In the following, we will only consider such regulators, and fix the hard function to be in the $\MS$ scheme, which then immediately fixes the scheme for the product of the TMDPDFs.
This applies to most modern regulators, but not to the formulations of \mycites{Soper:1979fq,Collins:1981uk,Ji:2004wu} where $H_{ij}$ and $f_i^\TMD$ depend on an additional parameter $\rho$.

For schemes considered here we can generically define UV and rapidity-renormalized beam and soft functions as
\begin{align} \label{eq:B_renorm}
 B_i(x, \bt, \mu, \zeta/\nu^2) &
 = \lim_{\substack{\epsilon\to 0 \\ \tau\to 0}} Z_B^i(b_T, \mu, \nu, \eps, \tau, x P^-) B_i(x, \bt, \eps, \tau, x P^-)
 \nn\\&
 = \lim_{\substack{\epsilon\to 0 \\ \tau\to 0}}Z_B^i(b_T, \mu, \nu, \eps, \tau, x P^-) \frac{B_i^\unsub(x, \bt, \eps, \tau, x P^-)}{S_i^0(b_T, \eps, \tau)}
\,, \\ \label{eq:S_renorm}
 S^i(b_T,\mu,\nu) &= \lim_{\substack{\epsilon\to 0 \\ \tau\to 0}} Z_S^i(b_T, \mu, \nu, \eps, \tau) S^i(b_T,\eps,\tau)
\,.\end{align}
Note that for simplicity we use the same notation for bare and renormalized soft functions $S^i$, since it will always be clear from the arguments or context which one we are discussing.
For the soft function, this is straightforward.
For the beam function, we note that the renormalized beam function depends on $\zeta \propto (x P^-)^2$, where the proportionality is scheme-dependent, i.e.\ depends on the precise definition of $\tau$.
Secondly, we note that $B_i$ is defined to only describe collinear radiation. In practical calculations, one often encounters an overlap with the soft function when the collinear radiation becomes soft.
To avoid double counting with the soft function, one has to subtract out this overlap from the unsubtracted beam function $B_i^\unsub$, and the subtraction factor is denoted as $S_i^0$ in \eq{B_renorm}.
(In SCET, this is referred to as soft zero-bin subtraction~\cite{Manohar:2006nz}.)
Since $S_i^0$ describes soft physics, it cannot depend on the large momentum $x P^-$.
As indicated by the notation for $S_i^0(b_T,\eps,\tau)$, this factor is often equivalent or closely related to the soft function $S^i(b_T,\epsilon,\tau)$ itself, but this is not always the case.
In particular, one can find rapidity renormalization schemes which
(a) have no zero-bin subtraction, $S_i^0 = 1$ \cite{Chiu:2012ir},
(b) where the zero-bin is equal to the soft function, $S_i^0 = S^i$ \cite{GarciaEchevarria:2011rb,Li:2016axz}, and
(c) where the subtraction is given by a combination of soft functions in different non-lightlike directions \cite{Collins:1350496}.

In practice, one often combines the beam and soft functions to yield two separate TMDPDFs $f^\TMD_i$ and $f^\TMD_j$,
as used in the factorized cross section in \eq{sigmab}.
This can be achieved either by combining the renormalized beam and soft functions,
\begin{align}
 \label{eq:tmdpdf1r}
 f^\TMD_{i}(x, \bt, \mu , \zeta)
 &= B_{i}\bigl(x, \bt, \mu, \zeta/\nu^2\bigr) \sqrt{S^i(b_T,\mu,\nu)}
\,,\end{align}
or by combining the bare functions and performing the UV renormalization afterwards,
\begin{align} \label{eq:tmdpdf1b}
 f^\TMD_{i}(x, \bt, \mu , \zeta) &
 =  \lim_{\substack{\epsilon\to 0 \\ \tau\to 0}} Z_{\rm uv}^i(\mu,\zeta,\epsilon)\, B_{i}^\unsub\bigl(x, \bt, \epsilon, \tau, x P^- \bigr) \frac{\sqrt{S^i(b_T,\epsilon,\tau)}}{S^0_i(b_T,\eps,\tau)}
 \nn\\&
 \equiv \lim_{\substack{\epsilon\to 0 \\ \tau\to 0}} Z_{\rm uv}^i(\mu,\zeta,\epsilon)\, B_{i}^\unsub\bigl(x, \bt, \epsilon, \tau, x P^- \bigr) \Delta_S^i(b_T,\epsilon,\tau)
\,.\end{align}
In \eq{tmdpdf1r}, the rapidity renormalization scale $\nu$ cancels between $B_i$ and $S^i$, leaving only a dependence on $\zeta$ in the TMDPDF.
Likewise, in \eq{tmdpdf1b} the dependence on the regulator $\tau$ cancels between $B_i$ and $S^i$.
Note that \eq{tmdpdf1b} has been written in terms of the unsubtracted beam function $B_i^\unsub$, as is common in the literature, and for ease of notation we have combined the zero-bin subtraction $S_i^0$ and the soft function $S^i$ to define the soft factor $\Delta_S^i = \sqrt{S^i} / S_i^0$.
The UV renormalization factor $Z_{\rm uv}$ in \eq{tmdpdf1b} is trivially related to those of $B_i$ and $S^i$ in \eqs{B_renorm}{S_renorm},
\begin{align} \label{eq:Z_uv_eta}
 Z_{\rm uv}(\mu, \zeta, \eps) = Z_B(b_T, \mu, \nu, \eps, \tau, x P^-) \sqrt{Z_S(b_T, \mu, \nu, \eps, \tau)}
\,.\end{align}
As before, the $\zeta$ scale arises as a rapidity-regulator dependent function with the general form $\zeta \propto (x P^-)^2$.

The need to regulate (and renormalize) the rapidity divergences in the beam and soft functions has led to several different definitions in the literature.
The original derivation by Collins and Soper used a non-lightlike axial gauge \cite{Soper:1979fq,Collins:1981uk}.
Since the same non-lightlike axial gauge has to be used in the calculation of the hard function,
this regulator does not fall in the class of rapidity regulators considered here.%
\footnote{Note that in their original work, the hard function was absorbed in the TMDPDFs.
The separation into process-dependent form factors and process-independent TMDPDFs was first noted in \mycite{Catani:2000vq}.}
In \mycite{Ji:2004wu} rapidity divergences are regulated by taking Wilson lines off the light cone.
Since the same Wilson lines enter the hard factor, this again falls outside the class of regulators we consider, as is explicitly visible by a dependence of their hard function and TMDPDFs on an extra parameter $\rho$.
This is distinct from Collin's new regulator \cite{Collins:1350496} where again Wilson lines are taken off the lightcone, but the lightlike limit is taken such that the hard function is independent of the regulator.
The regulators used in the SCET-based approaches are the analytic regulator acting on eikonal propagators~\cite{Beneke:2003pa, Chiu:2007yn, Becher:2011dz}, the $\eta$-regulator inserted into Wilson lines on the light-cone~\cite{Chiu:2011qc, Chiu:2012ir}, the $\delta$-regulator which adds  mass-like terms to eikonal propagators~\cite{Chiu:2009yx, GarciaEchevarria:2011rb}, and the exponential regulator inserted into the phase space~\cite{Li:2016axz}.
The definition of the TMDPDF in terms of bare beam and soft functions, \eq{tmdpdf1b}, is used in \mycites{Collins:1350496,Ji:2004wu,GarciaEchevarria:2011rb,Echevarria:2012js}, where renormalized beam and soft functions were not defined.
Both the bare and renormalized forms can be used in \mycites{Chiu:2012ir,Li:2016axz}.
In the analytic regulator approach of \mycites{Becher:2010tm,Becher:2012yn}, the soft function vanishes and hence one can only define a product of rapidity-finite TMDPDFs, not individual TMDPDFs.
We provide a more detailed discussion of these regulators in \app{overview_tmdpdfs}, including explicit one-loop results for the quark beam function illustrating the combination of beam and soft functions into the TMDPDF.

\begin{figure}[pt]
 \centering
 \begin{subfigure}{0.45\textwidth}
  \includegraphics[width=\textwidth]{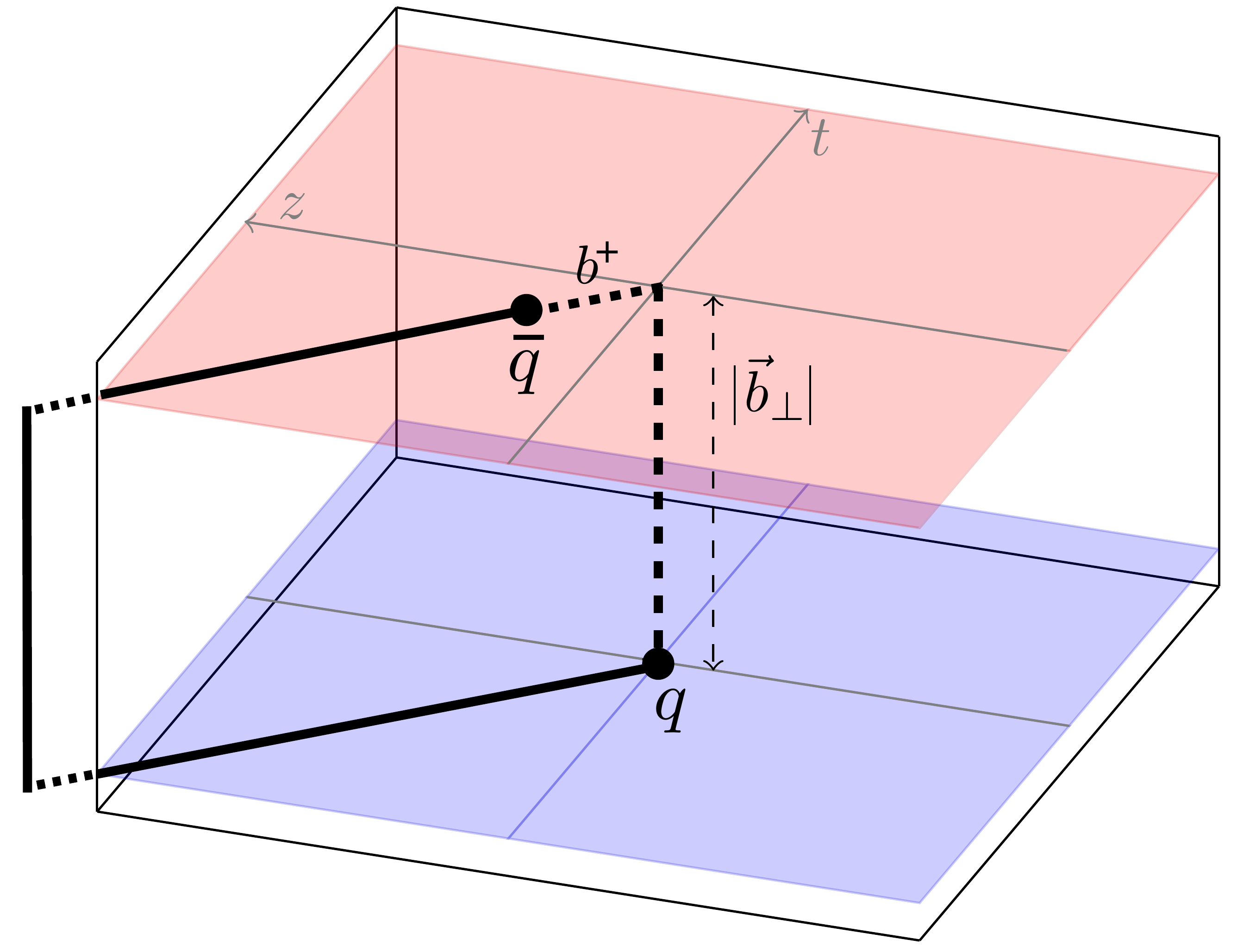}
  \caption{}
  \label{fig:wilson_beam}
 \end{subfigure}
 \hfill
 \begin{subfigure}{0.45\textwidth}
  \includegraphics[width=\textwidth]{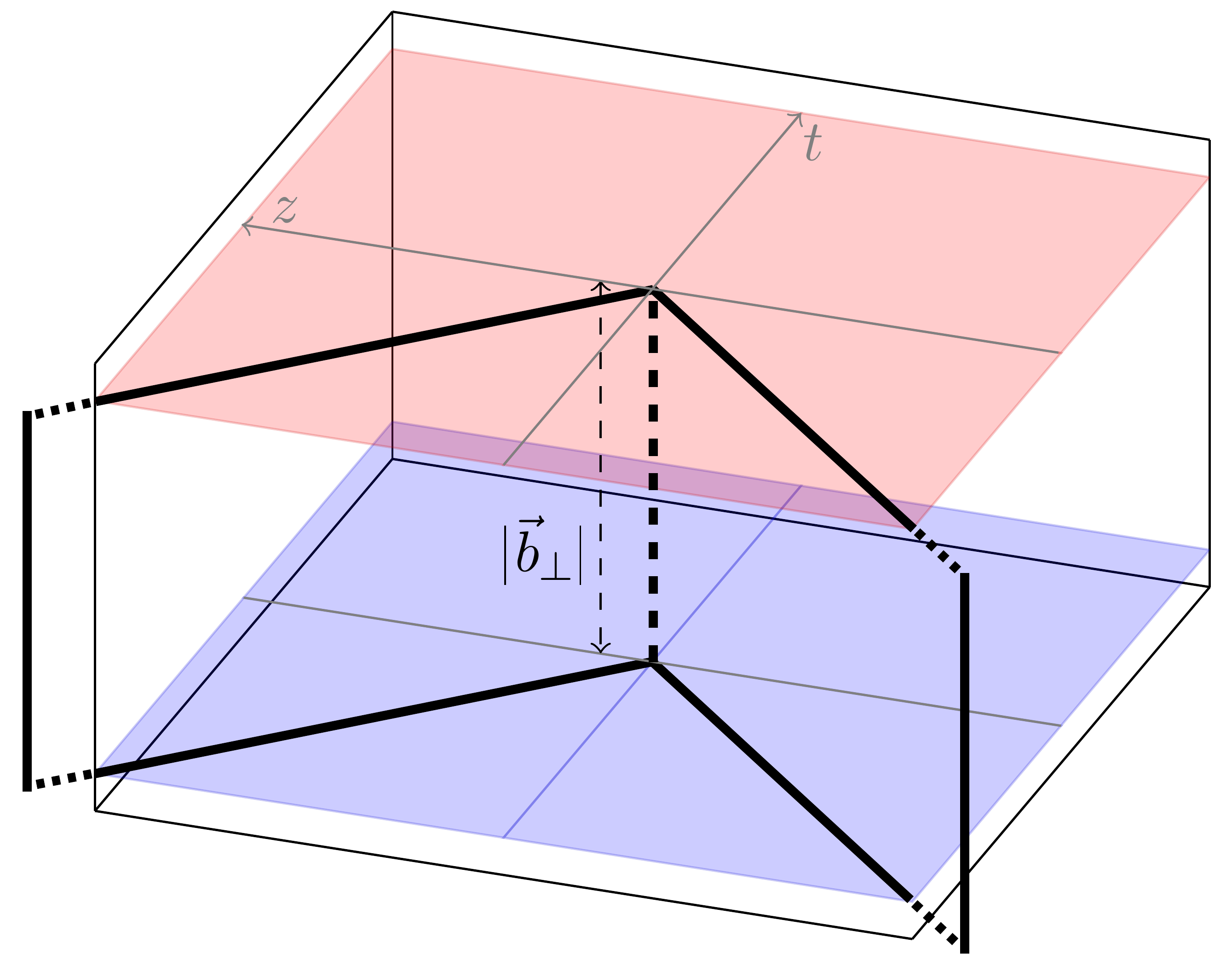}
  \caption{}
  \label{fig:wilson_soft}
 \end{subfigure}
 \caption{Graphs of the Wilson line structure of the $n$-collinear beam function $B_{q}$ (a)
 and the soft function $S^q$ (b), defined in \eqs{beamfunc}{softfunc}.
 The Wilson lines (solid) extend to infinity in the directions indicated.
 Adapted from \mycite{Li:2016axz}.}
 \label{fig:wilsonlines}
\end{figure}

In this work, we focus for simplicity only on the quark TMDPDF.
For a hadron $h$ moving along the $n$ direction with momentum $P$,
the bare beam and soft function are defined as
\begin{align} \label{eq:beamfunc}
 B_{q}(x,\bt,\eps,\tau,x P^-) &= \int\frac{\df b^+}{4\pi} e^{-\img \frac12 b^+ (x P^-)}
 \Bigl< h(P) \Bigr|  \Bigl[ \bar q(b^\mu)
 W(b^\mu) \frac{\gamma^-}{2}
 W_{T}\bigl(-\infty\bn;\vec b_T,\vec 0_T\bigr)
 \nn\\&\hspace{5.5cm}\times
 W^\dagger(0)  q(0) \Bigr]_\tau \Bigl| h(P) \Bigr>
,\\ \label{eq:softfunc}
 S^q(b_T,\eps,\tau) &= \frac{1}{N_c} \bigl< 0 \bigr| {\rm Tr} \bigl[ S^\dagger_n(\bt) S_\bn(\bt)
   S_{T}(-\infty \bn;\vec b_T,\vec 0_T)
 \nn\\&\hspace{2cm}\times
 S^\dagger_\bn(\vec 0_T) S_n(\vec 0_T)
 S_{T}^\dagger\bigl(-\infty n;\vec b_T,\vec 0_T\bigr) \bigr]_\tau
 \bigl|0 \bigr>
\,.\end{align}
where $b^\mu = b^+ \bn^\mu/2+b_T^\mu$. 
Here the bracket $[\cdots]_\tau$ denotes that the operator inside is considered by implementing the rapidity regulator $\tau$.
For the matrix element in \eq{beamfunc}, we note that diagrams that have no fields contracted with the states $|h(P)\rangle$ are excluded.
For clarity, we denote the Wilson lines in the $n$-collinear and soft matrix elements by $W$ and $S_X$, respectively, and both are defined by path-ordered exponentials. One needs both lines of infinite length along the light-cone,
\begin{align} \label{eq:Wilson_lines}
 W(x^\mu) &= P \exp\biggl[ -\img g \int_{-\infty}^0 \df s\, \bn \cdot \cA(x^\mu + s \bn^\mu) \biggr]
\,,\nn\\
 S_n(x^\mu) &= P \exp\biggl[ -\img g \int_{-\infty}^0 \df s\, n \cdot \cA(x^\mu + s n^\mu) \biggr]
\,,\end{align}
as well as finite-length gauge links with transverse paths, 
\begin{align}  \label{eq:Tgaugelinks}
  W_{T}(x^\mu;\vec b_T,\vec 0_T) &= 
   P \exp\left[ \img g \int_{\vec 0_T}^{\vec b_T} \df \vec s_T \cdot \vec \cA_T(x^\mu + s_T^\mu) \right]
  = 
  S_{T}(x^\mu;\vec b_T,\vec 0_T)
  \,.
\end{align}
The Wilson paths for matrix elements in \eqs{beamfunc}{softfunc} are shown in the $(b^+,b^-,\bt)$ plane in \fig{wilsonlines}.
Note that the transverse gauge links at light-cone infinity create a closed path for the soft function, and a connected path between quark fields for the beam function, thus yielding matrix elements that are gauge invariant for both $B_q$ and $S^q$. In nonsingular gauges
such as Feynman gauge where the gluon field strength vanishes at infinity, the transverse gauge links can often be neglected, but are known to be important in certain singular gauges, see e.g.\ \mycites{Ji:2002aa,Belitsky:2002sm,Idilbi:2010im,GarciaEchevarria:2011md}.
These gauge links are also important for constructing the analogues of
\eqs{beamfunc}{softfunc} on a  finite-size lattice.

Note that the inclusion of a rapidity regulator can in principle spoil the gauge invariance of the matrix elements in \eqs{beamfunc}{softfunc}.
Gauge invariance trivially holds for regulators only affecting the Wilson line paths, as for example Collins' regulator \cite{Collins:1350496}, the exponential regulator~\cite{Li:2016ctv}, or the finite-length Wilson lines to be introduced in \sec{towards_qtmd} for lattice calculations.
Gauge invariance has been explicitly shown to hold in the $\tau\to 0$ limit for the $\eta$ regulator \cite{Chiu:2012ir}, the analytic phase space regulator of \mycite{Becher:2011dz}, while it is known to be violated in for TMDs with the analytic regulator used in \mycite{Becher:2010tm}. 
Gauge invariance is also known to be more tricky for the $\Delta$ regulator, where the limit $\Delta\to 0$ is also required~\cite{Echevarria:2015usa,Echevarria:2015byo}, and individual beam function matrix elements may only be gauge invariant after including 0-bin subtractions~\cite{Chiu:2009yx}.

Lastly, note that extracting TMDPDFs from lattice QCD (or experiment)
is only necessary for nonperturbative $q_T \sim b_T^{-1} \sim \LQCD$.
For perturbative values, one can instead perform an operator-product expansions
to match the TMDPDF, or equivalently the beam function,
onto the collinear PDF \cite{Collins:1981uw,Collins:1984kg},
\begin{align} \label{eq:tmdpdf_matching}
f_i^\TMD(x, \bt, \mu , \zeta) = \sum_j \int_x^1 \frac{\df y}{y} C_{ij}\Bigl(\frac{x}{y}, \bt, \mu, \zeta\Bigr) f_j(y, \mu)
+ \cO(b_T \LQCD)
\,.\end{align}
Here, $C_{ij}$ are perturbative matching kernels that are known to NNLO
\cite{Catani:2011kr,Catani:2012qa,Gehrmann:2014yya,Luebbert:2016itl,Echevarria:2015byo,Echevarria:2016scs},
and even to N$^3$LO for the soft contribution \cite{Li:2016ctv}.
The nonperturbative input is given solely by the standard longitudinal PDFs.
Throughout this work, we will limit our discussion to the case $b_T \sim \LQCD^{-1}$,
where \eq{tmdpdf_matching} cannot be applied.

\subsection{Rapidity Divergences in TMDs}
\label{sec:rapidity_divergences}

\begin{figure}[tp]
 \centering
 \includegraphics[width=0.55\textwidth]{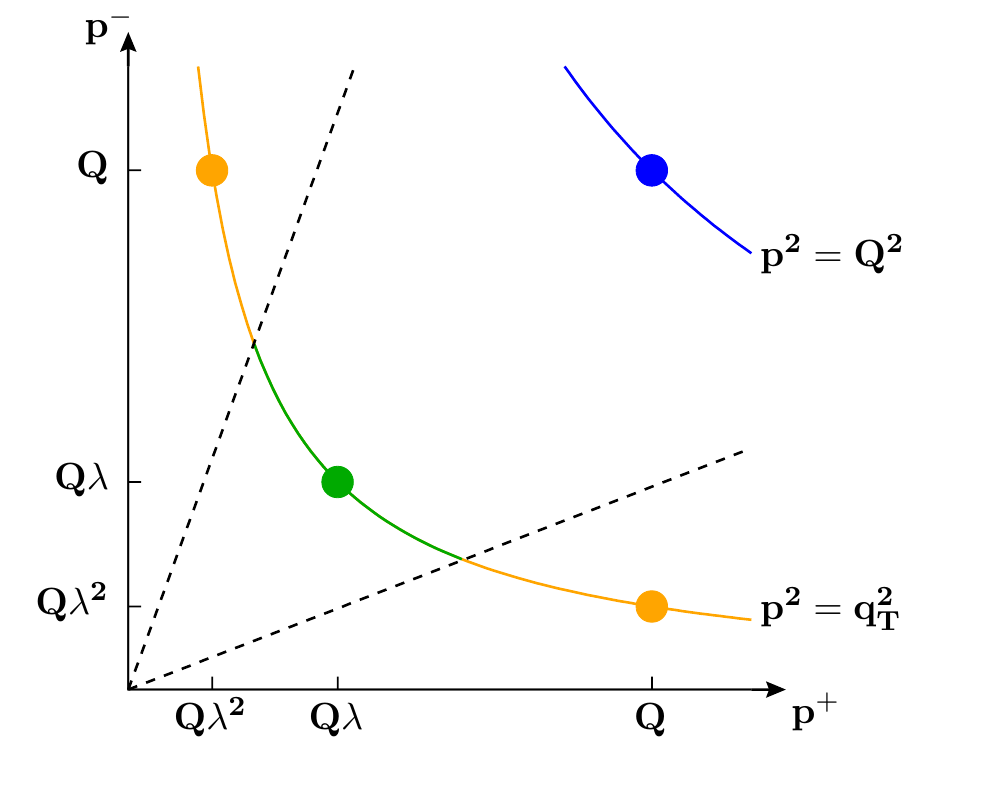}
 \caption{Illustration of the collinear (orange) and soft (green) modes contribution to the $\qt$ measurement.
  The hard modes (blue) describe offshell modes producing the energetic final state.
  The dashed lines indicate that the degeneracy in $p^+ p^-$ has to be resolved
  to properly define separate collinear and soft functions.}
 \label{fig:modes}
\end{figure}

Quantum corrections to the beam and soft function defined in \eqs{beamfunc}{softfunc}
lead to two types of divergences: ordinary UV and IR divergences that can be regulated
using dimensional regularization (and if desired a different IR regulator),
and so-called \emph{rapidity} divergences requiring a dedicated regulator
\cite{Collins:1981uk,Collins:1992tv,Manohar:2006nz,Collins:2008ht,Becher:2010tm,GarciaEchevarria:2011rb,Chiu:2011qc,Chiu:2012ir}.
Rapidity divergences arise because $n$- and $\bn$-collinear beam functions and the soft function,
or equivalently the collinear and soft Wilson lines, are defined to describe modes with momenta scaling as
\begin{align}
 B_n:& 
 &p_n& \sim Q (\lambda^2, 1, \lambda)
 \quad \text{so that} \quad
 &p_n^-& \gg p_{n \perp} \sim q_T \gg p_n^+
\,,\nn\\
 B_\bn:& 
 &p_\bn& \sim Q (1, \lambda^2, \lambda)
 \quad \text{so that} \quad
 &p_\bn^+& \gg p_{\bn \perp}  \sim q_T \gg p_{\bn}^-
\,,\nn\\
 S:&
 &p_s& \sim Q (\lambda, \lambda, \lambda)
 \quad \text{ so that } \quad
 &p_s^-& \sim\, p_{s \perp}  \,\sim q_T \sim p_s^+
\,,\end{align}
where we follow the usual SCET conventions, and use the light-cone notation $p = (p^+, p^-, p_T)$, see \app{conventions}, and $\lambda \sim q_T / Q\ll 1$.
These modes indicate approximations that are used to derive the corresponding operators and their Feynman rules.
This is illustrated in \fig{modes}, where the orange dots denote the dominant region for the $n$- and $\bn$-collinear modes,
and the green dots denotes that for the soft modes. 
Hard modes mediating the underlying hard process are shown in blue.
Although these dots indicate the dominant momentum region, in matrix elements the corresponding fields are still integrated over all momenta.
Since collinear and soft momenta have the same virtuality
\begin{equation}
 p_n^2 \sim p_\bn^2 \sim p_s^2 \sim q_T^2
\,,\end{equation}
which in \fig{modes} is shown by lying on the same hyperbola,
the soft and collinear momentum regions are only distinguished by their rapidity $y = 1/2 \ln(k^-/k^+)$.
In calculations using regulators such as dimensional regularization, which only regulate virtualities, one 
can therefore encounter additional rapidity divergences which arise in soft and collinear matrix elements when integrating over all $y$, 
and have to be resolved using a dedicated regulator.
This is indicated schematically by the dashed lines in \fig{modes} that split the hyperbola.
These divergences can also be understood as a conformal mapping from UV divergences in soft multi-parton scatterings \cite{Vladimirov:2016dll,Vladimirov:2017ksc},

To give a concrete example of how rapidity divergences appear in perturbative calculations,
consider the following integral that appears in calculations of the soft factor,
\begin{align} \label{eq:illustration_rapidity_div}
I_{\rm div} = \int \df k^+ \df k^- \frac{f(k^+ k^-)}{(k^+ k^-)^{1+\eps}}
 = \frac{1}{2} \int \frac{\df(k^-/k^+)}{k^-/k^+} \int \df (k^+ k^-) \frac{f(k^+ k^-)}{(k^+ k^-)^{1+\eps}}
\,.\end{align}
Here the integrand only depends on the product $k^+ k^-$.
Singularities as $k^+ k^- \to 0$ or $k^+ k^- \to \infty$ are clearly regulated by dimensional regularization,
while the integration over $k^-/k^+$ is unconstrained, leading to a divergence that requires a separate regulator.
Since $k^- / k^+ = e^{2 y}$ is directly related to the rapidity $y$ of $k$,
these are often referred to as rapidity divergences that require an additional rapidity regulator.
Alternatively, these are sometimes referred to as light-cone divergences,
as they can also be avoided by displacing the light-cone propagators $1/k^+$ and $1/k^-$ away from the light-cone.

To render the beam and soft functions well defined, integrals such as \eq{illustration_rapidity_div} require an additional regulator. A large variety of regulators has been suggested in the literature, see \mycites{Collins:1981uk,Collins:1350496,Ji:2004wu,Beneke:2003pa, Chiu:2007yn, Becher:2011dz,Chiu:2011qc, Chiu:2012ir,Chiu:2009yx, GarciaEchevarria:2011rb,Li:2016axz,Ebert:2018gsn}.
The key idea in all rapidity regulators is to regulate the behavior as $y = 1/2 \ln(k^-/k^+) \to \pm \infty$,
which in the example of \eq{illustration_rapidity_div} amounts to lifting the dependence on $k^+ k^-$ only.
For example, in the $\eta$ regulator of \mycites{Chiu:2011qc, Chiu:2012ir} one regulates the integral \eq{illustration_rapidity_div} by inserting a factor $|2k^z/\nu|^{-\eta}$, such that
\begin{align} \label{eq:illustration_rapidity_div_2}
I_{\rm div} = \int \df k^+ \df k^- \frac{f(k^+ k^-)}{(k^+ k^-)^{1+\eps}}
 \to \int \df k^+ \df k^- \frac{f(k^+ k^-)}{(k^+ k^-)^{1+\eps}} \biggl|\frac{k^- - k^+}{\nu}\biggr|^{-\eta}
\,.\end{align}
Divergences as $k^-/k^+ \to 0$ or $k^-/k^+\to\infty$ are thus made manifest as poles in $\eta$.
Similar to dimensional regularization, one introduces a new dimensionful scale $\nu$ to keep the regulator scaleless,
and a parameter $\eta$ to be taken to zero at the end of the calculation, analogous to the limit $\eps\to0$.
In this way, renormalized beam and soft functions acquire an additional scale dependence, namely $\nu$,
or equivalently the bare functions entering \eq{tmdpdf1b} depend on $\eta$.
The chosen rapidity regulator determines the precise form of this dependence on $\nu$ and $\eta$, respectively.
As discussed, this dependence cancels in the combination $f_i^\TMD = B_i \sqrt{S^i}$,
but the cancellation leaves a residual dependence on the scale $\zeta$.
Intuitively, the definitions of $\zeta_a$ and $\zeta_b$ in \eq{zeta} reflects how the hyperbola in \fig{modes}
is split between soft and collinear modes, see e.g.\ \mycite{Echevarria:2012js} for more details.

\subsection{Evolution of TMDPDFs}
\label{sec:evolution}

The cross section in \eq{sigma} must be independent of the unphysical scale $\mu$ and of the precise choice of $\zeta_a$ and $\zeta_b$ as long as $\zeta_a \zeta_b = Q^4$.
This induces renormalization group equations (RGEs) that encode the dependence of hard, beam and soft functions,
or equivalently hard functions and TMDPDF, on these scales.
Here, we only discuss the RGEs for the TMDPDF, which is the object of interest in this work.
Details on the separate evolution of beam and soft functions can be found e.g.\ in \mycites{Chiu:2012ir,Ebert:2016gcn} and in \app{eta_regulator}.
The RGEs for $f_i^\TMD$ read
\begin{align} \label{eq:TMD_RGEs}
\mu \frac{\df}{\df\mu} f^\TMD_i(x, \bt, \mu, \zeta) &
= \gamma^i_\mu(\mu,\zeta) f^\TMD_i(x, \bt, \mu, \zeta)
\,,\nn\\
\zeta \frac{\df}{\df\zeta} f^\TMD_i(x, \bt, \mu, \zeta) &
= \frac{1}{2} \gamma^i_\zeta(\mu,b_T) f^\TMD_i(x, \bt, \mu, \zeta)
\,,\nn\\
 \mu \frac{\df}{\df\mu} \gamma_\zeta^i(\mu,b_T)
 &= 2 \zeta \frac{\df}{\df\zeta} \gamma_\mu^i(\mu,\zeta) = -2 \GammaC^i[\as(\mu)]
\,,\end{align}
where $\GammaC^i[\as]$ is the cusp anomalous dimension.
The second equation in \eq{TMD_RGEs} is known as the Collins-Soper equation \cite{Collins:1981va,Collins:1981uk}.
The subscripts $\mu$ and $\zeta$ on the anomalous dimensions $\gamma_\mu^i$ and $\gamma_\zeta^i$ denote the scale evolution they govern, and the superscript $i = q,g$ distinguishes quarks from gluons.
These anomalous dimensions are defined as
\begin{align}
 \gamma_\mu^i(\mu,\zeta) &
 = \frac{\df\ln f_i^\TMD}{\df\ln\mu}
 = \frac{\df\ln B_i}{\df\ln\mu}
   + \frac12 \frac{\df\ln S^i}{\df\ln\mu}
\,,\nn\\ \label{eq:gamma_zeta}
 \gamma_\zeta^i(\mu,b_T) &
 = 2 \frac{\df\ln f_i^\TMD}{\df\ln\zeta}
 = - \frac{\df\ln B_i}{\df\ln\nu}
 = \frac12 \frac{\df\ln S^i}{\df\ln\nu}
\,.\end{align}
Here, we used that the $\nu$-dependence in $f^\TMD = B \sqrt{S}$ cancels 
and that $B$ only depends on the combination $\nu^2/\zeta$,
and thus $\gamma_\zeta^i$ can be equivalently obtained from either $f^\TMD$ or $B$ or $S$.
Since the definition of $S$, \eq{softfunc}, is independent of the hadron state entering the TMDPDF and the light quark flavor,
this immediately implies that $\gamma_\zeta^i$ is independent of the choice of hadron state and light quark flavor as well.

The all-order forms of the anomalous dimensions are given by
\begin{align} \label{eq:anom_dims}
 \gamma_\mu^i(\mu,\zeta) &= \GammaC^i[\as(\mu)] \ln\frac{\mu^2}{\zeta} + \gamma_\mu^i[\as(\mu)]
\,,\nn\\
 \gamma_\zeta^i(\mu,b_T) &= -2 \int_{1/b_T}^\mu \frac{\df\mu'}{\mu'} \GammaC^i[\as(\mu')] + \gamma_\zeta^i[\as(1/b_T)]
\,,\end{align}
where $\gamma^i[\as]$ denotes the noncusp piece.

Note that $\gamma_\zeta^i(\mu,b_T)$ has an intrinsically nonperturbative component for $b_T \sim \LQCD^{-1}$, independently of the scale $\mu$, as is clear from \eq{anom_dims}.
However, once it has been nonperturbatively determined at a scale $\mu_0 \gg \LQCD$, it can be perturbatively evolved to any other scale $\mu \gg \LQCD$ using
\begin{align}
 \gamma_\zeta^i(\mu,b_T) = \gamma_\zeta^i(\mu_0,b_T) - 2\int_{\mu_0}^\mu  \frac{\df\mu'}{\mu'} \GammaC^i[\as(\mu')]
\,.\end{align}
The boundary term $\gamma_\zeta^i(\mu_0,b_T)$ is known to three loops for perturbative $b_T$ and $\mu_0$ \cite{Li:2016axz,Li:2016ctv,Vladimirov:2016dll}.

Combining the solutions to \eq{TMD_RGEs}, the TMDPDF can be evolved from arbitrary initial scales $(\mu_0,\zeta_0)$ to the desired final scales $(\mu,\zeta)$ through
\begin{align} \label{eq:tmd_evolution}
 f_i^\TMD(x, \bt, \mu, \zeta) = f_i^\TMD(x, \bt, \mu_0, \zeta_0)
 \exp\biggl[ \int_{\mu_0}^\mu \frac{\df\mu'}{\mu'} \gamma_\mu^i(\mu',\zeta_0) \biggr]
 \exp\biggl[ \frac12 \gamma_\zeta^i(\mu,b_T) \ln\frac{\zeta}{\zeta_0} \biggr]
\,.\end{align}
Here, we have chosen to first evolve $\zeta_0 \to \zeta$ at fixed $\mu_0$, and then $\mu_0 \to \mu$.
Since the evolution must be path independent, other choices are also possible,
see also \mycites{Chiu:2012ir,Scimemi:2018xaf} for a discussion of this path independence.
Also note that due to the $b_T$ dependence of $\gamma_\zeta$, the TMD evolution is severely more complicated in momentum ($\qt$) space \cite{Ebert:2016gcn}, which is part of the reason why the factorization is commonly written in $\bt$ space.

\eq{tmd_evolution} is crucial to relate the TMDPDF at reference scales $(\mu_0,\zeta_0)$,
where they are either measured or determined from lattice QCD, to the phenomenological scales $(\mu,\zeta)$.

The boundary term $f_i^\TMD(x,\bt,\mu_0,\zeta_0)$ in \eq{tmd_evolution} is by definition nonperturbative.
For perturbative $b_T \ll \LQCD^{-1}$, one can can match it onto the collinear PDF, see \eq{tmdpdf_matching},
thereby reducing all the nonperturbative input to these more well-studied PDFs.
In this case, the RG-evolved TMDPDF \eq{tmd_evolution} serves to resum large logarithms $\ln(\mu b_T)$ and $\ln(\zeta b_T^2)$, which otherwise spoil the perturbative convergence of the matching kernel $C_{ij}$ in \eq{tmdpdf_matching}.

In this paper, we are instead interested in obtaining nonperturbative TMDPDFs valid also for the case $b_T\sim \Lambda_{\rm QCD}^{-1}$,
where the boundary term $f_i^\TMD(x,\bt,\mu_0,\zeta_0)$ in \eq{tmd_evolution} is obtained from lattice QCD
(or equivalently is measured from experiment).
In this case, the values $\mu_0$ and $\zeta_0$ are fixed by the lattice calculation (or measurement),
and the role of \eq{tmd_evolution} is to evolve $f_i^\TMD(x,\bt,\mu_0,\zeta_0)$ to the scales $(\mu,\zeta)$ required in the phenomenological application.
This is completely analogous to the determination of collinear PDFs, which are extracted at a reference scale $\mu_0$ and then evolved via the DGLAP evolution.
Similar to the DGLAP evolution, the $\mu$ evolution of the TMDPDF encoded in the first exponential in \eq{tmd_evolution} is perturbative as long as $\mu_0$ and $\mu$ are perturbative.
In contrast, the $\zeta$ evolution intrinsically contains a nonperturbative component $\gamma_\zeta^i(\mu,b_T)$ for $b_T \sim \LQCD^{-1}$,
even if the TMDPDF is extracted at perturbative $(\mu_0,\zeta_0)$.
Thus one needs to determine both $f_i^\TMD(x,\bt,\mu_0,\zeta_0)$ and $\gamma_\zeta(\mu, b_T)$ nonperturbatively.
In particular, the nonperturbative determination of $\gamma_\zeta(\mu,b_T)$ is a phenomenologically relevant task on its own, as it provides valuable information on its all-order structure.
For example, it is known to suffer from renormalons at large $b_T$ \cite{Scimemi:2016ffw}.
A dedicated discussion of the extraction of $\gamma_\zeta(\mu, b_T)$ from lattice QCD, exploiting some of the results discussed in this paper, has been given in \mycite{Ebert:2018gzl}.

\section{Towards Constructing quasi-TMDPDFs}
\label{sec:towards_qtmd}

The goal of this work is to define a quasi-TMDPDF $\tilde f^\TMD$ which involves matrix elements that are calculable with lattice QCD,
and which can be matched onto the TMDPDF $f^\TMD$ that is relevant for collider phenomenology.
As reviewed in \sec{tmd_review}, TMDPDFs are constructed from hadronic and vacuum matrix elements,
namely the beam function $B_i$ and the soft function $S^i$.  Both of these functions are sensitive to infrared physics for $b_T\sim \LQCD^{-1}$.
According to the LaMET approach, to calculate the TMDPDF in lattice QCD we need to construct quasi observables with the same infrared physics, which we refer to as a ``quasi beam function'' $\tilde B_i$ and ``quasi soft function'' $\tilde S^i$.

In general both beam and soft functions can be rapidity divergent.
On the lattice, we will see that the analog of rapidity divergences are regulated by a finite length $L$ of the Wilson lines, while in the lightlike case many regulators have been suggested (see \sec{tmd_review} and \app{overview_tmdpdfs} for more details).

In principle, one could envision separately matching the quasi beam and quasi soft functions onto the lightlike beam and lightlike soft functions, and then combining the matched results into the physical TMDPDF.
However, such individual matching results necessarily depend on the choice of rapidity regulators, and furthermore since these regulators break boost invariance, some of them will likely spoil the boost relation between quasi and lightlike functions. For the physical TMDPDF the choice of the rapidity regulator has a significant impact on the form of the beam and soft functions, 
including their infrared logarithms, so at minimum the individual matching would have to be worked out separately for different schemes.
Finally, due to the fact that $L$ plays the role of a rapidity regulator for quasi distributions on the lattice, there could be non-trivial $L$ dependence in intermediate stages of the matching result.
For these reasons taking an approach of individually matching beam and soft functions is not preferred.

As discussed in \sec{tmd_factorization}, it is equivalent to instead combine the unrenormalized beam and soft functions, in which case the rapidity divergences cancel, and then perform the UV renormalization.
This should hold for the quasi functions as well. Hence the more straightforward approach, adopted here, is to combine the unrenormalized quasi beam and quasi soft functions into a quasi-TMDPDF, in analogy to \eq{tmdpdf1b}.
This approach also has the advantage of canceling out all $L$ dependence in the quasi-TMDPDF calculation, up to power suppressed terms. Of course, if the resulting quasi-TMDPDF fails the infrared consistency test, by not yielding infrared logarithms that are consistent with those in the TMDPDF, then, as we will see, it will still be advantageous to examine the contributing quasi beam and quasi soft functions to determine where the issues lie.

An additional complication in the lattice computation is the appearance of power law divergences from Wilson line self energies that have to be subtracted.
Since the self energies are also proportional to the length of the Wilson lines, and $L$ dependence cancels when combining $\tilde B_i$ and $\tilde S^i$, this gives a $b^z$-dependent contribution to be removed by the UV renormalization. Here $b^z$ is the Fourier transform variable to $xP^z$.
Since this contribution is multiplicative in $b^z$ position space, it is most natural to perform the UV renormalization in position space.
We thus define the quasi-TMDPDF in the $\MS$ scheme analogous to \eq{tmdpdf1b} as
\begin{align} \label{eq:qtmdpdf}
 \tilde f_i^\TMD(x, \bt,\mu,P^z) = \int \frac{\df b^z}{2\pi} \, e^{\img b^z (x P^z)}\,
 &\tilde Z'_i(b^z,\mu,\tilde \mu) \tilde Z_{\rm uv}^i(b^z,\tilde \mu, a)
 \nn\\&\times
 \tilde B_i(b^z, \bt, a, L, P^z) \tilde\Delta_S^i(b_T, a, L)
\,.\end{align}
Here, $\tilde B_i \equiv \tilde B_i^\unsub $ and $\tilde \Delta_S^i$ are the quasi beam and quasi soft function, which remain to be constructed.
They are the analogs of the unsubtracted beam function and the soft factor in \eq{tmdpdf1b}.
We will always consider the unsubtracted beam function and absorb the zero-bin subtraction factor into $\tilde \Delta_S^i$, so for simplicity we drop the superscript ``(unsub)''.
$\tilde Z_{\rm uv}$ is the lattice renormalization constant, and $\tilde Z'$ converts from the lattice renormalization scheme to the $\MS$ scheme. These schemes are typically distinct, and hence we distinguish the lattice renormalization scale $\tilde \mu$ from the $\MS$ scale $\mu$.
The finite lattice spacing $a$ takes the role of $\eps$ as an UV regulator.
On the lattice, one also has to truncate the Wilson lines that enter $\tilde B_i$ and $\tilde \Delta_S^i$ at a finite length $L$.
As we will discuss in \sec{finite_L}, having a finite $L$ regulates the analog of rapidity divergences on lattice and thus the $\tau$ dependence is replaced by additional dependence on $L$.
The $L$ dependence associated with both the rapidity regularization and finite length of Wilson lines cancels between $\tilde B_i$ and $\tilde \Delta_S^i$, and hence $\tilde f_i^\TMD$ does not depend on $L$.
(In practice, there will be a finite $L$ dependence from power corrections that vanishes in the limit $L\to\infty$, and we suppress these in our notation.)
Finally, $\tilde f_i^\TMD$ also depends on the proton momentum $P^z$, which encodes short distance ultraviolet effects like in the quasi-PDF, and in addition acts as the analog of its $\zeta$ scale.

\subsection{Constraints on the matching relation}
\label{sec:schematic_matching}

The main focus of this paper is to give constructions of $\tilde B_i$ and $\tilde \Delta_S^i$ such that $\tilde f_i^\TMD$ can be perturbatively matched onto the TMDPDF $f_i^\TMD$. 
In this section we \emph{assume} such a perturbative matching exists in order to constrain the general form it has to take.

Given the physical scales present in the calculation,
the matching relation is expected to take the schematic form
\begin{align} \label{eq:matching_schematic}
 \tilde f_i^\TMD(x, \bt, \mu, P^z) &
 \sim \sum_j \int_{-1}^1 \frac{\df y}{|y|} \, C^\TMD_{ij}\bigl(x, y, \mu, P^z, \zeta \bigr) f^\TMD_{j}(y, \bt, \mu, \zeta)
 \nn\\&\quad
 + \cO\biggl(\frac{b_T}{L}, \frac{1}{b_T P^z}, \frac{1}{P^z L}\biggr)
\,,\end{align}
where $i,j=q,g$ are the parton flavors and $f^\TMD_{j}$ for negative $y$ corresponds to the antiquark distribution, $f^\TMD_{j}(-y,\ldots)=f^\TMD_{\bar j}(y,\ldots)$. 
The precise dependence of the matching kernel $C_{ij}^\TMD$ on its arguments is not obvious {\it a priori},
see for example \eq{fimatch} for the exact form for collinear PDFs.
In \eq{matching_schematic} we have kept things generic and allowed for dependence on $x$ and $y$ to match the parton momenta.
Since we have already converted the quasi-TMDPDF in \eq{qtmdpdf} into the $\MS$ scheme,
$C^\TMD_{ij}$ only depends on the $\MS$ scale $\mu$ and not on the lattice renormalization scale $\tilde\mu$.
The kernel also depends on the finite proton momentum $P^z$ and the $\zeta$ scale that enters the TMDPDF $f_j^\TMD$.

Given the physical picture underlying LaMET, we can also write down the
expected power corrections to the matching \eq{matching_schematic},
obtained from the assumed hierarchy of scales
\begin{equation}
 b^z \sim \frac{1}{P^z} \ll b_T \ll L
\,,\qquad
 b_T \sim \LQCD^{-1}
\,.\end{equation}
This scaling is motivated as follows:
First, the lattice box size should be large enough so that the Wilson line length $L$ is the largest length scale and finite $L$ effects and finite volume effects are suppressed.
Second, we assume $b_T \sim \LQCD^{-1} $ to be a nonperturbative scale
of the order of the characteristic size of transverse fluctuations of constituents in the hadron.
Lastly, LaMET assumes large $P^z$ to approximate a lightlike correlator by boosting an equal-time correlator, so $1/P^z$ should be the smallest length scale in the system other than the lattice spacing $a$. Since for power counting we take $x$ as $\cO(1)$ this also implies that $b^z\sim 1/P^z$ is small.

Note that for a matching formula like \eq{matching_schematic} to exist, $C^\TMD_{ij}$ must \emph{not} depend on $b_T$, since $b_T$ encodes infrared physics and is assumed to be a nonperturbative scale.
This immediately implies the necessary condition that in perturbation theory,
the $b_T$ dependence of both quasi-TMDPDF and TMDPDF must agree, up to power corrections,
which will be our most stringent consistency test in the one-loop study in \sec{nlo_results}.

We can deduce further constraints on the matching relation from the Collins-Soper equation, and thus arrive at an improved schematic form for the matching relation. To do so we assume that $\zeta$ and $x P^z$ are independent variables, which could be achieved for example by considering a different hadronic momentum $\tilde P^z$ for the quasi-TMDPDF than the momentum $P^z$ used for the  TMDPDF.
In \eq{matching_schematic}, the $\zeta$ dependence must then cancel between $C^\TMD_{ij}$ and $f_j^\TMD$
to yield a $\zeta$-independent quasi-TMDPDF $\tilde f_i^\TMD$, which implies that
\begin{align}
\zeta \frac{\df}{\df \zeta} \ln C^\TMD_{ij}\bigl(x, y, \mu, \tilde P^z, \zeta \bigr)
\stackrel{?}{=} -\zeta \frac{\df}{\df \zeta} \ln f^\TMD_{j}(y, \bt, \mu, \zeta)
=-\frac12 \gamma_\zeta^j(\mu,b_T)
\,.\end{align}
This clearly violates the requirement that $C^\TMD_{ij}$ must be independent of $b_T$ to carry out short distance matching. 
This mismatch occurs because the rapidity evolution governed by $\gamma_\zeta^j(\mu,b_T)$ is also nonperturbative. To correct for this we therefore must consider the modified matching relation
\begin{align} \label{eq:matching_schematic2}
\tilde f_{i}^\TMD(x, \bt, \mu, \tilde P^z) &
\sim \sum_j \int_{-1}^1 \frac{\df y}{|y|} \, C^\TMD_{ij}\bigl[x, y, \mu, \tilde P^z, \tilde \zeta(x,\tilde P^z)\bigr]
\exp\biggl[\frac12 \gamma_\zeta^j(\mu,b_T) \ln\frac{\tilde \zeta(x,\tilde P^z)}{\zeta} \biggr]
\nn\\&\hspace{2cm}\times
f^\TMD_{j}(y, \bt, \mu, \zeta)
\,,\end{align}
where for brevity we suppress the power corrections which are the same as in \eq{matching_schematic}.
In \eq{matching_schematic2}, the $\zeta$ dependence cancels between the exponential and $f^\TMD$,
at the cost that the kernel $C_{ij}^\TMD$ now depends on the auxiliary scale $\tilde \zeta = \tilde \zeta(x, \tilde P^z)$.
As indicated, this scale must be fixed in terms of $x$ and $\tilde P^z$, the relevant scales that $\tilde f^\TMD$ depends on, and hence does not technically add additional functional dependence to $C_{ij}^\TMD$.  

In order to interpret \eq{matching_schematic2} as a true matching equation without any renormalization group evolution, it must be possible to make the exponential in \eq{matching_schematic2} vanish, to yield a purely perturbative relation between $\tilde f^\TMD$ and $f^\TMD$.
Thus, a perturbative matching can only be possible if one can choose $\tilde\zeta(x,\tilde P^z) = \zeta$ to cancel the Collins-Soper evolution to all orders in perturbation theory.  From \eq{zeta} for $f_j^\TMD(y,\bt,\mu,\zeta)$  we have $\zeta = (2 y P^z)^2$ (choosing $y_n=0$ without loss of generality). Accounting for dimensions we see that we must choose $\tilde P^z \propto P^z$, and since at tree level $C_{ij}^\TMD[x,y,\ldots] = \delta(1-x/y)$ this fixes the constant of proportionality so that $\tilde\zeta = (2 x P^z)^2$. With $\zeta\propto y^2$ and $\tilde\zeta=\tilde\zeta(x)$, the only possibility that allows for perturbative matching is to take $\tilde P^z=P^z$ and have $C_{ij}^\TMD[x,y,\ldots] \propto \delta(1-x/y)$ to all orders. (We will confirm this explicitly at one loop in \sec{nlo_results} below.)
With this constraint the schematic relationship between quasi-TMDPDF and TMDPDF becomes multiplicative in $x$ space,
\begin{align} \label{eq:matching_result0}
\tilde f_{i}^\TMD(x, \bt, \mu, \tilde P^z) &
 = \sum_j  C^\TMD_{ij}\bigl[x, \mu, \tilde P^z, \tilde \zeta(x,\tilde P^z)\bigr]
\exp\biggl[\frac12 \gamma_\zeta^j(\mu,b_T) \ln\frac{\tilde \zeta(x,\tilde P^z)}{\zeta} \biggr]
\nn\\&\hspace{2cm}\times
f^\TMD_{j}(x, \bt, \mu, \zeta)
\nn\\&\quad
 + \cO\biggl(\frac{b_T}{L}, \frac{1}{b_T P^z}, \frac{1}{P^z L}\biggr)
\,.\end{align}
To derive a perturbative formula for $C^\TMD_{ij}$ then requires choosing $\tilde P^z=P^z$ and $\zeta = \tilde \zeta = (2 x P^z)^2$. For this choice the effect of changing $\zeta$ in $f^\TMD_j$ is exactly balanced by a corresponding change of $P^z$ in $\tilde f_i^\TMD$. Note that the lack of an integral over $y$ in \eq{matching_result0} is analogous to the fact that no such integral appears in the renormalization group equations for the TMDPDF, see \eq{TMD_RGEs}. 

We will demonstrate that a quasi-TMDPDF can be defined such that the use of \eq{matching_result0} and matching with a short distance $C^\TMD_{qq}$ is  fully satisfied at one loop for quark quasi-TMDPDF and TMDPDFs.
However this is not automatic, and indeed we find that the most naive definition of a quasi-TMDPDF does not agree with \eq{matching_result0}.
An all orders derivation of a formula like \eq{matching_result0} would be  required to completely address this relationship, and is left for future work.

\subsection{Impact of finite-length Wilson lines}
\label{sec:finite_L}

\begin{figure}[tp]
 \centering
 \begin{subfigure}{0.45\textwidth}
  \includegraphics[width=\textwidth]{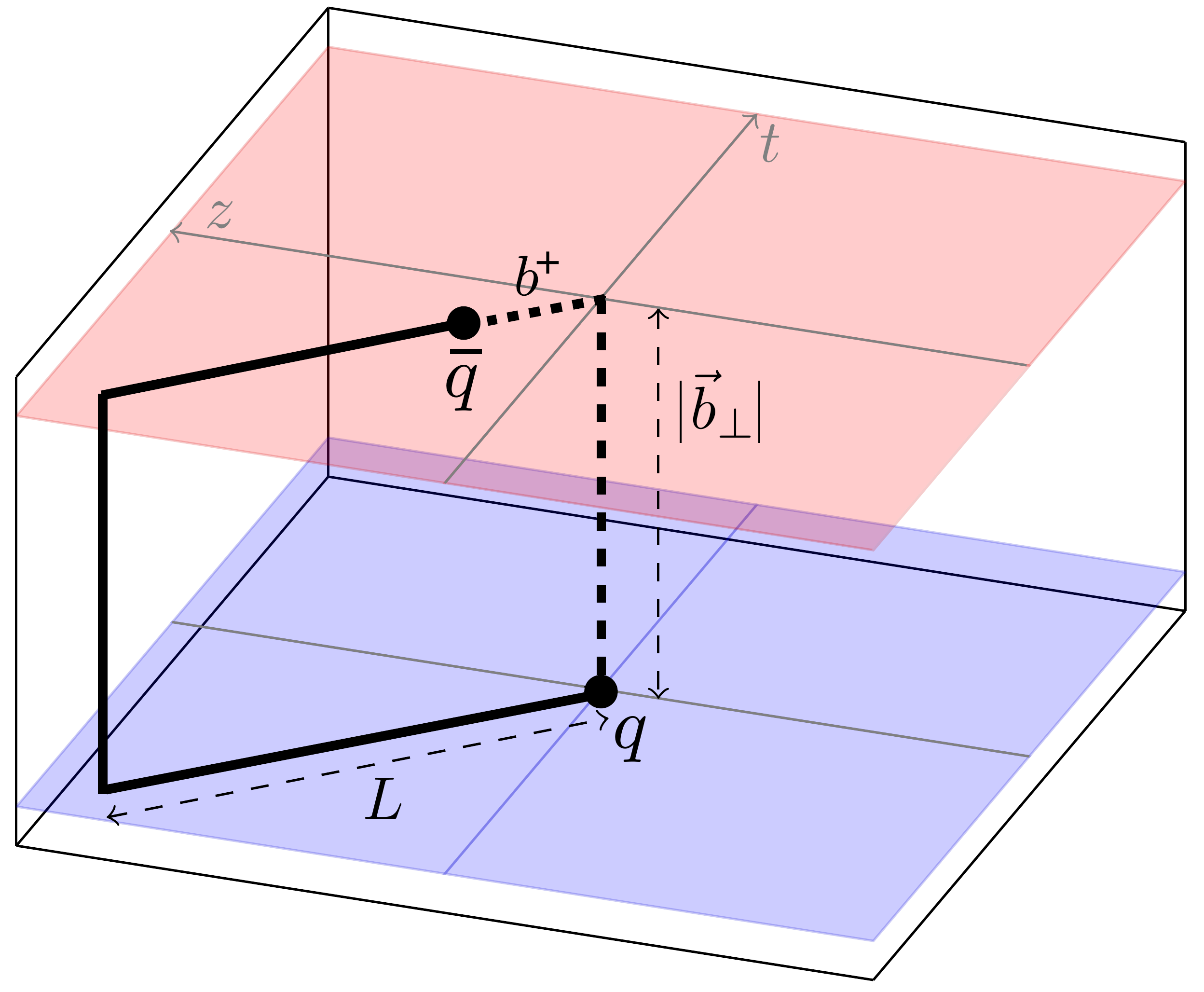}
  \caption{}
  \label{fig:wilsonlines_truncated_beam}
 \end{subfigure}
 \hfill
 \begin{subfigure}{0.45\textwidth}
  \includegraphics[width=\textwidth]{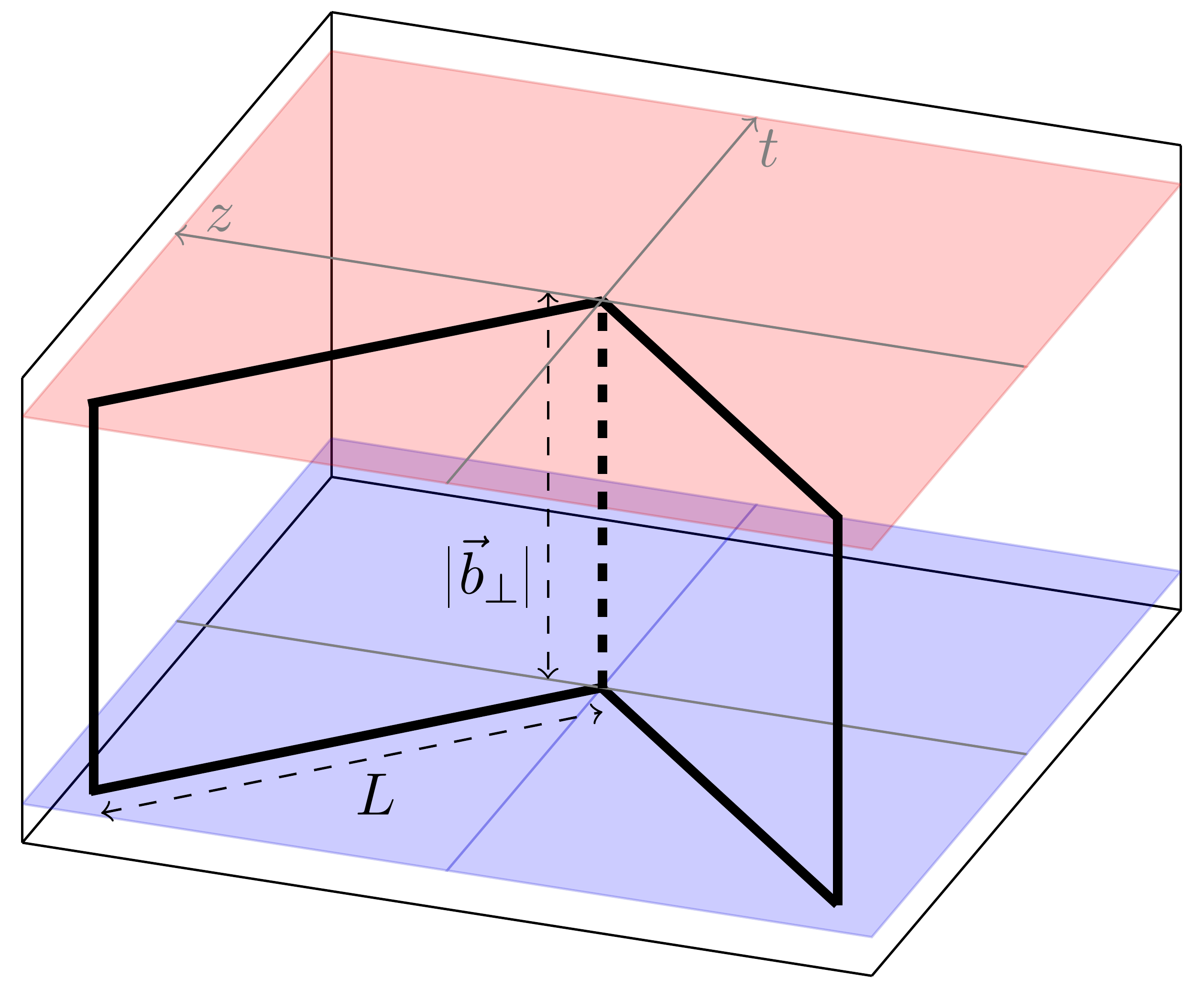}
  \caption{}
  \label{fig:wilsonlines_truncated_soft}
 \end{subfigure}
 \caption{Illustration of the Wilson line structure of the $n$-collinear beam function $B$ (a)
 and the soft function $S$ (b), with Wilson lines truncated at some finite length $L$.
 The corresponding Wilson line paths with infinite-long Wilson lines is shown in \fig{wilsonlines}.}
 \label{fig:wilsonlines_truncated}
\end{figure}

Before constructing quasi beam and soft functions, we discuss the impact of having infinitely-long Wilson lines
in the definitions \eqs{beamfunc}{softfunc} of the lightlike beam and soft functions.
On the lattice, the finite lattice size prevents infinitely-long Wilson lines.
Hence one has to truncate them at some finite length $L$, as illustrated in \fig{wilsonlines_truncated}.
Here, it is important to include transverse gauge links as required by the factorization theorem,
which ensures gauge invariance of the soft and collinear matrix elements.

Naively, one might assume that any effect of $L$ is suppressed as $b_T/L$ or $b^+/L$ for sufficiently large $L$.
In practice, the finite $L$ also regulates the analog of rapidity divergences on the lattice,
and hence yields divergences as $L\to\infty$.
While this has the advantage of not having to implement a dedicated rapidity regulator in the lattice calculation,
the drawback is that one cannot easily disentangle finite-$L$ effects from rapidity divergences.

To show that finite $L$ is a sufficient rapidity regulator, we first note that at leading power in $q_T^2/Q^2$, all emissions arise from Wilson lines and thus rapidity divergences can be regulated by modifying Wilson lines alone, as is done in most regulators in the literature.
(This does not hold at subleading power, where regulating Wilson lines alone is insufficient, see~\cite{Ebert:2018gsn}.)
Intuitively, by modifying the Wilson lines in such a way that boost invariance is broken,
for example by adjusting its geometry (taking it off the light cone or restricting its length)
or explicitly regulating the momentum flowing into the Wilson line,
one distinguishes collinear and soft modes and thus regulates the rapidity divergences.
To concretely show how this is achieved for finite $L$, consider the one-gluon Feynman rule for a Wilson line of size $L$ stretching along the $n$ direction, compared to its $L\to\infty$ limit,
\begin{align}
 g_s t^a n^\mu \frac{1 - e^{\img n \cdot k L}}{n \cdot k}
 \quad\stackrel{L\to\infty}{\longrightarrow}\quad
 g_s t^a n^\mu \frac{1}{n \cdot k + \img 0}
\,.\end{align}
For $n\cdot k\to0$, the limit where the emission of momentum $k^\mu$ becomes collinear to the Wilson line, i.e.\ it approaches infinite rapidity, is clearly regulated by exponential phase, whereas it is unregulated for $L\to\infty$.
This pattern continues for multiple gluon emission since the eikonal denominators are always in one-to-one correspondence with the regulating factors involving $L$ in the numerator.
The opposite limit is regulated analogously by an $\bn$-collinear Wilson of finite length.
For example, in \sec{tmd_review} a generic example of rapidity-divergent integral was discussed, \eq{illustration_rapidity_div}.
For finite $L$, the example integral changes to
\begin{align} \label{eq:illustration_rapidity_div_3}
 I_{\rm div} = \int \df k^+ \df k^- \frac{f(k^+ k^-)}{(k^+ k^-)^{1+\eps}}
 \to \int \df k^+ \df k^- \frac{f(k^+ k^-)}{(k^+ k^-)^{\eps}}
     \frac{1 - e^{\img k^+ L}}{k^+} \frac{1 - e^{- \img k^- L}}{k^-}
\,.\end{align}
Here we see that possible divergences as either $k^\pm \to 0$,
corresponding to the rapidity $y = \frac12 \ln(k^-/k^+) \to \pm \infty$, are regulated by having finite $L$,
and the leftover logarithmic divergence as either $k^\pm\to\infty$ is taken care of by dimensional regularization.

In our construction of the quasi functions on lattice, we will replace the lightlike Wilson lines by spacelike Wilson lines,
which affects the eikonal propagator, so the analog of \eq{illustration_rapidity_div_3} is
\begin{align} \label{eq:illustration_rapidity_div_4}
 \tilde I_{\rm div} &
 = \int \df k_0 \, \df k_z \frac{f(k_0^2 - k_z^2)}{(k_0^2-k_z^2)^{\eps}} \frac{1}{k_z^2}
 \to \int \df k_0 \, \df k_z \frac{f(k_0^2 - k_z^2)}{(k_0^2-k_z^2)^{\eps}}
 \frac{1 - e^{\img k^z L}}{k^z} \frac{1 - e^{-\img k^z L}}{k^z}
\,.\end{align}
Clearly, the exponentials regulate a possible divergence as $k_z \to 0$,
and thus play a similar role as in the lightlike case.
However, \eq{illustration_rapidity_div_4} contains a quadratic dependence on $k_z$ in the denominator, rather than the linear dependence on $k^+$ and $k^-$ in \eq{illustration_rapidity_div_3}.
Thus, we can also encounter linear divergences in $L$, as opposed to having only logarithmic divergences $\ln(L)$ in the lightlike case.

\subsubsection{Example: Lightlike soft function at NLO}
\label{sec:soft_L_NLO}

\begin{figure}[pt]
\centering
 \begin{subfigure}[t]{0.24\textwidth}
  \includegraphics[width=\textwidth]{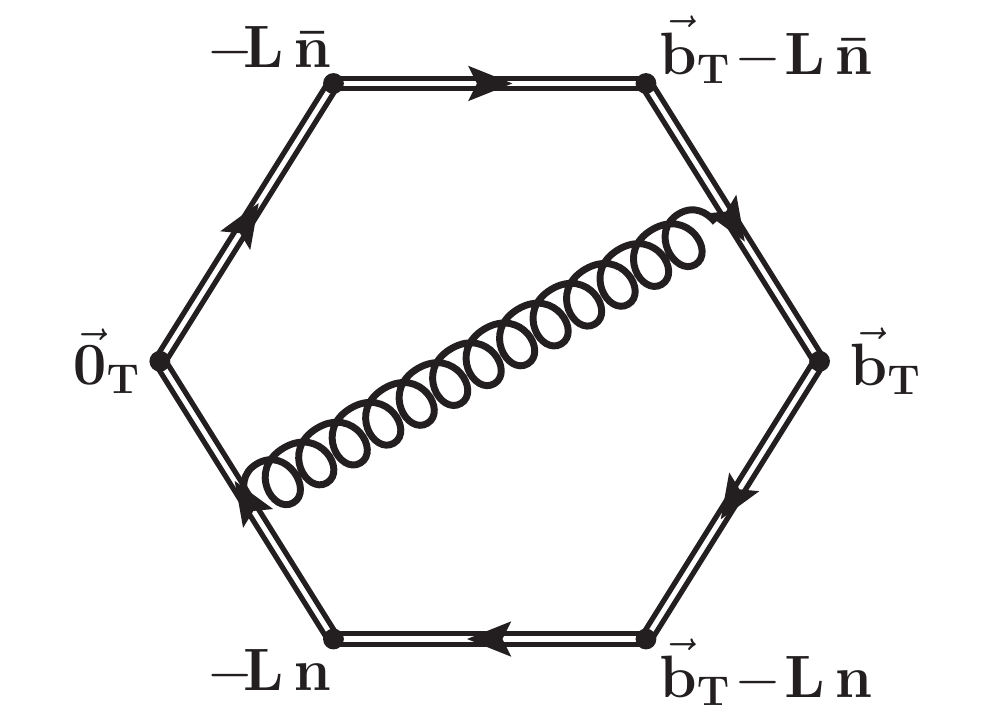}
  \caption{}
  \label{fig:soft_L_nlo_a}
 \end{subfigure}
 \begin{subfigure}[t]{0.24\textwidth}
  \includegraphics[width=\textwidth]{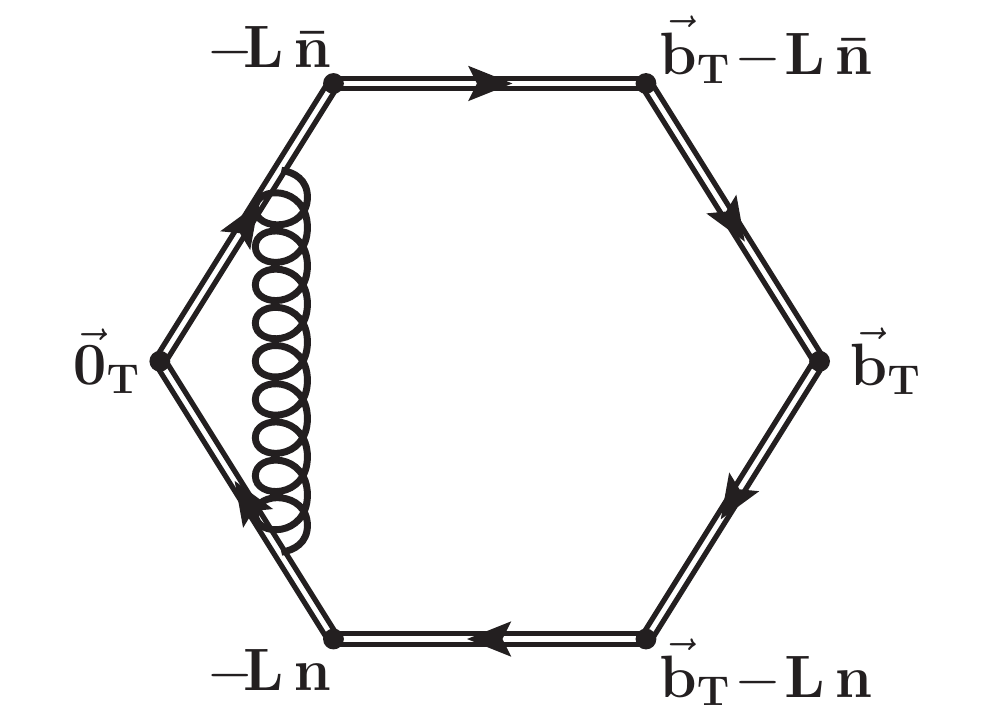}
  \caption{}
  \label{fig:soft_L_nlo_b}
 \end{subfigure}
 \begin{subfigure}[t]{0.24\textwidth}
  \includegraphics[width=\textwidth]{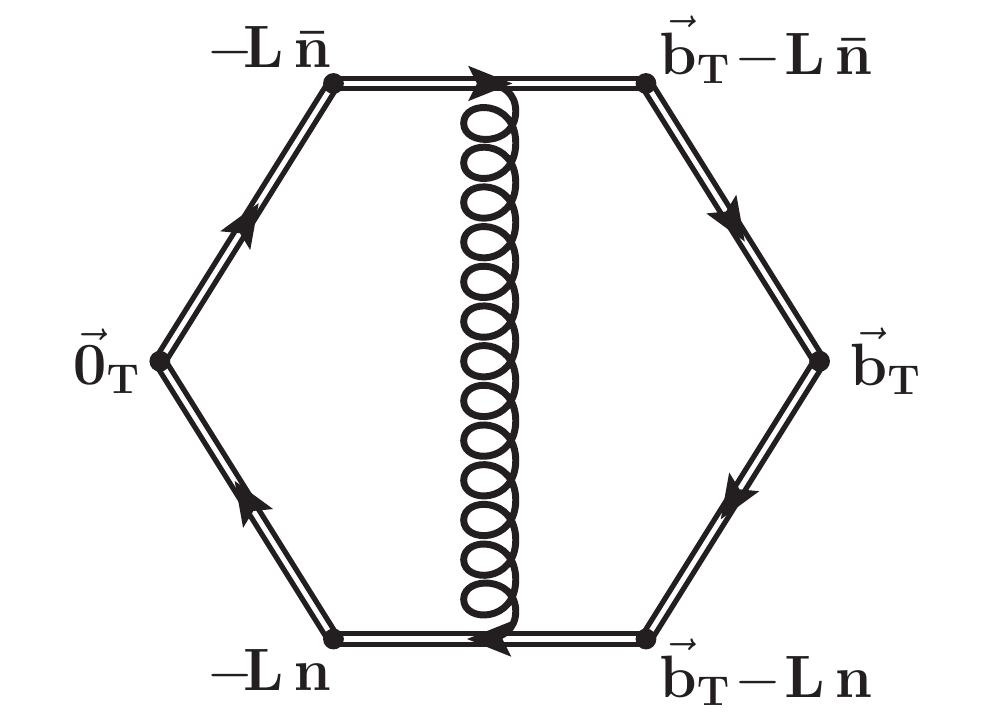}
  \caption{}
  \label{fig:soft_L_nlo_c}
 \end{subfigure}
 \begin{subfigure}[t]{0.24\textwidth}
  \includegraphics[width=\textwidth]{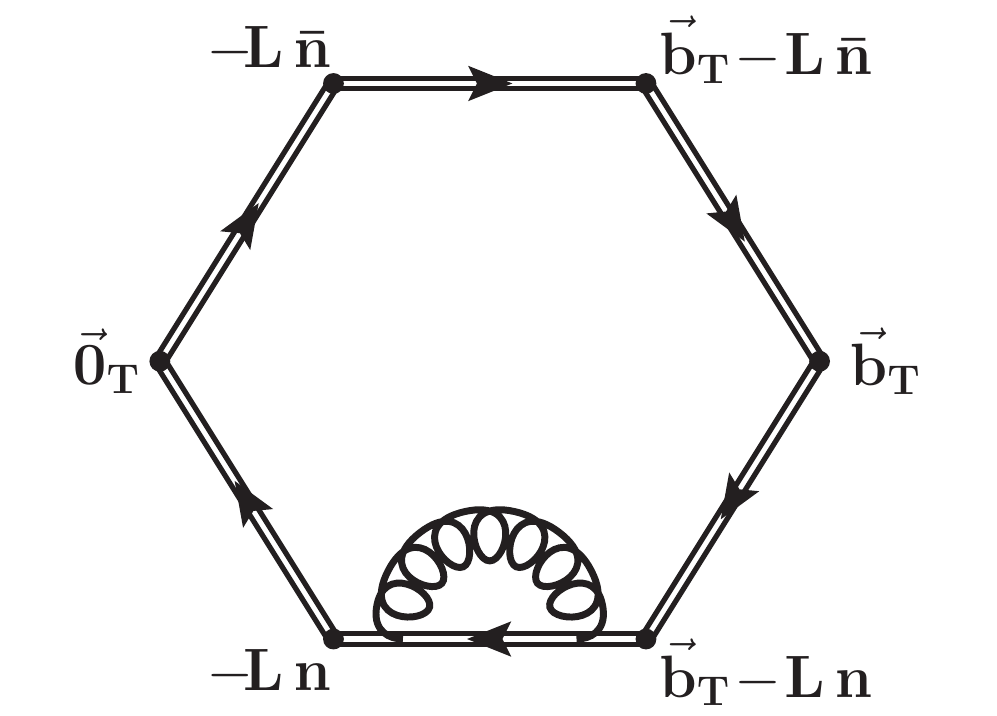}
  \caption{}
  \label{fig:soft_L_nlo_d}
 \end{subfigure}
 \caption{One loop diagrams for the TMD soft function with finite-length Wilson lines
  in Feynman gauge, up to mirror diagrams. The labels indicate the Wilson line paths in position space.}
 \label{fig:soft_L_nlo}
\end{figure}

To give a concrete example of the effect of finite $L$,
we consider in detail the lightlike soft function, defined in \eq{softfunc}, at one loop.
To account for the effect of finite lattice size, the Wilson lines along the $n$ and $\bn$ directions
are truncated at $L n$ and $L \bn$, respectively, and transverse gauge links are included,
as shown in \fig{wilsonlines_truncated_soft}.
In Feynman gauge, there are four relevant diagrams, shown in \fig{soft_L_nlo},
of which only (a) and (b) have rapidity divergences, while (c) and (d) do not.

\paragraph{Rapidity-divergent diagrams.}
Let us first discuss \fig{soft_L_nlo_a}, where a gluon connects the Wilson lines separated by the transverse displacement $\bt$.
Together with its mirror diagram, it is given by
\begin{align} \label{eq:soft_L_a}
 S_a(b_T,\eps,L) &
 =  2 g^2 C_F (n\cdot\bn) \mu_0^{2\eps} \int\frac{\df^d k}{(2\pi)^d}
   \frac{1 - e^{\img n \cdot k L}}{n \cdot k} \frac{1 - e^{- \img \bn \cdot k L}}{-\bn \cdot k} e^{-\img \bt \cdot \kt}
   \frac{-\img}{k^2+\img0}
\nn\\&
 = - \frac{g^2 C_F  \mu_0^{2\eps}}{(2\pi)^{2-2\eps}}
   \int_0^\infty \df k^+ \int_0^\infty \df k^- \, \frac{J_0\bigl(b_T \sqrt{k^+ k^-}\bigr)}{(k^+ k^-)^{\eps}}
   \frac{1 - e^{\img k^+ L}}{k^+} \frac{1 - e^{- \img k^- L}}{k^-}
\nn\\&
 = \frac{\as C_F}{2\pi} \biggl[ \ln^2\frac{b_T^2}{4 L^2}
   + 2 \Li{2}\biggl(\!-\frac{b_T^2}{4 L^2}\biggr) + \frac{\pi^2}{3} \biggr]
\,.\end{align}
In the second line, one can see how keeping $L<\infty$ regulates divergences
as the light-cone coordinates $k^\pm$ approach zero, thereby regulating the whole integral.
In the final result, the rapidity divergences are then reflected as a double logarithm in $b_T/L$.

The second rapidity-divergent diagram, \fig{soft_L_nlo_b} and its mirror diagram are independent of $b_T$, and are given by
\begin{align} \label{eq:soft_L_b}
 S_b(b_T,\eps,L) &
 = \frac{\as C_F}{2\pi} \biggl[ -\frac{2}{\eps^2} - \frac{2}{\eps} \ln\frac{\mu^2 L^2}{e^{-2\gamma_E}}
      - \ln^2\frac{\mu^2 L^2}{e^{-2\gamma_E}} - \frac{\pi^2}{6} \biggr]
\,.\end{align}
Together, we obtain
\begin{align}
 S_{a+b}(b_T,\eps,L)
 = \frac{\as C_F}{2\pi} \biggl[& - \frac{2}{\eps^2}
   - 2 \left(\frac{1}{\eps} + \Lb{}\right) \ln\frac{\mu^2 L^2}{e^{-2\gamma_E}}
   + \Lb{2} + \frac{\pi^2}{6}
   \nn\\&
   + \underbrace{2 \Li{2}\biggl(\!-\frac{b_T^2}{4 L^2}\biggr)}_{\stackrel{L\gg b_T}{\longrightarrow} 0}
   \biggr]
\,,\end{align}
where we have defined $b_0=2 e^{-\gamma_E}$.
The rapidity logarithms $\ln(\mu L)$ are manifest, while the last term is an example of a finite-$L$ contribution
that vanishes in the limit $b_T \ll L$.
Let us also compare this to the soft function using the $\delta$ regulator \cite{Echevarria:2012js} [see also \eq{soft_nlo_delta}]
\begin{align}
 S^{(1)}\bigl(b_T,\eps, \sqrt{\delta^+ \delta^-}\bigr) &= \frac{\as C_F}{2\pi} \biggl[
  - \frac{2}{\eps^2} - 2 \left(\frac{1}{\eps} + \Lb{}\right) \ln\frac{\mu^2}{\delta^+\delta^-}
  + \Lb{2} + \frac{\pi^2}{6}  \biggr]
\,.\end{align}
The two expression agree upon identifying $1/L^2 = e^{2\gamma_E} \delta^+ \delta^-$,
showing that for this function the two regulators are closely related.

\paragraph{Diagrams involving transverse gauge links.}
The Feynman rule for the transverse gauge links at offsets $n L$ and $\bn L$
are given by
\begin{align}
 g_s t^a n_\perp^\mu \frac{1 - e^{\img \bt \cdot \kt}}{\vec n_\perp \cdot \kt} e^{\img n \cdot k L}
\,,\quad
 g_s t^a n_\perp^\mu \frac{1 - e^{\img \bt \cdot \kt}}{\vec n_\perp \cdot \kt} e^{\img \bn \cdot k L}
\,,\end{align}
where, ${\vec n}_\perp = \bt / |\bt|$.
In Feynman gauge, this vertex can thus be neglected for $L\to\infty$, and one would not consider \fig{soft_L_nlo_c}.
For finite $L$, we instead obtain for \fig{soft_L_nlo_c} 
\begin{align} \label{eq:soft_L_c}
 S_c^{(L)} &
 = \frac{\as C_F}{2\pi} \biggl[ \frac{2 b_T}{L} \arctan\frac{b_T}{2L} - 2\ln\left(1 + \frac{b_T^2}{4L^2}\right) \biggr]
\,.\end{align}
This result vanishes for $L \to \infty$ (or $b_T \ll L$) as expected.

The situation is more intricate for \fig{soft_L_nlo_d}.
Again, for $L\to\infty$ the diagram would not be considered.
For finite $L$ instead, the $L$ dependence drops out and one obtains
\begin{align} \label{eq:soft_L_d}
 S_d^{(L)} &
 = g_s^2 C_F n_\perp^2 \int\frac{\df^d k}{(2\pi)^d} \frac{-\img}{k^2 + \img0}
    \frac{1 - e^{+\img \bt \cdot \kt}}{\vec n_\perp \cdot \kt} e^{\img n \cdot k L}
    \frac{1 - e^{-\img \bt \cdot \kt}}{-\vec n_\perp \cdot \kt} e^{-\img n \cdot k L}
\nn\\&
 = \frac{\as C_F}{2\pi} \biggl[ \frac{2}{\eps} + 2 \ln\frac{\mu^2 b_T^2}{b_0^2} + 4 \biggr]
\,.\end{align}
Here, the symmetry factor $1/2$ is compensated by the mirror diagram of \fig{soft_L_nlo_d}.
The relative minus sign of the momentum $k$ in the vertices arises because $k$ is incoming into one vertex and outgoing from the other.
Due to this relative sign, the exponential factors cancel, yielding a nonvanishing result of the diagram.
In particular, it is independent of $L$ and thus does not vanish as $L\to\infty$.

In order to obtain a result consistent with the known $L\to\infty$ limit, where this diagram does not contribute,
the transverse self-energy has to cancel with other diagrams to not give a physical contribution to the cross section.
Indeed, when combining the unsubtracted beam function with the soft function and zero-bin subtraction into $f^\TMD$ as in \eq{tmdpdf1b}, we find that these transverse self-energies will exactly cancel.

Lastly, we remark that we have explicitly checked that the diagrams with
the transverse gauge links are indeed necessary to ensure gauge invariance using a $R_\xi$ gauge,
but the calculation is otherwise not instructive and hence not presented here.

\paragraph{Full result.}
The full result for the soft function with Wilson lines of finite length $L$ is given by adding
Eqs.\ \eqref{eq:soft_L_a}, \eqref{eq:soft_L_b}, \eqref{eq:soft_L_c} and \eqref{eq:soft_L_d},
\begin{align} \label{eq:S_L_nlo}
 S^{(1)}(b_T,\eps,L)
 = \frac{\as C_F}{2\pi} \biggl[& - \frac{2}{\eps^2} + \frac{2}{\eps}
   - 2 \left(\frac{1}{\eps} + \Lb{}\right) \ln\frac{\mu^2 L^2}{e^{-2\gamma_E}}
   \nn\\&
   + \Lb{2} + 2 \Lb{} + \frac{\pi^2}{6} + 4 \biggr]
\,.\end{align}

\subsection{Construction of the quasi beam function}
\label{sec:qbeam}

Recall the definition \eq{beamfunc} of the light-cone beam function,
\begin{align} \label{eq:beam2}
 &B_{q}(x,\bt,\eps,\tau,x P^-) = \int\frac{\df b^+}{4\pi} e^{-\img \frac12 b^+ (x P^-)}
 \Bigl< h(P) \Bigr|  \Bigl[ \bar q(b^\mu)
 W(b^\mu) \frac{\gamma^-}{2}
 W_{T}\bigl(-\infty\bn;\vec b_T,\vec 0_T\bigr)
 \nn\\&\hspace{8.5cm}
 W^\dagger(0)  q(0) \Bigr]_\tau \Bigl| h(P) \Bigr>
\,,\end{align}
where $b^\mu = b^+ \bn^\mu/2 + b_\perp^\mu$. 
The Wilson lines $W$ extend to light-cone infinity, where they are closed by $W_T$ in the transverse direction, see \fig{wilson_beam}.

\begin{figure}[tp]
 \centering
 \begin{subfigure}{0.5\textwidth}
  \includegraphics[width=\textwidth]{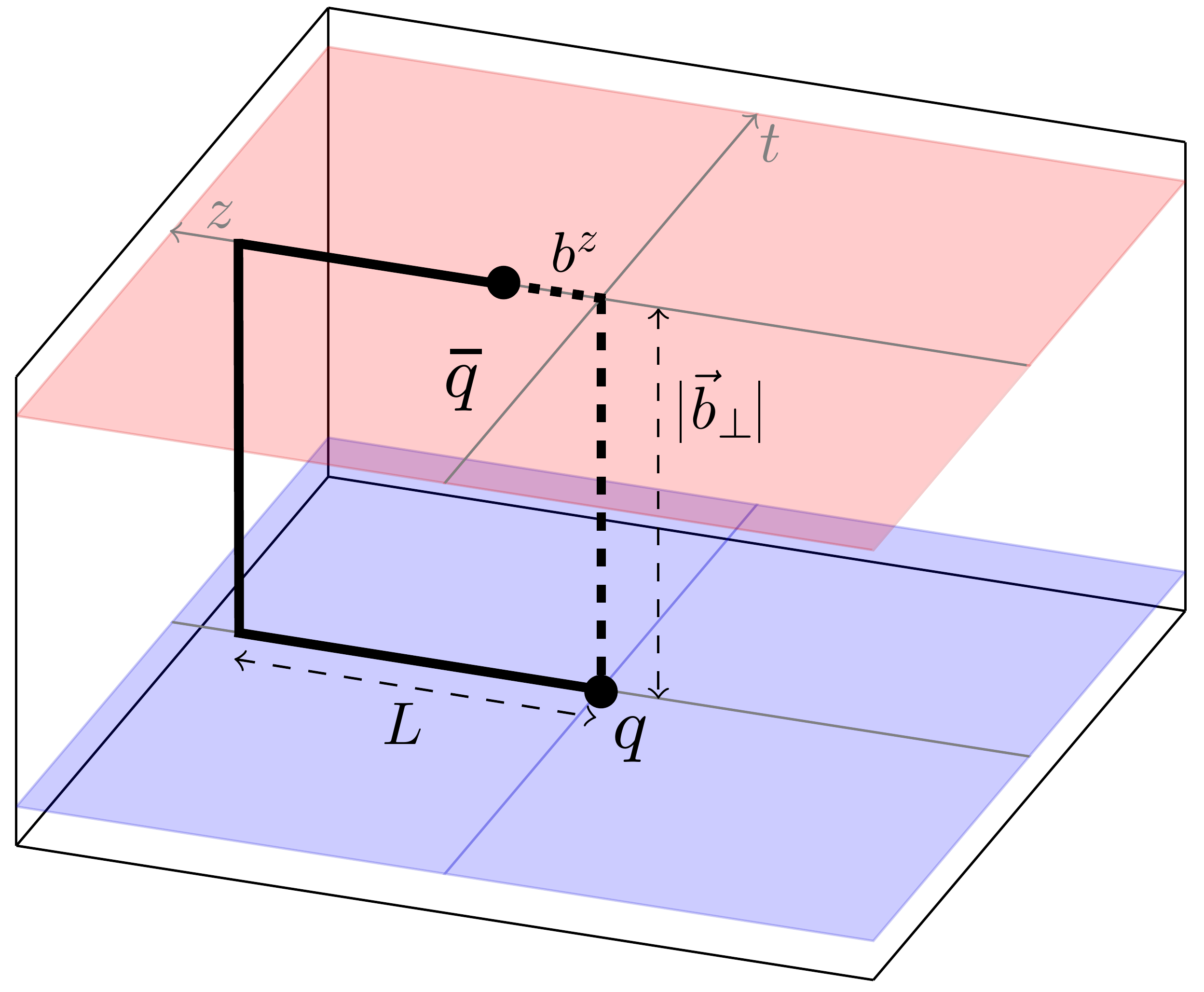}
  \caption{}
 \label{fig:qbeam_Wilson_lines}
 \end{subfigure}
 \qquad
 \begin{subfigure}{0.4\textwidth}
  \includegraphics[width=\textwidth]{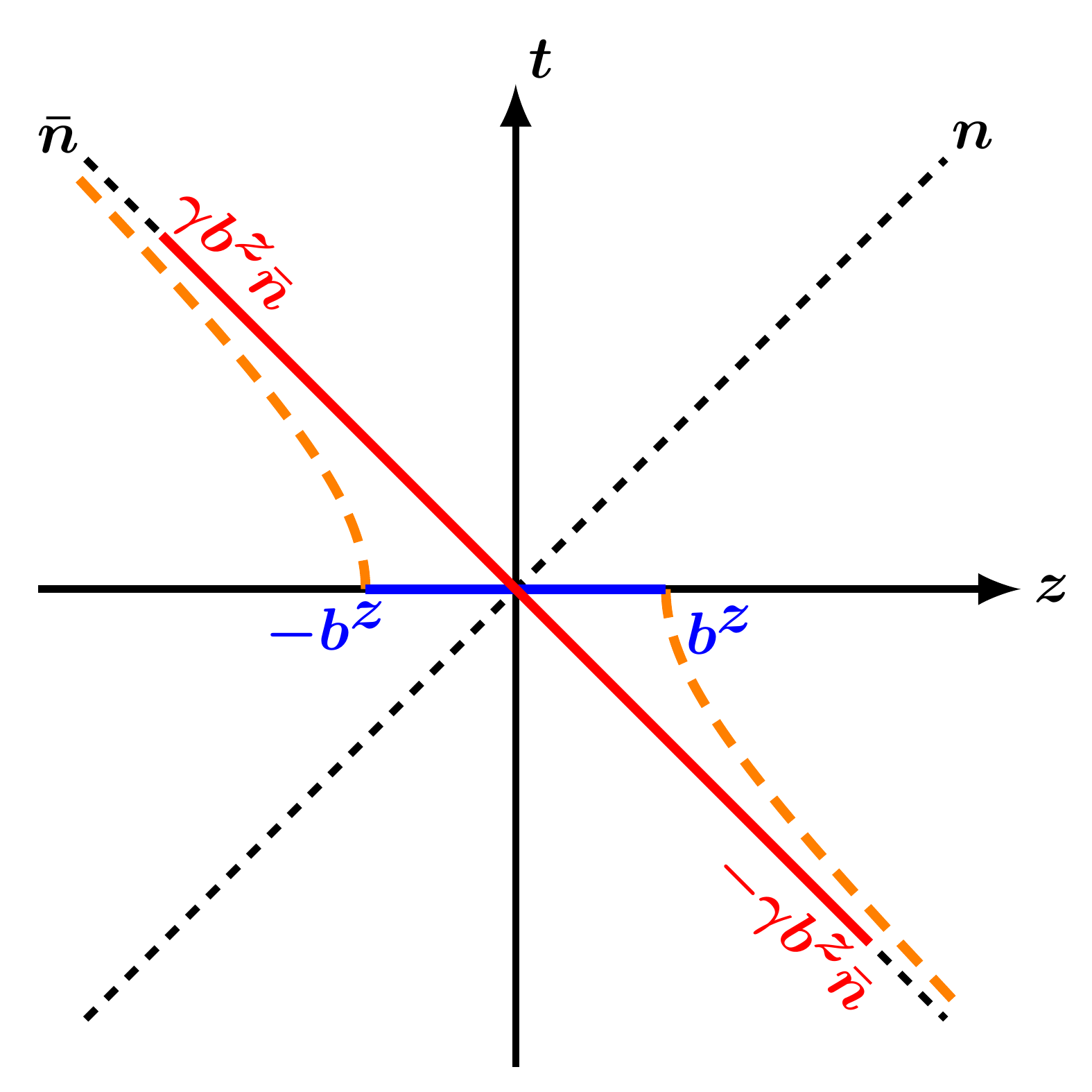}
  \caption{}
 \label{fig:qbeam_boost}
 \end{subfigure}
 \caption{Illustration of the Wilson line structure of the quasi beam function (a),
 and the behavior of the longitudinal separation under a Lorentz boost along the $z$ direction (b).}
 \label{fig:qbeam}
\end{figure}

Following the standard LaMET procedure, we define the quasi beam function analogous to the beam function
by replacing the light-cone correlator with an equal-time correlator,
which in particular includes replacing the $n$-collinear Wilson lines by Wilson lines along the $\hat z$ direction.
Due to the finite lattice size, they must also be truncated at a length $L$,
where one needs to include transverse Wilson lines to ensure gauge invariance.
The resulting Wilson line path is illustrated in \fig{qbeam_Wilson_lines}.
The definition of the bare quasi beam function in position space thus reads
\begin{align} \label{eq:qbeam}
 \tilde B_{q}(b^z, \bt, a,L,P^z) =
 \Bigl< h(P) \Big| &\bar q(b^\mu) W_{\hat z}(b^\mu;L-b^z) \frac{\Gamma}{2}
 W_T(L \hat z; \bt, \vec{0}_T) W^\dagger_{\hat z}(0;L) q(0) \Big| h(P) \Bigr>
\,,\end{align}
where $b^\mu = (0, \bt, b^z)$.
Here, we also replaced $\gamma^-$ by the Dirac structure $\Gamma$, which can be chosen as either $\Gamma = \gamma^0$ or $\Gamma = \gamma^z$, as booth can be boosted onto $\gamma^-$.
The Wilson lines of length $L$ are defined by
\begin{align} \label{eq:coll_Wilson_L}
 W_{\hat z}(x^\mu;L) &= P \exp\left[ \img g \int_{L}^0 \df s \, \cA^z(x^\mu + s \hat z) \right]
\,.\end{align}
The transverse gauge links are given by \eq{Tgaugelinks}.   Again in \eq{qbeam}, diagrams that have no fields contracted with the states $|h(P)\rangle$ are excluded.

A crucial feature of \eq{qbeam} is that it is an equal-time correlation function,
i.e.\ it neither depends on the time-dependent light-cone separation $b^+$ nor on the lightlike directions $n$ and $\bn$ as \eq{beam2}, which makes it computable on lattice.
The physical picture underlying this specific Wilson line structure is that boosting a purely spatial separation
along the $\hat z$ direction yields an almost lightlike separation.
Concretely, for a Lorentz boost along the $\hat z$ direction with velocity $v$ and $\gamma=1/\sqrt{1-v^2}$, one obtains
\begin{align} \label{eq:boost}
 \hat z = \left(\begin{matrix} 0 \\ 0 \\ 0 \\ 1 \end{matrix}\right)
 \quad\rightarrow\quad
 {1\over \sqrt{1-v^2}}\left(\begin{matrix} v \\ 0 \\ 0 \\  1 \end{matrix}\right)
 \stackrel{v\to-1}{\approx} - \gamma \bn^\mu
\,.\end{align}
This is illustrated in \fig{qbeam_boost}: The pure spatial separation (blue)
is boosted along the orange-dotted trajectory, thereby approaching the lightlike
separation (red).
Note that regardless of whether $b^z$ is positive or negative,
it is always boosted onto the same lightlike axis $\bn^\mu$, as required for the $n$-collinear PDF.
To boost onto $n^\mu$, one needs to reverse the boost, $v>0$, as is appropriate for the $\bn$-collinear PDF,
since the corresponding proton is moving into the opposite direction.
It is easy to check that by applying the Lorentz boost \eq{boost} to \eq{qbeam}, one recovers the matrix element in \eq{beam2}.

When evaluated in a large-momentum nucleon state, the quasi beam function defined in
\eq{qbeam} is equivalent to the matrix element of an almost-lightlike correlator 
in a static nucleon state. According to LaMET~\cite{Ji:2013dva,Ji:2014gla}, the quasi beam function
is related to the beam function in \eq{beam2} through a factorization formula which includes
perturbative matching and nonperturbative power corrections determined by the large nucleon momentum.
This has been proven rigorously for the collinear PDF \cite{Ma:2014jla,Ma:2014jga,Izubuchi:2018srq}. For the TMDPDF  the physical boost picture shown in \fig{qbeam_boost} implies a relation for the bare operators.  However we must test the extent to which this relation survives when rapidity regulators are implemented, which we will do in \sec{qbeam_nlo}.  For the TMDPDF such regulators are known to have a significant effect, and therefore we do not expect a simple short distance matching relation between the quasi-beam function and beam function alone. This expectation is also consistent with the known importance of the soft region in the TMD factorization theorem.

\subsection{Construction of the quasi soft function}
\label{sec:qsoft}

\begin{figure}[tp]
 \centering
 \begin{subfigure}{0.45\textwidth}
  \includegraphics[width=\textwidth]{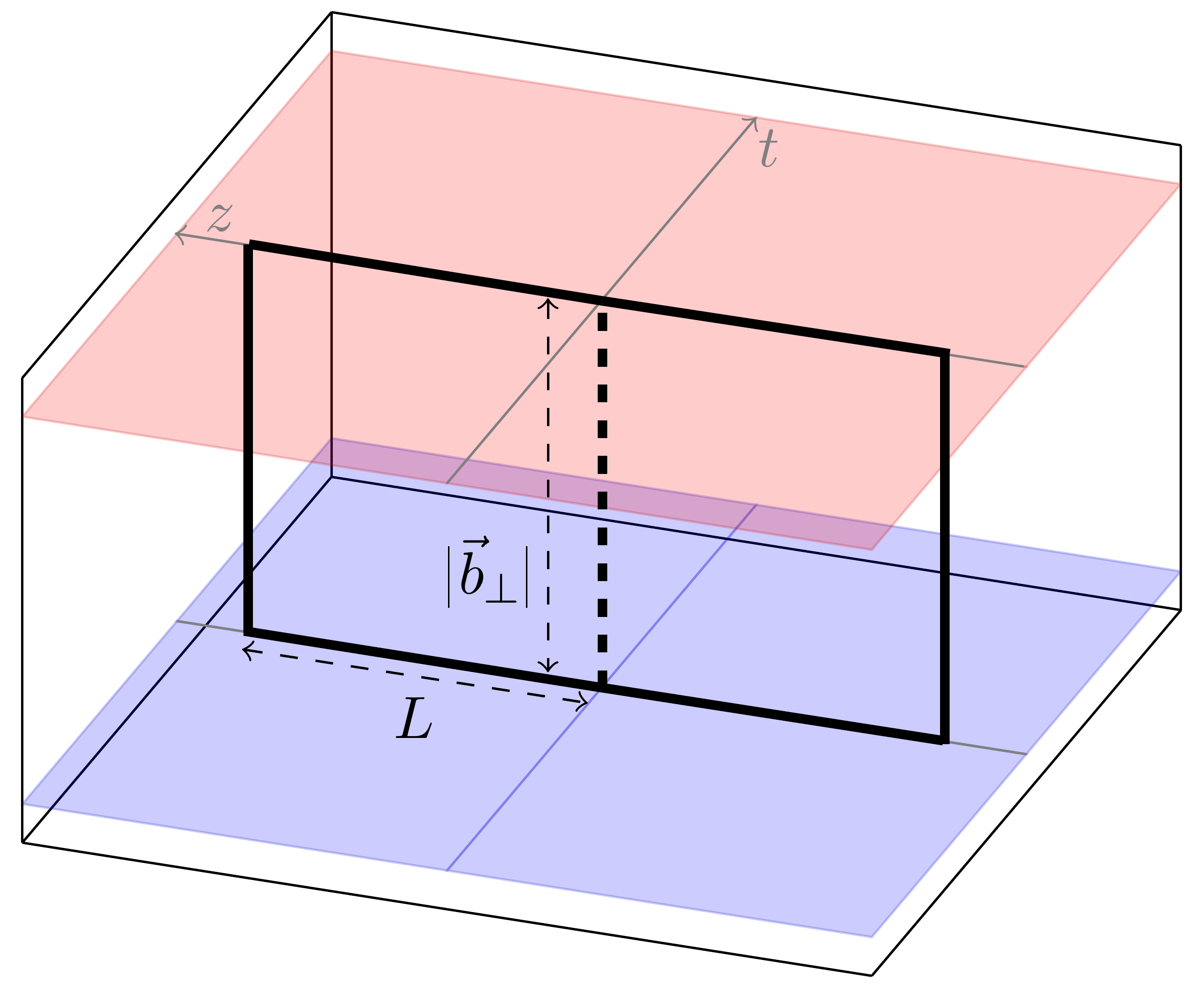}
  \caption{}
 \label{fig:qsoft_Wilson_lines}
 \end{subfigure}
 \qquad
 \begin{subfigure}{0.4\textwidth}
  \includegraphics[width=\textwidth]{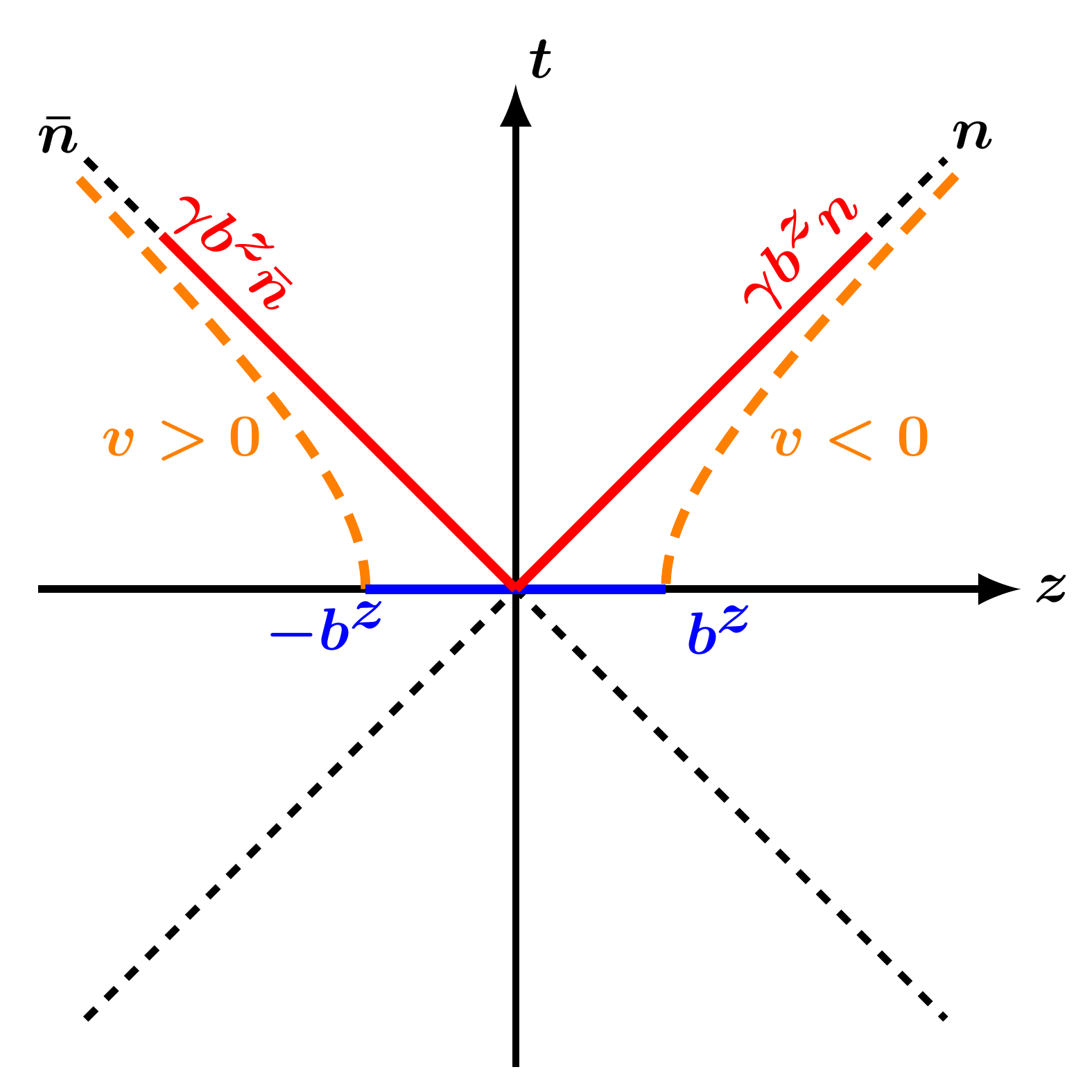}
  \caption{}
 \label{fig:qsoft_boost}
 \end{subfigure}
 \caption{Illustration of the Wilson line structure of the naive quasi soft function (a),
 and the behavior of the longitudinal separation under a Lorentz boost along the $z$ direction (b).
 $v > 0$ and $v < 0$ denote that the required Lorentz boosts have opposite signs.}
 \label{fig:qsoft}
\end{figure}

Recall the definition \eq{softfunc} of the bare TMD soft function,
\begin{align}
 S^q(b_T,\eps,\tau) = \frac{1}{N_c} \bigl< 0 \bigr| {\rm Tr} \bigl[ S^\dagger_n(\bt) S_\bn(\bt)
   S_{T}(-\infty \bn;\vec b_T,\vec 0_T)
   S^\dagger_\bn(\vec 0_T) S_n(\vec 0_T)
   S_{T}^\dagger\bigl(-\infty n;\vec b_T,\vec 0_T\bigr) \bigr]_\tau
 \bigl|0 \bigr>
\,.\end{align}
Note that this vacuum matrix element has no explicit time dependence,
in contrast to the collinear matrix element \eq{beam2}.
Time dependence only enters indirectly through the lightlike directions of the Wilson lines $S_n$ and $S_{\bn}$, which on its own prohibits a direct computation on lattice.
To obtain a lattice-computable quasi soft function, it thus seems reasonable to follow the same logic as above and replace
\begin{align}
 n^\mu = (1,0,0,1) ~\to~ \hat z^\mu
\,,\qquad
 \bn^\mu = (1,0,0,-1) ~\to~ {-}\hat z^\mu
\,.\end{align}
As before, the lattice computation also requires to truncate the Wilson lines at a length $L$, where they are joined by transverse gauge links.
The most naive attempt of constructing a quasi version of the soft function \eq{softfunc} thus takes the form
\begin{align} \label{eq:qsoft}
 \tilde S^q(b_T, a, L) &= \frac{1}{N_c} \bigl< 0 \big|
  \Tr\bigl\{ S^\dagger_{\hat z}(\bt;L) S_{-\hat z}(\bt;L)
   S_{T}(L \hat z;\vec b_T,\vec 0_T)
   \nn\\&\hspace{2cm}\times
 S^\dagger_{-\hat z}(\vec 0_T;L) S_{\hat z}(\vec 0_T;L)
 S_{T}^\dagger\bigl(-L \hat z;\vec b_T,\vec 0_T\bigr) \bigr\}
 \bigl|0 \bigr>
\,,\end{align}
where the soft Wilson lines of finite length are given by
\begin{align} \label{eq:soft_Wilson_L}
 S_{\pm \hat z}(x^\mu; L) &= P \exp\biggl[ \pm \img g \int_{-L}^0 \df s \, \cA^z(x^\mu \pm s \hat z^\mu) \biggr]
\,.\end{align}
The resulting Wilson line path is illustrated in \fig{qsoft_Wilson_lines}.

Unfortunately, the physical boost argument that allowed us to relate spatial Wilson lines to lightlike Wilson lines
in the quasi-PDF [see \eq{boost}] does not apply to the quasi soft function.
Since the soft function involves both light-cone directions $n$ and $\bn$,
it is necessary to simultaneously obtain them from boosting $\pm \hat z$.
However, this requires boosts of opposite signs, as illustrated in \fig{qsoft_boost}.
Note that if this were possible with a single boost, it would directly violate the boost
argument for $\tilde B$, where it is essential that both positive
and negative $b^z$'s are boosted onto the same light-cone direction.

Despite the simple boost picture breaking down, one can still test whether the matching for the obtained quasi-TMDPDF in the form of \eq{matching_result0} is possible, and we study this in the next section at NLO.
Indeed, we find that for the naive quasi soft function the matching is spoiled by the structure of infrared $b_T$ dependence.
In \sec{qsoft_bent_nlo}, we will suggest a modified quasi soft function that yields a quasi-TMDPDF which has the correct infrared $b_T$ dependence at one loop. Given the absence of an intuitive boost relation, a rigorous all orders proof for any such quasi-TMDPDF proposal will certainly be required before full confidence can be obtained.

\section{One Loop Results}
\label{sec:nlo_results}

In this section we present explicit one-loop results for the TMDPDF,
the quasi beam and naive quasi soft function, and their combination into the quasi-TMDPDF.
Here, we work in the $\MS$ scheme, as opposed to considering renormalization schemes appropriate for lattice calculations as discussed in \sec{towards_qtmd}, since a pure $\MS$ calculation is fully sufficient to perturbatively test the matching relation.
Furthermore, we limit ourselves to the quark PDF and neglect mixing with gluons for simplicity, which corresponds to considering a non-singlet flavor combination.
All results are calculated by evaluating the appropriate matrix elements for the (quasi) beam function with an on-shell external quark with momentum $P^\mu = (P^z,0,0,P^z)$.

\subsection{Lightcone TMDPDF at one loop}
\label{sec:tmdpdf_nlo}
The unrenormalized result for the TMDPDF at one loop is given by
\begin{align} \label{eq:tmd_nlo}
 f_q^{\TMD\,(1)}(x, \bt, \eps, \zeta) &
 = \frac{\as C_F}{2\pi} \biggl[ - \left(\frac{1}{\eps_\IR} + \Lb{} \right) P_{qq}(x) + (1-x) \biggr]_+^1\Theta(1-x) \Theta(x)
 \nn\\&
  + \frac{\as C_F}{2\pi} \delta(1-x) \Biggl[ \frac{1}{\eps_\UV^2}
   + \frac{1}{\eps_\UV} \left(\frac{3}{2} + \ln\frac{\mu^2}{\zeta^2} \right)
   + \frac{1}{2} - \frac{\pi^2}{12} \Biggr]
 \nn\\&
  + \frac{\as C_F}{2\pi} \delta(1-x) \Biggl[
  - \frac{1}{2} \Lb{2} + \frac{3}{2} \Lb{}
  + \Lb{} \ln\frac{\mu^2}{\zeta^2} \Biggr]
\,,\end{align}
where $P_{qq}(x) = (1+x^2)/(1-x)$ is the quark-to-quark splitting function
and the subscript $_+$ denotes a plus distribution such that $\int_0^1 \df x \, [f(x)]_+^1 = 0$.
In \app{overview_tmdpdfs}, we show that \eq{tmd_nlo} agrees with the vast majority of regulators used to define TMDPDFs.

As indicated, the divergence in the first line in \eq{tmd_nlo} is of infrared origin
and matches precisely the IR divergence in the collinear PDF,
which is crucial for matching the TMDPDF onto the PDF for perturbative $b_T \ll \LQCD^{-1}$, see \eq{tmdpdf_matching}.
Likewise, it must be exactly reproduced by the quasi-TMDPDF for a matching relation to exist.
The second line in \eq{tmd_nlo} contains UV poles and constants,
and the last line contains the $b_T$ dependence.
Similar to the IR pole, the $b_T$ dependence must be identical in the quasi-TMDPDF
in order for a perturbative matching for $b_T \sim \LQCD^{-1}$ to exist.

\subsection{Quasi beam function}
\label{sec:qbeam_nlo}

\begin{figure*}
 \centering
 \begin{subfigure}{0.4\textwidth}
  \includegraphics[width=\textwidth]{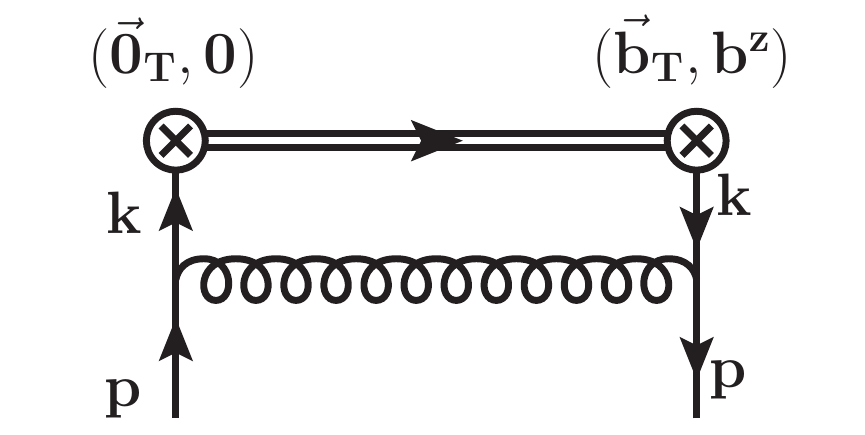}
  \caption{Vertex diagram}
  \label{fig:qbeam_a}
 \end{subfigure}
 \quad
 \begin{subfigure}{0.4\textwidth}
  \includegraphics[width=\textwidth]{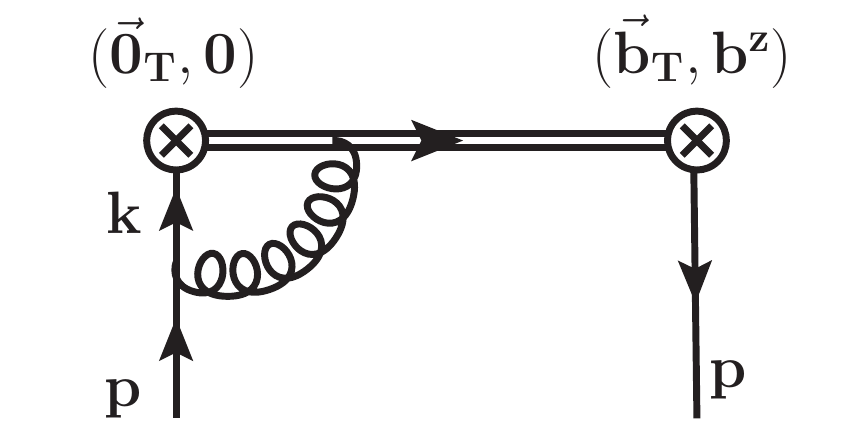}
  \caption{Sail topology}
  \label{fig:qbeam_b}
 \end{subfigure}
\\\vspace{1ex}
 \begin{subfigure}{0.4\textwidth}
  \includegraphics[width=\textwidth]{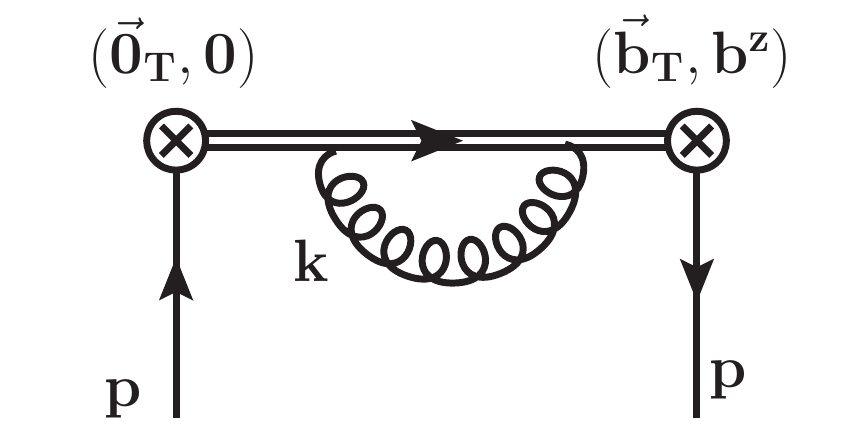}
  \caption{Wilson line self energy (tadpole)}
  \label{fig:qbeam_c}
 \end{subfigure}
 \quad
 \begin{subfigure}{0.4\textwidth}
  \includegraphics[width=\textwidth]{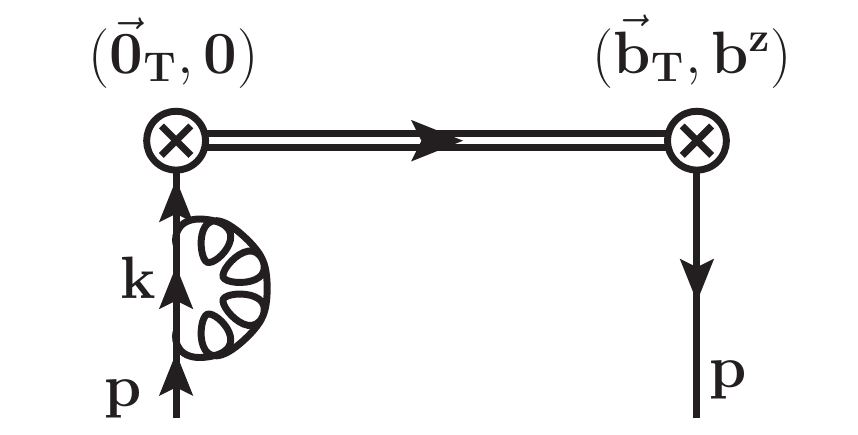}
  \caption{Wave function renormalization}
  \label{fig:qbeam_d}
 \end{subfigure}
 \caption{One-loop diagrams contributing to the quasi beam function
          in Feynman gauge, up to mirror diagrams.}
 \label{fig:qbeam_nlo}
\end{figure*}

We first calculate the quasi beam function defined in \eq{qbeam} by evaluating the operator in a quark state with on-shell momentum $P^2 = 0$.
Working in pure dimensional regularization and taking the physical limit $L \to \infty, P^z \to \infty$,
we can directly Fourier transform the result into $x$ space.
At one loop, there are four contributions, shown in \fig{qbeam_nlo}.
The calculation is quite lengthy and shown in detail in \app{qbeamfunc}.
The result is given by
\begin{align} \label{eq:qbeam_nlo}
 \tilde B_{q}^{(1)}(x, \bt, \eps, L, P^z) &
 = \frac{\as C_F}{2\pi} \biggl[ - \left(\frac{1}{\eps_\IR} + \Lb{} \right) P_{qq}(x)
   +  (1-x) \biggr]_+^1 \Theta(1-x)\Theta(x)
 \\\nn&
 + \frac{\as C_F}{2\pi} \delta(1-x) \biggl[
  \frac{7}{2} \frac{1}{\eps_\UV} - \frac{1}{2} \Lb{2} + \frac{9}{2} \Lb{} + \frac{1}{2} + \frac{2\pi L}{b_T}
  \\\nn&\hspace{3cm}
  - \Lb{} \ln\frac{(2P^z)^2}{\mu^2}
  - \frac{1}{2} \ln^2\frac{(2P^z)^2}{\mu^2} + \ln\frac{(2P^z)^2}{\mu^2}
  \biggr]
\,.\end{align}
As anticipated, it contains the same IR divergence as the TMDPDF, \eq{tmd_nlo}.
Note the presence of a linear divergence in $L/b_T$, which we interpret as the analog of a rapidity divergence.
As discussed below \eq{illustration_rapidity_div_4}, these divergences appear as power-law divergences.
Thus, after regularization they yield a linear dependence on the regulator $L$,
rather than a logarithmic dependence $\ln(L/b_T)$.
This linear $L/b_T$ term will exactly cancel with a similar term in the quasi soft factor when combining $\tilde B_q$ and $\tilde \Delta_S^q$ as in \eq{qtmdpdf}.

In order to directly match this quasi beam function onto the lightlike beam function,
we require that the logarithmic $b_T$ dependence, arising from IR physics, must be equal between them.
However, the $b_T$ dependence does not agree with any beam function known in the literature, see the results in \app{overview_tmdpdfs} which are summarized in \tbl{LogB} below.
In particular, only in Collins' scheme with Wilson lines off the light-cone one has the correct double-logarithm $-\frac12 \ln^2(b_T^2\mu^2/b_0^2)$, while in all schemes with Wilson lines on the light-cone this double logarithm is (at least partially) contained in the soft function. Even in Collins' scheme the single $\ln(b_T^2\mu^2/b_0^2)$ does not match up with the corresponding single logarithm in the quasi beam function.\footnote{We have also checked that this problem is not simply due to the contribution from the transverse Wilson line self energy diagram.}
Hence, for all the rapidity regulators used in the literature, which yield the same universal TMDPDF defined in \sec{tmd_review}, none are in agreement with
the simple physical picture of relating beam function and quasi beam function.  
The Lorentz boost relation is spoiled by the presence of a rapidity regulator, which by construction is not boost invariant.
Since it is well known that the choice of rapidity regulator can modify the logarithms of $b_T$,
one may still hope to find a regulator for the beam function which yields the same IR structure as the naive quasi beam function and thus yields a perturbative matching that agrees with the boost relation. 
However, the more important test is whether the quasi-TMDPDF can be matched to the TMDPDF, in which case the regulator dependence cancels. This requires considering the quasi soft function.

\subsection{Naive quasi soft function}
\label{sec:qsoft_naive_nlo}

\begin{figure*}
\centering
 \begin{subfigure}[t]{0.32\textwidth}
  \includegraphics[width=\textwidth]{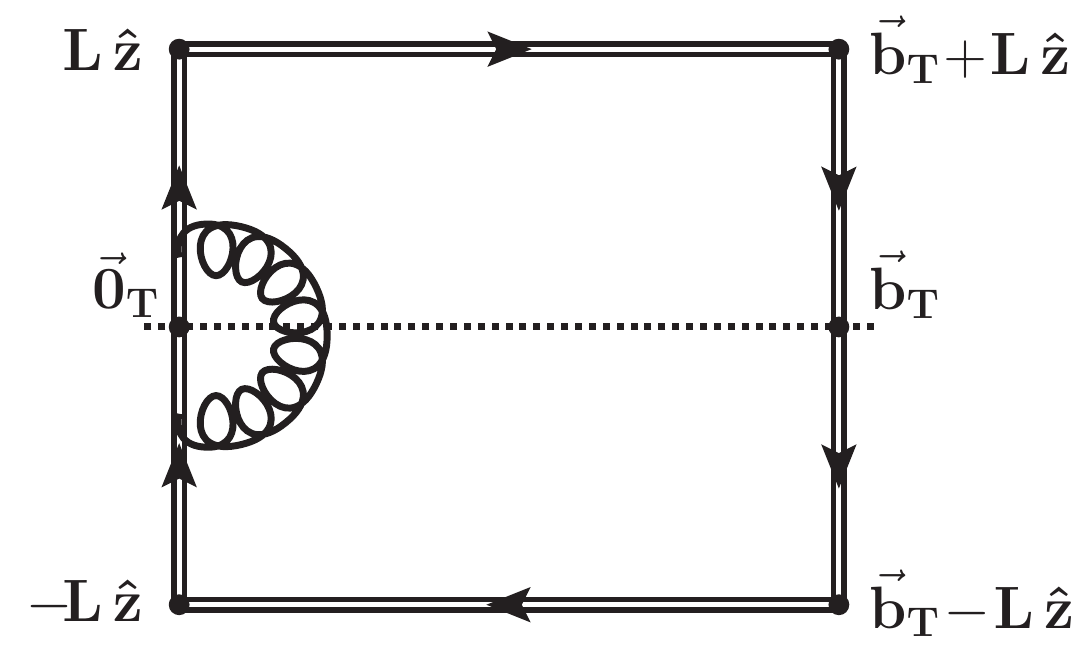}
  \caption{}
 \label{fig:qsoft_nlo_a}
 \end{subfigure}
 \begin{subfigure}[t]{0.32\textwidth}
  \includegraphics[width=\textwidth]{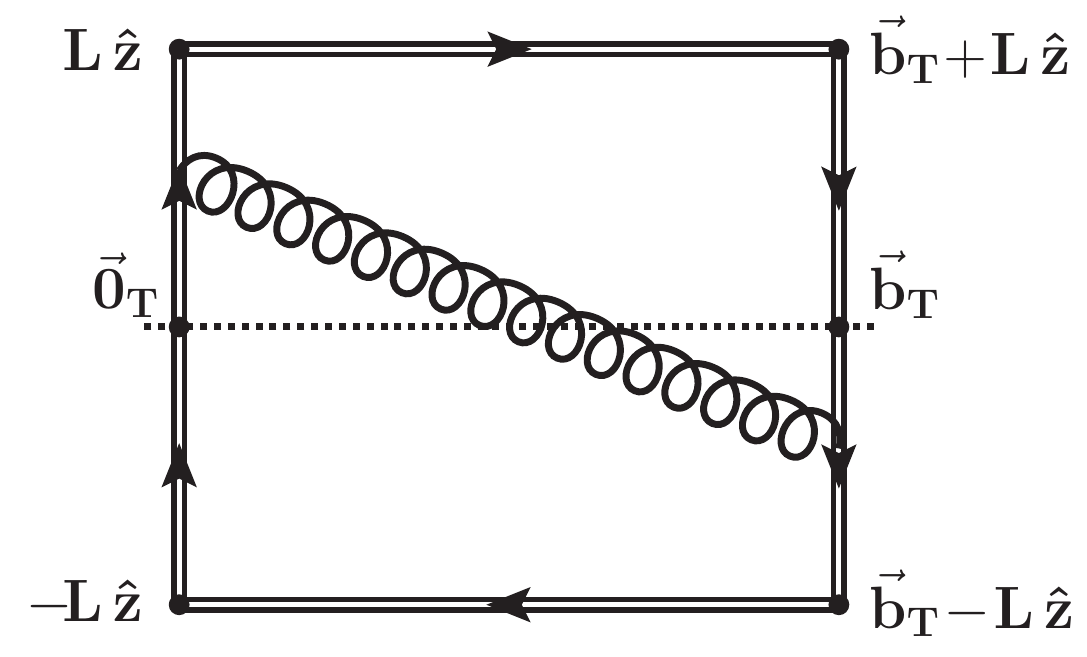}
  \caption{}
 \label{fig:qsoft_nlo_b}
 \end{subfigure}
 \begin{subfigure}[t]{0.32\textwidth}
  \includegraphics[width=\textwidth]{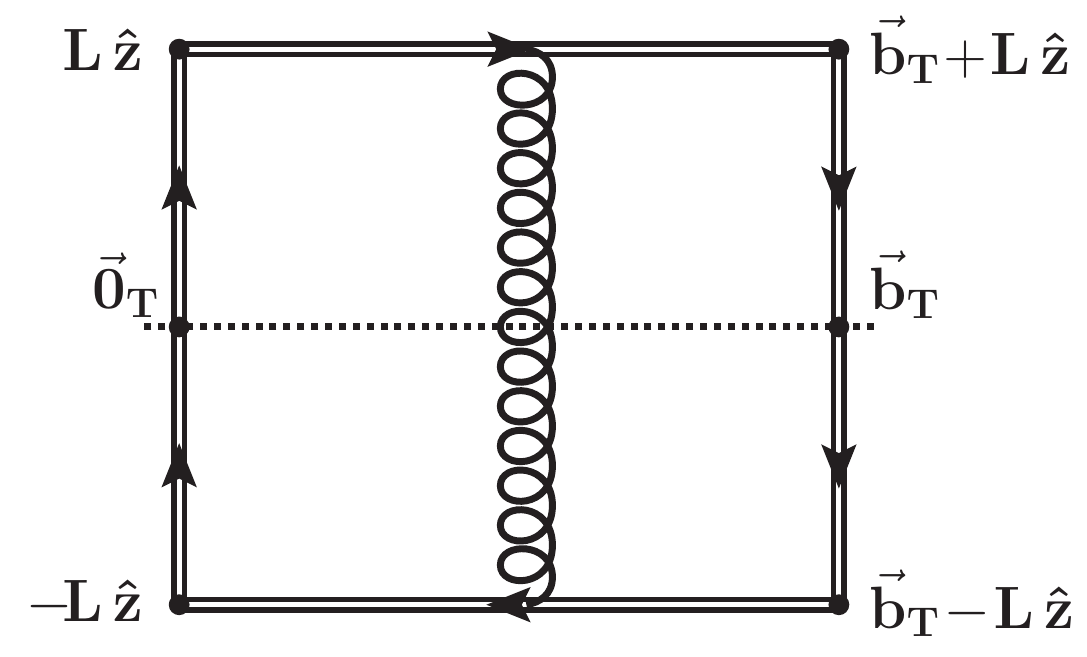}
  \caption{}
 \label{fig:qsoft_nlo_c}
 \end{subfigure}
\\
 \begin{subfigure}[t]{0.32\textwidth}
  \includegraphics[width=\textwidth]{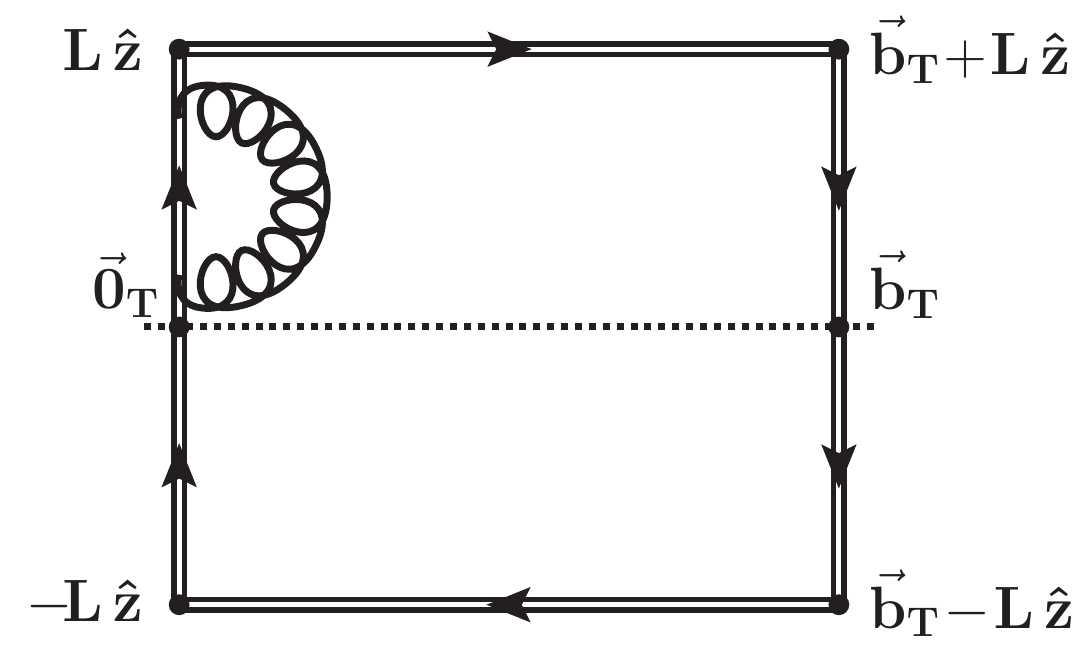}
  \caption{}
 \label{fig:qsoft_nlo_f}
 \end{subfigure}
 \begin{subfigure}[t]{0.32\textwidth}
  \includegraphics[width=\textwidth]{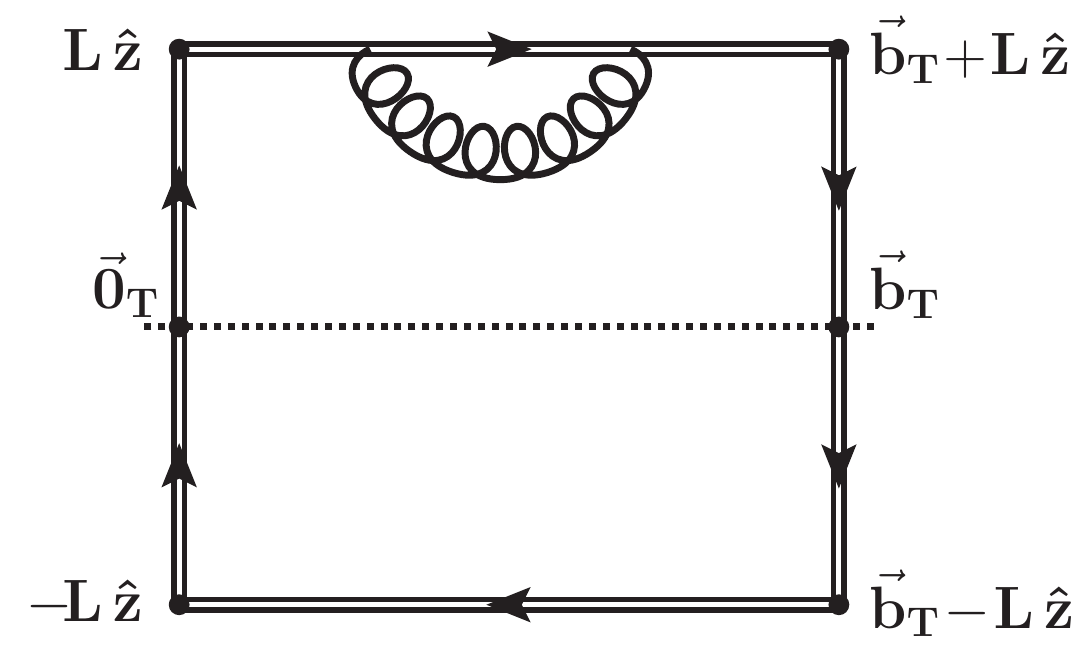}
  \caption{}
 \label{fig:qsoft_nlo_e}
 \end{subfigure}
 \begin{subfigure}[t]{0.32\textwidth}
  \includegraphics[width=\textwidth]{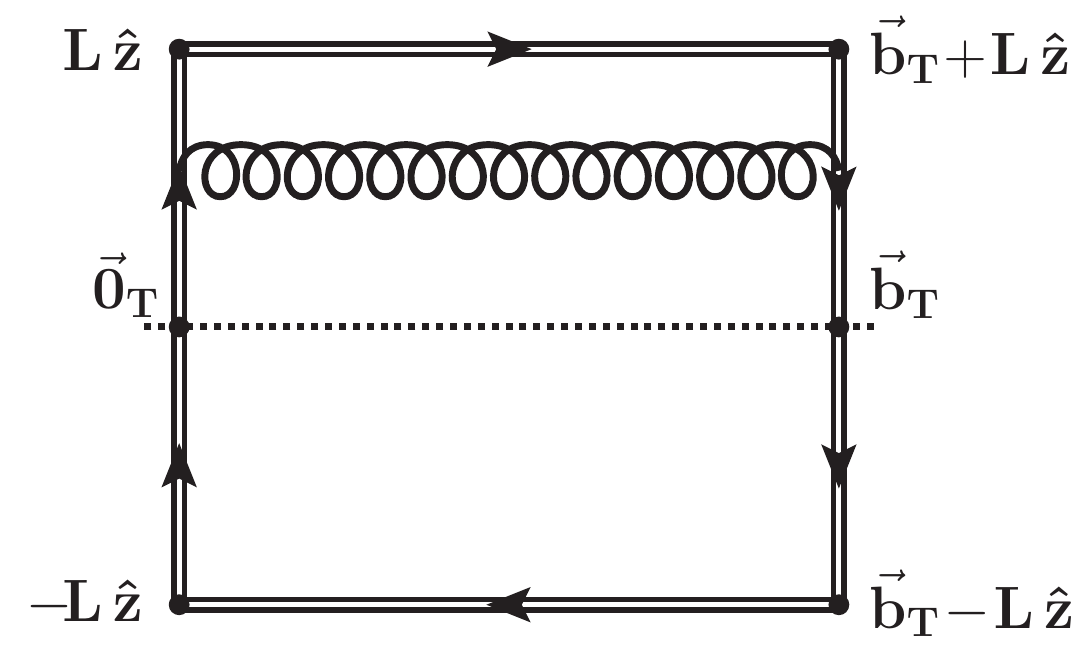}
  \caption{}
 \label{fig:qsoft_nlo_d}
 \end{subfigure}
 \caption{One loop diagrams for the quasi soft function in Feynman gauge, up to mirror diagrams.
 The dashed line indicates that we consider the upper part of the diagrams separate from the lower part,
 such that diagrams in the top row only contains a gluon exchange from the $\hat z$ to $-\hat z$ sector,
 while the bottom row only contains corrections within the $\hat z$ sector.
 Similar diagrams for $-\hat z$ exchanges are now shown.}
 \label{fig:qsoftfunc_nlo}
\end{figure*}

Next, we calculate the naive quasi soft function defined in \eq{qsoft}.
Working in Feynman gauge, there are six diagrams that contribute at NLO,
shown in \fig{qsoftfunc_nlo}, where double lines represent Wilson lines
and the labels denote the endpoints of the Wilson lines in position space.
For later convenience, we distinguish diagrams where the gluon is exchanged
between the $+\hat z$ and $-\hat z$ Wilson lines (upper row)
and diagrams where the gluon is emitted between Wilson lines of the $+\hat z$ sector alone (lower row).
The latter is identical to the result for gluons exchanged within the $- \hat z$ sector.

In Feynman gauge, the generic expression for a one-loop diagram in the $\MS$ scheme,
parameterized by the spatial paths $\gamma_1$ and $\gamma_2$ of the Wilson lines, is
\begin{align} \label{eq:soft_nlo}
 \tilde S^{(1)}[\gamma_1,\gamma_2] &
 = - g_s^2 C_F \mu_0^{2\eps} \int_0^1 \df s \,\gamma_1'(s)^\mu \, \int_0^1 \df t \, \gamma_2'(t)^\nu \,
   \int\frac{\df^{4-2\eps}k}{(2\pi)^{4-2\eps}}
   \frac{-\img g_{\mu\nu}}{k^2 + \img0} e^{-\img k \cdot [\gamma_1(s) - \gamma_2(t)]}
\nn\\&
 = \frac{\as C_F}{\pi} \biggl(\frac{\mu^2}{b_0^2}\biggr)^\eps
   \frac{\Gamma(1-\eps)}{e^{\eps \gamma_E}}
   \int_0^1 \!\df s \int_0^1 \!\df t \, \frac{\gamma_1'(s) \cdot \gamma_2'(t)}{\bigl[-(\gamma_1(s)-\gamma_2(t))^2 \bigr]^{1-\eps}}
\,.\end{align}
Note that diagrams with the gluon attaching to the same line
have an additional symmetry factor of $1/2$.

For example, for \fig{qsoft_nlo_a} one reads off the paths
\begin{equation}
 \gamma_1(s) = s L \hat z
\,,\quad
 \gamma_2(t) = (t - 1) L \hat z
\,.\end{equation}
Together with its mirror diagram, this gives
\begin{align} \label{eq:qsoft_nlo_a}
 \tilde S_a^{(1)}(b_T,\eps,L) &
 = 2 \frac{\as C_F}{\pi} \biggl(\frac{\mu^2}{b_0^2}\biggr)^\eps \frac{\Gamma(1-\eps)}{e^{\eps \gamma_E}}
   \int_0^1 \!\df s \int_0^1 \!\df t \, \frac{(L \hat z) \cdot (L \hat z)}{\bigl[- L^2 (s -t + 1)^2 \hat z^2 \bigr]^{1-\eps}}
\nn\\&
 = 2 \frac{\as C_F}{\pi} \biggl(\frac{\mu^2}{b_0^2}\biggr)^\eps \frac{\Gamma(1-\eps)}{e^{\eps \gamma_E}}
   \frac{ L^{2\eps} (2 - 4^\eps)}{2 \eps (2\eps - 1)}
\nn\\&
 = \frac{\as C_F}{2\pi} \biggl[ -\frac{2}{\eps} - 2 \ln\frac{\mu^2 L^2}{b_0^2} + 4(\ln2 - 1) \biggr]
\,.\end{align}
Note that this result also contains a divergence at $\eps=1/2$, which signals a power-law divergence $\propto \mu L$ in dimensional regularization, which is expected for the Wilson-line self energy.
In pure dimensional regularization, these divergences are not visible when expanding at $\eps=0$,
but on the lattice they explicitly arise and have to be canceled.
For this reason, the beam function renormalization in \eq{qtmdpdf} contains a $b^z$-dependent counter term to absorb the divergence associated with the self energy of the Wilson-line segment of length $L-b^z$ [see \eq{qtmd_c_3}],
while the other self-energies will cancel against the soft factor.

The diagrams in \figs{qsoft_nlo_b}{qsoft_nlo_c} yield
\begin{align}
 \label{eq:qsoft_nlo_b}
 \tilde S_b^{(1)}(b_T,\eps,L) &
 = \frac{\as C_F}{2\pi} \biggl[ 2 \ln\frac{(b_T^2 + L^2)^2}{b_T^2 ( b_T^2 + 4L^2)}
  - 8 \frac{L}{b_T} \arctan\frac{L}{b_T} + 8 \frac{L}{b_T} \arctan\frac{2L}{b_T} \biggr]
\,,\\
 \label{eq:qsoft_nlo_c}
 \tilde S_c^{(1)}(b_T,\eps,L) &
 = \frac{\as C_F}{2\pi} \biggl[ \frac{2 b_T}{L} \arctan\frac{b_T}{2L} - 2 \ln\frac{b_T^2 + 4 L^2}{4 L^2} \biggr]
\,.\end{align}
Summing Eqs.~\eqref{eq:qsoft_nlo_a}--\eqref{eq:qsoft_nlo_c}, we obtain
the soft contribution from interactions between the two collinear sectors,
\begin{align} \label{eq:qsoft_nlo_zz}
 \tilde S_{z/-z}^{(1)}(b_T,\eps,L)
 =~&\frac{\as C_F}{2\pi} \biggl[
  - \frac{2}{\eps} - 2 \ln\frac{\mu^2 L^2}{b_0^2} - 2 \ln\frac{b_T^2 + 4 L^2}{4 L^2}
  + 2\ln\frac{(b_T^2 + L^2)^2}{b_T^2(b_T^2 + 4L^2)}
  + 4 (\ln2-1)
  \nn\\&\qquad\qquad
  + \frac{2 b_T}{L} \arctan\frac{b_T}{2L}
  - 8 \frac{L}{b_T} \arctan\frac{L}{b_T} + 8 \frac{L}{b_T} \arctan\frac{2L}{b_T}  \biggr]
\nn\\
 =
 ~&\frac{\as C_F}{2\pi} \biggl[ - \frac{2}{\eps} - 2 \Lb{} \biggr] + \cO\biggl(\frac{b_T^2}{L^2}\biggr)
\,,\end{align}
where we have also taken the limit $L\gg b_T$ for illustration.
Interestingly, in this limit all dependence on $L$ drops out,
leaving only a pure logarithm in $b_T$, the physical observable.

The three remaining diagrams in Figs.~\ref{fig:qsoft_nlo_f}--\ref{fig:qsoft_nlo_d} yield
\begin{align}
 \label{eq:qsoft_nlo_d}
 \tilde S_d^{(1)}(b_T,\eps,L) &
 =  \frac{\as C_F}{2 \pi} \biggl[ \frac{2}{\eps} + 2\ln\frac{\mu^2 L^2}{b_0^2} + 4 \biggr]
\\
 \label{eq:qsoft_nlo_e}
 \tilde S_e^{(1)}(b_T,\eps,L) &
 =  \frac{\as C_F}{2 \pi} \biggl[ \frac{1}{\eps} + \Lb{} + 2 \biggr]
\\
 \label{eq:qsoft_nlo_f}
 \tilde S_f^{(1)}(b_T,\eps,L) &
 = \frac{\as C_F}{2\pi} \biggl[ \frac{4 L}{b_T} \arctan\frac{L}{b_T} - 2 \ln\frac{b_T^2+L^2}{b_T^2} \biggr]
\,,\end{align}
and their sum gives the contribution to the soft function from
interactions within the $\hat z$ sector alone.
The same result is obtained for the $-\hat z$ sector, so we have
\begin{align} \label{eq:qsoft_nlo_z}
 \tilde S_{z/z}^{(1)}(b_T,\eps,L) &= \tilde S_{-z/-z}^{(1)}(b_T,\eps,L)
 \nn\\&
 = \frac{\as C_F}{2\pi} \biggl[
  \frac{3}{\eps} + 2\ln\frac{\mu^2 L^2}{b_0^2} + \ln\frac{\mu^2 b_T^2}{b_0^2} + 6
  - 2 \ln\frac{b_T^2+L^2}{b_T^2} + \frac{4 L}{b_T} \arctan\frac{L}{b_T}
  \biggr]
 \nn\\& =
 \frac{\as C_F}{2\pi} \biggl[
  \frac{3}{\eps} + 3\Lb{} + 2 + \frac{2\pi L}{b_T} \biggr] + \cO\biggl(\frac{b_T^2}{L^2}\biggr)
\,.\end{align}
Note that this contains the same linear divergence in $L/b_T$ as the quasi beam function,
\eq{qbeam_nlo}.
Interestingly, the logarithmic dependence on $L$ cancels within each collinear sector of the quasi soft function.

The full bare quasi soft function is then obtained by summing the contributions from \eqs{qsoft_nlo_zz}{qsoft_nlo_z},
\begin{align} \label{eq:qsoft_nlo}
 \tilde S^{(1)}(b_T, \eps, L)
 =~& \frac{\as C_F}{2\pi} \biggl[
  \frac{4}{\eps} + 4\ln\frac{\mu^2 b_T^2}{b_0^2} + 8
  -4 \ln\frac{b_T^2 + 4 L^2}{4 L^2}
  \nn\\&\qquad\quad
  + \frac{2 b_T}{L} \arctan\frac{b_T}{2L}
   + 8 \frac{L}{b_T} \arctan\frac{2L}{b_T}  \biggr]
\nn\\
 =
 ~&\frac{\as C_F}{2\pi} \biggl[ \frac{4}{\eps} + 4 \ln\frac{\mu^2 b_T^2}{b_0^2} + 4
  + \frac{4 \pi L}{b_T} \biggr] + \cO\biggl(\frac{b_T^2}{L^2}\biggr)
\,.\end{align}
We observe that the logarithmic $b_T$ dependence of the naive quasi soft function does not match that of any soft function in the literature, regardless of the employed rapidity regulator, see the comparison made below in \tbl{LogB}.
Here, this is not as surprising as for the comparisons for the beam function in \sec{qbeam_nlo}, as there is no relation between soft and quasi soft function through a Lorentz boost even at the bare level, see \sec{qsoft}.

\subsection{Failure of the naive quasi-TMDPDF}
\label{sec:qtmd_nlo}
We can now attempt to combine the quasi beam and naive quasi soft function to construct a quasi-TMDPDF in the form of \eq{qtmdpdf}.
To fix the form of the soft factor $\tilde\Delta_S^q$, we require that all divergences in $L/b_T$ cancel when combining $\tilde B_q$ and $\tilde\Delta_S^q$, analogous to the cancellation of rapidity divergences in the TMDPDF.
Comparing the one-loop results in \eqs{qbeam_nlo}{qsoft_nlo},
we deduce this to be $\tilde\Delta_S^q = 1/\sqrt{\tilde S^q}$, so
\begin{align} \label{eq:qtmdpdf2}
 \tilde f_{q}^\TMD(x, \bt,\mu,P^z) = \int \frac{\df b^z}{2\pi} \, e^{\img b^z (x P^z)}\,
 &\tilde Z'(b^z,\mu,\tilde \mu) \tilde Z_{\rm uv}(b^z,\tilde \mu, a)
 \frac{\tilde B_{q}(b^z, \bt, a, L, P^z)}{\sqrt{\tilde S^q(b_T, a, L)}}
\,.\end{align}
Note that this is similar to the $\delta$ regulator for the TMDPDF, where one has $\Delta_S^q = 1/\sqrt{S^q}$.
Our result $\tilde\Delta_S^q = 1/\sqrt{\tilde S^q}$ for the quasi-TMDPDF is consistent with this, considering that we showed in \sec{soft_L_NLO} that the lightlike soft function using the finite-$L$ regulator yields the same soft function as the $\delta$ regulator.

Here, we work purely in the $\MS$ scheme and in the physical limit, where the product $\tilde Z' \tilde Z_{\rm uv}$ is $b^z$-independent, so for our particular one-loop study we can equivalently work with the longitudinal momentum space formula
\begin{align} \label{eq:qtmd_naive}
 \tilde f^\TMD_q(x, \bt, \mu , P^z) &
 = \tilde Z_{\rm uv}^q(\mu,P^z,\epsilon)\, \frac{\tilde B_q\bigl(x, \bt, \epsilon, L, P^z \bigr)}{\sqrt{\tilde S^q(b_T,\epsilon,L)}}
\,.\end{align}
Combining \eqs{qbeam_nlo}{qsoft_nlo} according to \eq{qtmd_naive}, we obtain the NLO result for the quasi-TMDPDF evaluated in an on-shell quark state,
\begin{align} \label{eq:qtmd_nlo}
 \tilde f_{q,P^z}^{\TMD\,(1)}(x,\bt,\eps,P^z) &
 = \tilde B_{q}^{(1)}(x, \bt, \eps, L, P^z) - \frac{1}{2} \tilde S^{(1)}(b_T,\eps,L)
\\\nn&
 = \frac{\as C_F}{2\pi} \biggl[ - \left(\frac{1}{\eps_\IR} + \Lb{} \right) P_{qq}(x)
   +  (1-x) \biggr]_+^1 \Theta(1-x)\Theta(x)
 \\\nn&
 + \frac{\as C_F}{2\pi} \delta(1-x) \biggl[ \frac{3}{2} \frac{1}{\eps_\UV}
   - \frac{1}{2} \ln^2\frac{\mu^2}{(2P^z)^2} - \ln\frac{\mu^2}{(2P^z)^2} - \frac{3}{2} \biggr]
 \\\nn&
 + \frac{\as C_F}{2\pi} \delta(1-x) \biggl[
  - \frac{1}{2} \Lb{2} + \frac{5}{2} \Lb{} + \Lb{} \ln\frac{\mu^2}{(2P^z)^2}
   \biggr]
\,.\end{align}
Although our method of calculation is quite different, we note that our result in \eq{qtmd_nlo} fully agrees with the one loop calculation in \mycite{Ji:2018hvs} (up to trivial differences in our respective conventions for the $\overline{\rm MS}$ scheme).

There is an important subtlety concerning the logarithmic dependence $\ln(2P^z)^2$ in \eqs{tmd_nlo}{qtmd_nlo},
which arises from calculating matrix elements with an on-shell external quark of momentum $P^\mu = (P^z, 0, 0, P^z)$.
In the ratio of the actual TMDPDF evaluated in a proton state, we have to replace this by $P^z\to x P^z$, where $x P^z$ is the momentum  of the struck parton (see also the discussion in \mycite{Izubuchi:2018srq}), so we obtain
\begin{align} \label{eq:qtmd_nloxP}
 \tilde f_{q}^{\TMD\,(1)}(x,\bt,\eps,P^z) &
 = \frac{\as C_F}{2\pi} \biggl[ - \left(\frac{1}{\eps_\IR} + \Lb{} \right) P_{qq}(x)
   +  (1-x) \biggr]_+^1 \Theta(1-x)\Theta(x)
 \nn\\&\quad
 + \frac{\as C_F}{2\pi} \delta(1-x) \biggl[ \frac{3}{2} \frac{1}{\eps_\UV}
   - \frac{1}{2} \ln^2\frac{\mu^2}{(2xP^z)^2} - \ln\frac{\mu^2}{(2xP^z)^2} - \frac{3}{2} \biggr]
 \nn\\&\quad
 + \frac{\as C_F}{2\pi} \delta(1-x) \biggl[
  - \frac{1}{2} \Lb{2} + \frac{5}{2} \Lb{} + \Lb{} \ln\frac{\mu^2}{(2xP^z)^2}
   \biggr]
\,.\end{align}
This is our final result for the quasi-TMDPDF which uses the natural quasi beam function and the naive quasi soft function.

To test whether a perturbative matching between $\tilde f_q^\TMD$ and $f_q^\TMD$ is possible,
we need to UV renormalize both \eqs{tmd_nlo}{qtmd_nloxP} and study their difference:
\begin{align} \label{eq:relation_nlo}
 \frac{\tilde f_{q}^{\TMD}(x, \bt, \mu, P^z)}{f_{q}^{\TMD}(x, \bt, \mu, \zeta)}
 = 1 + \frac{\as C_F}{2\pi} \biggl[& \Lb{} - \Lb{} \ln\frac{(2xP^z)^2}{\zeta}
   \\\nn&
   - \frac{1}{2} \ln^2\frac{(2xP^z)^2}{\mu^2}  + \ln\frac{(2xP^z)^2}{\mu^2}
   - 2 + \frac{\pi^2}{12}
   \biggr] + \cO(\as^2)
\,.\end{align}
As expected, the explicit infrared poles in $\eps_\IR$ have canceled, as they arise entirely from the quasi beam function, which can be related to the beam function through a boost.
However, the $b_T$ dependence of $\tilde f_q^\TMD$ and $f_q^\TMD$ does not agree, leaving two uncanceled logarithms in \eq{relation_nlo}.
The second one multiplies a logarithm of $\zeta$, and in fact is exactly the one-loop expansion of the Collins-Soper kernel in \eq{tmd_evolution},
\begin{align}
 \exp\biggl[ \frac12 \ln\frac{(2xP^z)^2}{\zeta} \gamma_\zeta(\mu,b_T) \biggr]
 = 1 -  \frac{\as C_F}{2\pi} \Lb{} \ln\frac{(2xP^z)^2}{\zeta} + \cO(\as^2)
\,.\end{align}
This confirms our argument in \sec{schematic_matching} that the Collins-Soper equation prohibits a perturbative matching, unless $\zeta$ is fixed in terms of $P^z$ such that the $\zeta$-dependence of $\tilde f_q^\TMD(x, \bt, \tilde\mu, P^z)$ and $f_q^\TMD[x, \bt, \mu, \zeta(P^z)]$ exactly cancel.
From \eq{relation_nlo}, this is fulfilled with
\begin{align}
 \zeta = (2 x P^z)^2
\,,\end{align}
as expected. This leaves
\begin{align} \label{eq:relation_nlo_fixzeta}
 \frac{\tilde f_{q}^{\TMD}(x, \bt, \mu, P^z)}{f_{q}^{\TMD}\big(x, \bt, \mu, \zeta\!=\!(2xP^z)^2\big)}
 = 1 &+ \frac{\as C_F}{2\pi} \biggl[ \Lb{} 
   - \frac{1}{2} \ln^2\frac{(2xP^z)^2}{\mu^2}  + \ln\frac{(2xP^z)^2}{\mu^2}
   - 2 + \frac{\pi^2}{12}
   \biggr] 
  \nn \\
   &+ \cO(\as^2)
   \,.
\end{align}
The key problem with \eq{relation_nlo_fixzeta} is that it still contains a single infrared logarithm $\ln(b_T^2 \mu^2/b_0^2)$ which is not associated with the Collins-Soper evolution, and thus cannot be eliminated in a similar fashion.
Curiously, choosing $\zeta = (2xP^z)^2 / e$ would simultaneously cancel both logarithms in $b_T$ in \eq{relation_nlo}, but the term involving a $\ln\zeta$ is clearly related to the Collins-Soper kernel, whereas there is no clear relationship of the term $\ln(b_T^2 \mu^2/b_0^2)$ with this evolution. 
Therefore we deem this choice with an extra $e$ to be something that works at one loop by construction, but not a valid choice since it is very unlikely to continue to work at higher loop orders (unless there happens to be some unknown deep relationship).
The presence of this extra $\ln(b_T^2 \mu^2/b_0^2)$ thus indicates a failure of the naive quasi-TMDPDF to reproduce the same infrared physics at one-loop as required for the physical TMDPDF.

Our results can also be compared to those in \mycite{Ji:2018hvs},
where the soft factor was not calculated separately, but immediately combined with the beam function to yield the quasi-TMDPDF.
The final result obtained was a relation between the quasi-TMDPDF and TMDPDF at $\mu^2 = \zeta = (2xP^z)^2$ which was given as
\begin{align} \label{eq:matching_JJYZZ}
 \tilde f_{q}^\TMD (x,\bt,\mu=2xP^z, P^z) =
 &\exp\biggl[ \int_{(b_0/b_T)^2}^{\zeta} \frac{\df\mu'^2}{\mu'^2} \frac{\as(\mu') C_F}{2\pi} \biggr]
 \, \biggl(1 - \frac{\as C_F}{\pi} \biggr)
 \nn\\&\times
 \, f_{q}^\TMD\bigl(x, \bt, \mu=\sqrt\zeta, \zeta=(2xP^z)^2\bigr)
\,.\end{align}
If we take our result in \eq{relation_nlo} and set $\mu^2=\zeta=(2xP^z)^2$, then we obtain
\begin{align} \label{eq:relation_nlo_fixzeta_fixmu}
 \frac{\tilde f_{q}^{\TMD}(x, \bt, \mu=\sqrt{\zeta}, P^z)}{f_{q}^{\TMD}\big(x, \bt, \mu=\sqrt{\zeta}, \zeta\!=\!(2xP^z)^2\big)}
 = 1 &+ \frac{\as C_F}{2\pi} \biggl[ \Lb{} 
   - 2 + \frac{\pi^2}{12}
   \biggr] 
   + \cO(\as^2)
   \,.
\end{align}
This agrees with expanding \eq{matching_JJYZZ} to ${\cal O}(\alpha_s)$, where the single infrared logarithm is generated by the exponential term. (There is a
trivial mismatch from the constant $\pi^2/12$ term which arises because \mycite{Ji:2018hvs} uses a different definition of the $\MS$ scheme, see \app{conventions}.) In Ref.~\cite{Ji:2018hvs} \eq{matching_JJYZZ} was interpreted as being a valid matching formula between quasi-TMDPDF and TMDPDF.
However, for nonperturbative $b_T$ the exponential in \eq{matching_JJYZZ} becomes a nonperturbative function and cannot be included in a short distance matching coefficient, in agreement with our conclusions.

We conclude this section with an overview of the dependence of the one-loop coefficients of beam function $B_q$, soft subtraction $\Delta_S^q$ and TMDPDF $f_q^\TMD$ on the logarithm $L_b = \ln(b_T^2 \mu^2 / b_0^2)$ for the different regulators in the literature, as shown in \tbl{LogB}. The dependence of the quasi constructions $\tilde B_q$, $\tilde\Delta_S^q$ and $\tilde f_q^\TMD$ on $L_b$ is also shown in the lower part of the table.
Here we only show the dependence on standalone factors of $L_b$ (providing references for the full expressions in the table caption). 
As discussed previously, the quasi functions do not match their lightlike counterparts in any of the regulators.
In particular, the double-logarithm $L_b^2$ only agrees with Collins' regulator. All the other regulators involve Wilson lines with light-like directions, and here the double logarithm is part of the soft function (it is split between these two in the exponential regulator).\footnote{Note that in none of these cases does one include the transverse self energy, which would add a single logarithm $L_b$ to $B_q$ and $-L_b$ to $\Delta_S^q=1/\sqrt{S^q}$, see \eq{soft_L_d}.
Even after taking this into account, the mismatch persists.}

{
\renewcommand{\arraystretch}{1.4}
\begin{table}[t!]
 \centering
 \begin{tabular}{|c|c|c|c|}
  \hline
  Regulator & Beam function $B_q$ & Soft factor $\Delta_S^q$ & TMDPDF $f_q^\TMD = B_q \Delta_S^q$
  \\ \hline
  Collins \cite{Collins:1350496}
  & $-\frac12 L_b^2 \,,\, \frac52 L_b$
  & $- L_b$
  & $-\frac12 L_b^2 \,,\, \frac32 L_b$
  \\ \hline
  $\delta$ regulator \cite{GarciaEchevarria:2011rb,Echevarria:2012js}
  & $\frac32 L_b$
  & $-\frac12 L_b^2$
  & $-\frac12 L_b^2 \,,\, \frac32 L_b$
  \\ \hline
  $\eta$ regulator \cite{Chiu:2011qc,Chiu:2012ir}
  & $\frac32 L_b$
  & $-\frac12 L_b^2$
  & $-\frac12 L_b^2 \,,\, \frac32 L_b$
  \\ \hline
  Exp.~regulator \cite{Li:2016axz}
  & $-L_b^2 \,,\, \frac32 L_b$
  & $\frac12 L_b^2$
  & $-\frac12 L_b^2 \,,\, \frac32 L_b$
  \\ \hline
  & quasi $\tilde B_q$ & quasi $\tilde\Delta_S^q$ & quasi $\tilde f_q^\TMD = \tilde B_q \tilde \Delta_S^q$
  \\ \hline
  Finite $L$, naive $\tilde \Delta_S^q$
  & $-\frac12 L_b^2 \,,\, \frac92 L_b$
  & $-2 L_b$
  & $-\frac12 L_b^2 \,,\, \frac52 L_b$
  \\ \hline
  Finite $L$, bent $\tilde \Delta_S^q$
  & $-\frac12 L_b^2 \,,\, \frac92 L_b$
  & $-3 L_b$
  & $-\frac12 L_b^2 \,,\, \frac32 L_b$
  \\ \hline
 \end{tabular}
 \caption{Dependence of unsubtracted beam function $B_q \equiv B_q^\unsub$, soft subtraction $\Delta_S^q$ and TMDPDF $f_q^\TMD$ (upper part) and their quasi constructions $\tilde B_q$, $\tilde\Delta_S^q$, $\tilde f_q^\TMD$ (lower part) on the logarithm $L_b = \ln(b_T^2 \mu^2 / b_0^2)$ in  various rapidity regularization schemes. Results are shown for terms from the one loop matrix elements, pulling out an overall $\as/(2\pi)$.
 Only pure $L_b$ terms are shown. The full functional form for $B_q$, $\Delta_S^q$, and $f_q^\TMD$ can be found in \app{overview_tmdpdfs} for all regulators.
 The corresponding results are given in Eqs.\ \eqref{eq:qbeam_nlo}, \eqref{eq:qsoft_nlo} and \eqref{eq:qtmd_nlo} for the naive quasi-TMDPDF, and in Eqs.\ \eqref{eq:qbeam_nlo}, \eqref{eq:S_bent_nlo} and \eqref{eq:qtmd_nloxP_bent} for the quasi-TMDPDF using the bent soft function.
 }
 \label{tbl:LogB}
\end{table}
}

\subsection{Quasi-TMDPDF using a bent soft function}
\label{sec:qsoft_bent_nlo}

The construction of the quasi beam function in \sec{qbeam} was motivated by the physical picture of boosting a spatial to a lightlike correlation function, while the (naive) quasi soft factor construction in \sec{qsoft} was simply the most straightforward attempt.  This lead to a quasi-TMDPDF whose IR logarithms do not match those of the TMDPDF. 
However, there is significant freedom in constructing quasi functions on lattice, so we can consider alternate definitions with the goal of finding one which has the same infrared physics as the TMDPDF. 
When the quasi beam function and quasi soft function are combined, any dependence related to the method of regulating rapidity divergences (such as finite $L$) cancels. 
Since it was only the presence of rapidity regulators that causes problems for the physical boost argument for the beam function, one may infer that this issue is alleviated when considering the matching for the quasi-TMDPDF and TMDPDF. For this reason we will not try to adjust the definition of the quasi beam function here. 
However, the quasi soft function was not constructed based on a boost argument, and hence seems like the most likely culprit for the failure to match infrared logarithms. For the soft factors contained in the TMDPDF there are always two different spatial directions involved in the Wilson lines, while for our naive quasi soft factor there was only the $z$-direction. This motivates us to consider in this section a different ``bent'' quasi soft function which involves two spatial directions. 

This need for this type of bent quasi soft function can also be motivated by 
studying the failure of the naive quasi soft function to reproduce the IR physics needed for the TMDPDF in more detail. In particular, we can split the calculation of the naive quasi soft function into three distinct pieces,
arising from gluon exchanges either within the $\hat z$ or $-\hat z$ sector,
or between them, as done in \sec{qsoft_naive_nlo},
\begin{align}
 \tilde S^{(1)} = \tilde S_{z/z}^{(1)} + \tilde S_{-z/-z}^{(1)} + \tilde S_{z/-z}^{(1)}
\,.\end{align}
Physically, the first term is correctly boosted towards a $n/n$ contribution
by boosting with $P^z > 0$, and likewise the second term is boosted towards $\bn/\bn$
for $P^z < 0$. In practice, they are identical due to invariance under $z \leftrightarrow -z$.
At one loop, the quasi-TMDPDF in \eq{qtmd_naive} hence can be written as
\begin{align}
 \tilde f_{q}^{\TMD (1)} &
 = \tilde B^{(1)}_{q} - \tilde S_{z/z}^{(1)} - \frac{1}{2} \tilde S_{z/-z}^{(1)}
\,.\end{align}
Next, recall the Wilson line structures of quasi beam and quasi soft functions in \eqs{qbeam}{qsoft}, see also \figs{qbeam_Wilson_lines}{qsoft_Wilson_lines}.
Taking the soft limit of the quasi beam function, $b^z = 0$,
clearly gives the same Wilson lines as half of the quasi soft function,
and hence subtracting $\tilde S_{z/z}^{(1)}$ exactly cancels the soft limit
of the tadpole correction to the beam function. This can easily be verified
by comparing \eqs{qsoft_nlo_z}{qtmd_c}, from which one also sees
that this subtraction cancels the divergence in $L/b_T$.
It remains to consider $\tilde S_{z/-z}^{(1)}$, given in \eq{qsoft_nlo_zz}.
Its subtraction from $\tilde B_{q}$ adds a single $\ln(b_T^2\mu^2/b_0^2)$, which is exactly the leftover logarithm found in the relation between the naive quasi-TMDPDF and TMDPDF in \eq{relation_nlo}.
In conclusion, it thus appears to be precisely the interaction between the $+\hat z$ and $-\hat z$
part of the naive soft function that is spoiling the matching of infrared logarithms. 

\begin{figure}[pt]
 \centering
 \begin{subfigure}{0.4\textwidth}
  \includegraphics[height=6cm]{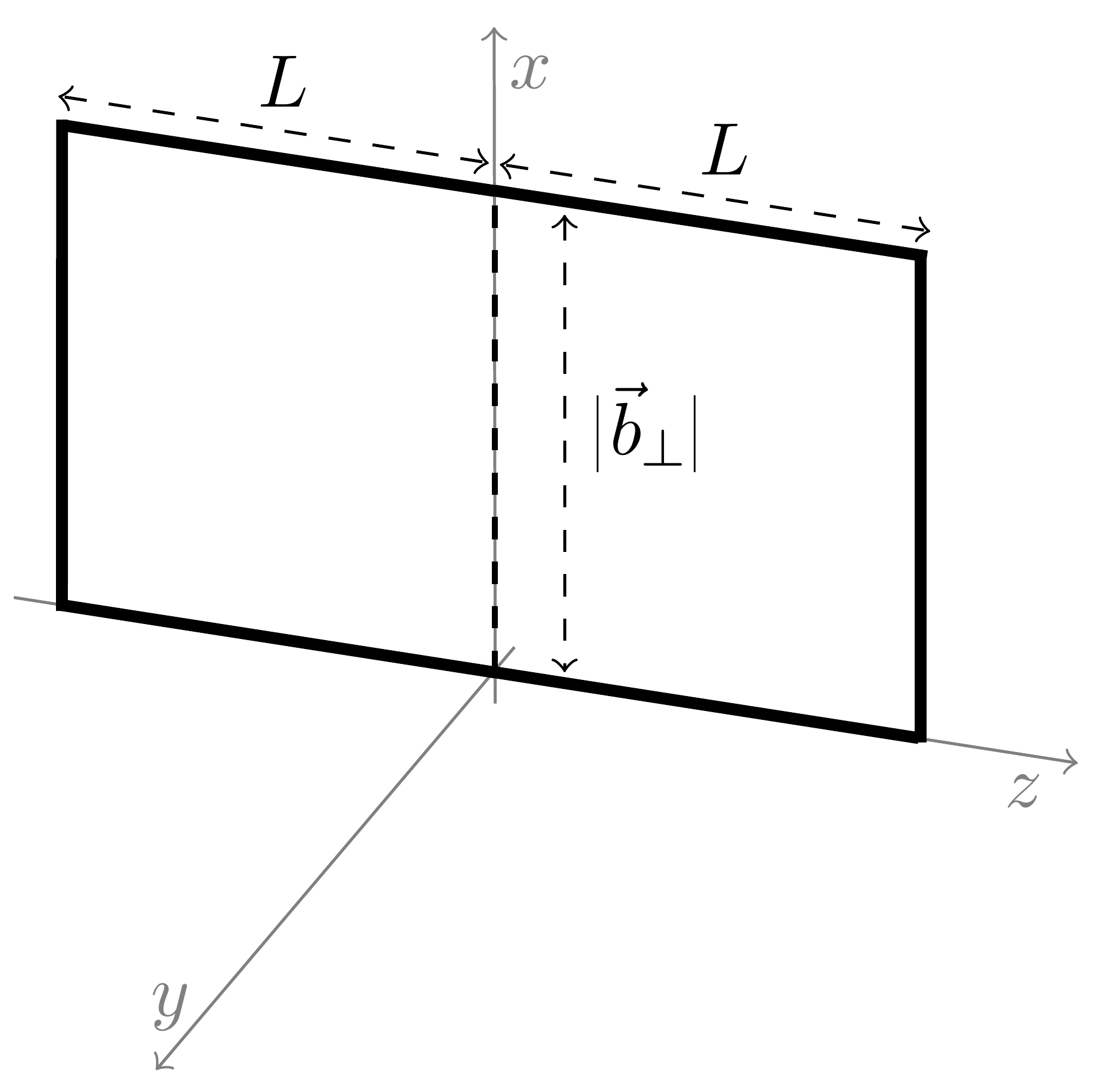}
  \caption{Naive quasi soft function}
  \label{fig:bent_soft_a}
 \end{subfigure}
 \qquad
 \begin{subfigure}{0.4\textwidth}
  \includegraphics[height=6cm]{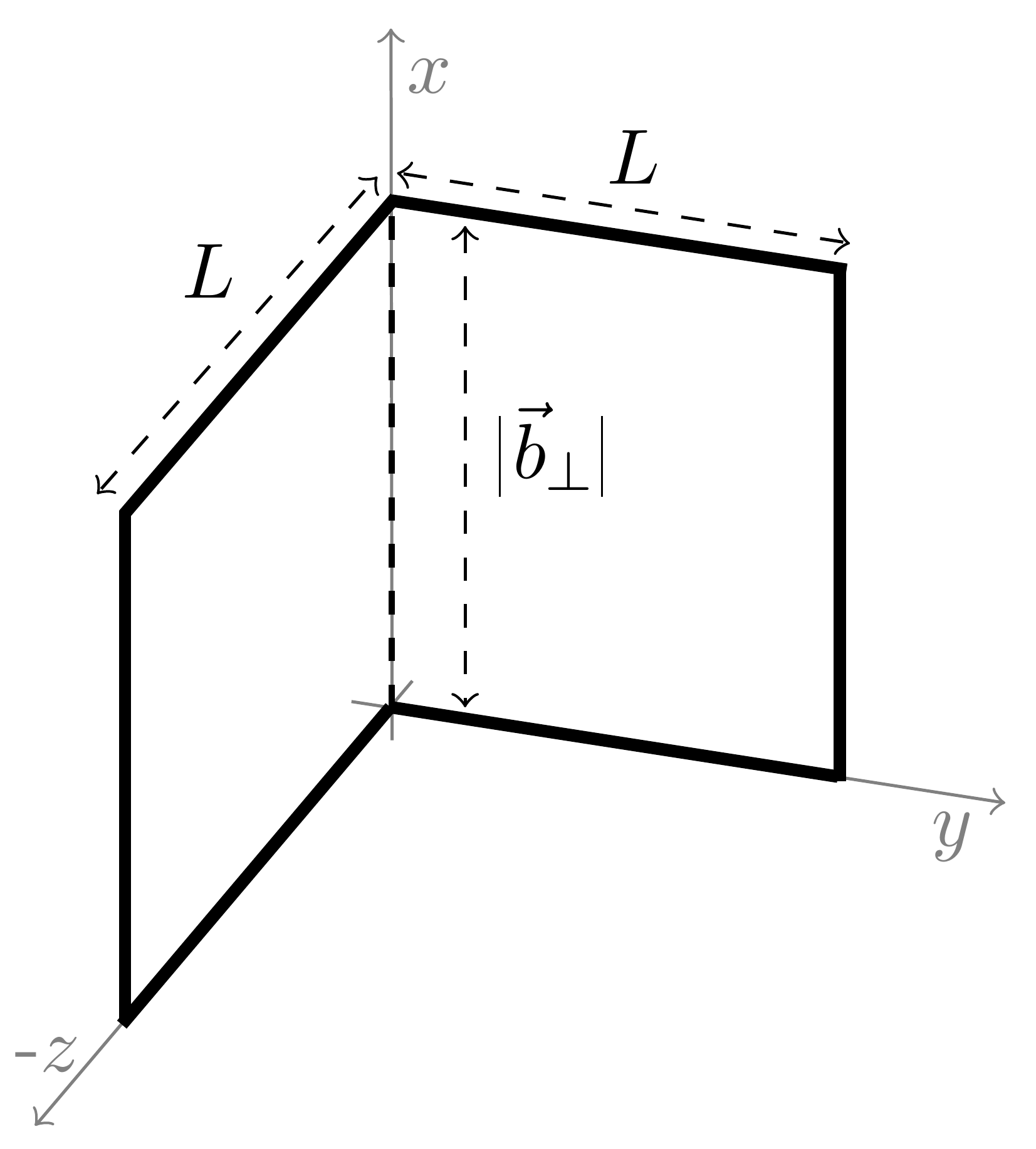}
  \caption{Bent quasi soft function}
  \label{fig:bent_soft_b}
 \end{subfigure}
 \caption{The bent quasi soft function. The impact parameter $\bt$ is aligned with the $x$ axis.}
 \label{fig:bent_soft}
\end{figure}

With these motivations and observations we can define a bent quasi soft function which gives a valid matching result between quasi-TMDPDF and TMDPDF at one-loop order.  Crucially, it must still cancel the $L/b_T$ divergence in the quasi beam function and after combination with the quasi beam function produce the same logarithms in $b_T$ as the TMDPDF.
More concretely, we can demand the Wilson line structure in the $z$ sector to match the soft-expanded ($b^z = 0$) structure of the naive quasi beam function
to ensure the cancellation of rapidity divergences.
Given these restrictions we define the ``bent'' soft function as
\begin{align} \label{eq:qsoft_bent}
 \tilde S_{\rm bent}(b_T, a, L) &= \frac{1}{N_c} \bigl< 0 \big|
  \Tr\bigl\{ S^\dagger_{\hat z}(\bt;L) S_{-\bn_\perp}\!(\bt;L)
   S_{T}(L\bn_\perp;\vec b_T,\vec 0_T)
   \nn\\&\hspace{2cm}\times
 S^\dagger_{-\bn_\perp}\!(\vec 0_T;L) S_{\hat z}(\vec 0_T;L)
 S_{T}^\dagger\bigl(-L \hat z;\vec b_T,\vec 0_T\bigr) \bigr\}
 \bigl|0 \bigr>
\,,\end{align}
where $\bn_\perp^\mu$ is the transverse unit vector orthogonal to $n_\perp^\mu = b_\perp^\mu / b_T$ and $\hat z$.
\fig{bent_soft} illustrates the Wilson line path in \eq{qsoft_bent} and compares it to the path for naive quasi soft function defined in \eq{qsoft}.

Above we deduced that the failure of a perturbative one-loop matching between quasi-TMDPDF and TMDPDF could be traced to soft diagrams mediating exchange between Wilson lines along the $+\hat z$ and $-\hat z$ directions.
These diagrams precisely vanish for the bent soft function due to $\bn_\perp \cdot n_\perp = \bn_\perp \cdot \hat z = 0$,
while all other diagrams are not affected by the new Wilson line paths.
Hence the bent soft function at one loop yields
\begin{align} \label{eq:S_bent_nlo}
 \tilde S_{\rm bent}^{(1)}(b_T,\eps,L) =
 \frac{\as C_F}{2\pi} \biggl[ \frac{6}{\eps} + 6\Lb{} + \frac{4\pi L}{b_T}  + 4\biggr]
\,.\end{align}
As before, it is related to the soft subtration through $\tilde\Delta_S^q = 1/\sqrt{\tilde S_{\rm bent}}$. This bent soft factor has precisely the infrared logarithms that are desired at one loop.\footnote{
A soft factor with more than one transverse directions was also used in \mycite{Ji:2014hxa} where the goal was to express the physical factorization theorem directly in terms of quasi objects. They define a soft factor
\begin{align} \label{eq:qsoft_JSXY}
 \tilde \Delta_S^{q\,{\rm JSXY}}(b_T)
 = \sqrt{ \frac{\tilde S_{n_x,n_y}(b_T)}{\tilde S_{n_x,n_z}(b_T) \tilde S_{n_z,n_y}(b_T)} }
\,,\end{align}
where $\tilde S_{n_1,n_2}(b_T)$ is the same as the result obtained from our naive quasi soft matrix element, \eq{qsoft}, by replacing $\hat z \to n_1$ and $-\hat z \to n_2$ 
but using infinitely-long Wilson lines ($L\to \infty$). 
Taking $b_T^\mu=b_Tn_x^\mu$ and using instead finite Wilson lines, we have checked that the resulting combination of terms in $\tilde \Delta_S^{q}(b_T, L)$ gives the same result at one-loop as that of our bent soft function.
}

Using \eq{qtmd_naive} to combine the bent quasi soft function from \eq{S_bent_nlo} together with the natural quasi beam function from \eq{qbeam_nlo} we obtain a new quasi-TMDPDF
\begin{align} \label{eq:qtmd_nloxP_bent}
 \tilde f_{q}^{\TMD\,(1)}(x,\bt,\eps,P^z)&
 = \frac{\as C_F}{2\pi} \biggl[ - \left(\frac{1}{\eps_\IR} + \Lb{} \right) P_{qq}(x)
   +  (1-x) \biggr]_+^1 \Theta(1-x)\Theta(x)
 \nn\\&\quad
 + \frac{\as C_F}{2\pi} \delta(1-x) \biggl[ \frac{1}{2} \frac{1}{\eps_\UV}
   - \frac{1}{2} \ln^2\frac{\mu^2}{(2xP^z)^2} - \ln\frac{\mu^2}{(2xP^z)^2} - \frac{3}{2} \biggr]
 \nn\\&\quad
 + \frac{\as C_F}{2\pi} \delta(1-x) \biggl[
  - \frac{1}{2} \Lb{2} + \frac{3}{2} \Lb{} + \Lb{} \ln\frac{\mu^2}{(2xP^z)^2}
   \biggr]
\,.\end{align}
Comparing this result to the TMDPDF at one loop yields
\begin{align} \label{eq:goodrelation_nlo_fixzeta}
 \frac{\tilde f_{q}^{\TMD}(x, \bt, \mu, P^z)}{f_{q}^{\TMD}\big(x, \bt, \mu, \zeta\!=\!(2xP^z)^2\big)}
 = 1 &+ \frac{\as C_F}{2\pi} \biggl[
   - \frac{1}{2} \ln^2\frac{(2xP^z)^2}{\mu^2}  + \ln\frac{(2xP^z)^2}{\mu^2}
   - 2 + \frac{\pi^2}{12}
   \biggr] 
  \nn \\
   &+ \cO(\as^2)
   \,,
\end{align}
where have again fixed $\zeta=(2xP^z)^2$ as explained previously.
Since there is no $b_T$ dependence on the RHS of \eq{goodrelation_nlo_fixzeta}, we see that all infrared logarithms of the TMDPDF are correctly reproduced by this quasi-TMDPDF construction at one loop. Thus this construction obeys the matching relation given in \eq{matching_result0} with a one loop result for the matching coefficient that is given by 
\begin{align} \label{eq:C_nlo}
 C_{qq'}^\TMD\bigl(\mu, x P^z\bigr) &
 = \delta_{q q'} \biggl[1 + \frac{\as C_F}{4\pi} \biggl( - \ln^2\frac{(2 x P^z)^2}{\mu^2}
                       + 2 \ln\frac{(2 x P^z)^2}{\mu^2} -4 + \frac{\pi^2}{6} \biggr) + \cO(\as^2) \biggr]
\,.
\end{align}
Here, we ignore possible mixing of quarks with gluons. Then since mixing of quark flavors can first arise at two loops, the one-loop coefficient is proportional to $\delta_{qq'}$.
This result provides a valid one-loop perturbative matching coefficient, which only depends on the hard scale of the struck parton, $xP^z$. 

Assuming the validity of this quasi-TMDPDF construction beyond one loop, \eq{C_nlo} can be used to match the lattice quasi-TMDPDF to the TMDPDF. To obtain the required input for this result one combines lattice calculations of the natural quasi beam function and bent quasi soft function to obtain a lattice quasi-TMDPDF, which is then converted into the $\overline{\rm MS}$ scheme. Results for matching in more lattice friendly renormalization schemes should be straightforward to derive following a similar approach to the one used here (see e.g.~\cite{Constantinou:2017sej,Stewart:2017tvs}).

\section{Results and Outlook}
\label{sec:results}

In this section, we briefly summarize the impact of our calculations in the previous sections for the matching between quasi-TMDPDF and TMDPDF, and what questions remain open for further study. Without relying on the existence of a quasi soft function that yields the correct infrared physics for a quasi-TMDPDF, we also discuss precisely what constraints on TMDPDFs can still be rigorously derived from lattice calculations.

\subsection{Matching relation between quasi-TMDPDF and TMDPDF}
\label{sec:relation_final}

The goal of this work was to establish a matching relation between the quasi-TMDPDF and TMDPDF analogous to the collinear PDF, where LaMET gives such a perturbative relation.  However, the physical picture for the existence of such a matching relation is much more complicated than in the PDF case. For the beam function, the need for a non-trivial rapidity regulator on the TMDPDF side appears to spoil the simple boost correspondence between hadronic quasi and non-quasi matrix elements. We have confirmed that this is the case at one loop in \sec{qbeam_nlo} and \tbl{LogB} by making comparisons of the most natural quasi beam function with all modern TMDPDF definitions for the beam function. For the soft function, the vacuum matrix elements that appear necessarily involve two directions, and hence does not satisfy a simple boost relation to a quasi soft function even at the bare level. In \sec{qtmd_nlo} we computed the most naive quasi soft function at one loop, and showed that it yields a quasi-TMDPDF which does not have infrared logarithms that agree with those of the TMDPDF.  Then in \sec{qsoft_bent_nlo} we considered a modified bent quasi soft function at one loop, which when combined with the natural beam function gives our  preferred quasi-TMDPDF definition. Its infrared logarithms at one loop properly agree with those of the TMDPDF, thus leading to a consistent matching formula at one loop.

The full utility of our preferred quasi-TMDPDF definition will depend on whether the correspondence between infrared logarithms continues to hold at higher orders in perturbation theory.  
For example, at two loops one can recouple the $-\hat z$ and $+\hat y$ sectors in the bent soft function, so these contributions will have to give contributions that match up correctly with corresponding contribution in the two loop soft function once its combined into the TMDPDF. Such calculations should be considered in the near future. 
More rigorously, one needs to show that our bent quasi soft function together with the natural quasi beam function yields a quasi-TMDPDF with the same infrared nonperturbative physics as the lightlike soft function, either nonperturbatively or at least to all orders in perturbation theory.
This is obviously also an important avenue for future work.

With lack of further information, in order to proceed at the current time one can do one of two things, i) make the assumption that 3our preferred quasi-TMPDF holds nonperturbatively, or ii) assume that our bent quasi soft factor may not work to all orders, and see if interesting constraints can still be derived. We discuss these in turn.

In the case of i) we have the matching formula derived in \sec{schematic_matching}, which for a non-singlet ($\ns$) quark flavor combination where there is no mixing reads:
\begin{align} \label{eq:matching_result}
\tilde f_{\ns}^\TMD(x, \bt, \mu, \tilde P^z) &
 = C^\TMD_{\ns}\bigl[x, \mu, \tilde P^z, \tilde \zeta(x,\tilde P^z)\bigr]
\exp\biggl[\frac12 \gamma_\zeta^q(\mu,b_T) \ln\frac{\tilde \zeta(x,\tilde P^z)}{\zeta} \biggr]
\nn\\&\quad\times
f^\TMD_{\ns}(x, \bt, \mu, \zeta)
\,.\end{align}
If the infrared structure of $\tilde f_{\ns}^\TMD(x, \bt, \mu, P^z)$ and $f^\TMD_{\ns}(x, \bt, \mu, \zeta=(2xP^z)^2)$ match to all orders, then our one loop calculation in \sec{qsoft_bent_nlo} provides a valid result for $C^\TMD_{\ns}$. For the matching from the $\MS$ renormalized quasi-TMDPDF it gives
\begin{align} \label{eq:kernels_nlo_ns}
 C^\TMD_{\ns}\bigl(\mu, x P^z\bigr) &
 \equiv C_\ns^\TMD\bigl[x, \mu, P^z, \tilde \zeta(x,P^z)=(2xP^z)^2\bigr]
 \nn\\&
 = 1 + \frac{\as C_F}{4\pi} \biggl( - \ln^2\frac{(2 x P^z)^2}{\mu^2}
                       + 2 \ln\frac{(2 x P^z)^2}{\mu^2} -4 + \frac{\pi^2}{6} \biggr) + \cO(\as^2) 
\,.\end{align}
At this order, the kernel is diagonal in the quark flavors, as flavor mixing can first appear at two loops. Under the same assumptions one can proceed to carry out calculations for our preferred quasi-TMDPDF on the lattice, and with suitable renormalization and scheme changes, then use the matching relation in \eq{kernels_nlo_ns} to obtain the physical TMDPDF for the nonsinglet case. 
 
In the case of ii) we assume that the matching between quasi-TMDPDF and TMDPDF is spoiled by a mismatch between the soft factors $\tilde\Delta_S^q$ and $\Delta_S^q$ at higher orders. In this case, although we do not have a matching relation, we can still write down a formula relating the quasi-TMDPDF and TMDPDF. For the nonsinglet case without mixing it reads:
\begin{align} \label{eq:relation_final_g}
 \tilde f_{\ns}^\TMD(x, \bt, \mu, P^z)
 = &~C^\TMD_{\ns}\bigl(\mu, x P^z\bigr)
   g^S_{q}(b_T, \mu)
   \exp\biggl[ \frac12  \gamma_\zeta^q(\mu, b_T) \ln\frac{(2 x P^z)^2}{\zeta} \biggr]
 \nn\\&\times
 f^\TMD_{\ns}(x, \bt, \mu, \zeta)
\,.\end{align}
Here, $C_\ns^\TMD$ is by definition still a perturbative function. On the other hand, the function $g^S_{q}(b_T,\mu)$ is nonperturbative and is associated with the failure of constructing a proper quasi soft function. It corrects the mismatch in the infrared physics.
Since the quasi soft factor and soft factor which differ are flavor independent, there are at most two different $g^S_i(b_T,\mu)$'s, namely for quarks $i=q$ and gluons $i=g$, and only $g^S_{q}$ shows up for the non-singlet flavor case. Note that \eq{relation_final_g} is also satisfied if we use the naive quasi soft function construction. Summarizing our results obtained with the naive and bent quasi soft functions we have
\begin{align}
 g^S_{\ns}(b_T,\mu) \Big|_{\text{naive quasi soft}} &= 1 + \frac{\as C_F}{2\pi} \Lb{} + \cO(\as^2)
 \,, \nn\\
 g^S_{\ns}(b_T,\mu) \Big|_{\text{bent quasi soft}} &= 1  + \cO(\as^2)
 \,.
\end{align}
Interestingly, even with the less strong assumptions present in case ii) we can still extract non-trivial information about the TMDPDF by using ratios of distribution functions where the $g_q^S$ factors cancel out. We discuss this further in the next section.

\subsection{Ratios of TMDPDFs}
\label{sec:ratios}
While the presence of the soft sector seems to prohibit a straightforward computation of TMDPDFs on lattice,
one can employ the fact that $g_{i}^S(b_T,\mu)$ only differs for quarks and gluons but otherwise is flavor blind
to try to construct ratios of quasi-TMDPDFs where $g_{i}^S(b_T,\mu)$ cancels.
To avoid possible mixing between quarks and gluons, here we only consider nonsinglet quasi-TMDPDFs such as $\ns=u{-}d$,
where \eq{relation_final_g} applies.
The soft factor $g_q^S$ cancels in any ratio of two (quasi) TMDPDFs with the same choice of $b_T$ and $\mu$, so we have
\begin{align} \label{eq:qTMD_ratio}
 \frac{\tilde f_\ns^\TMD(x_1, \bt, \mu, P_1^z)}{\tilde f_\ns^\TMD(x_2, \bt, \mu, P_2^z)}
 &= \frac{C^\TMD_\ns\bigl(\mu, x_1 P_1^z\bigr)}{C^\TMD_\ns\bigl(\mu, x_2 P_2^z\bigr)}
   \exp\biggl[\frac12\gamma_\zeta^q(\mu, b_T) \ln\biggl(\frac{(2 x_1 P_1^z)^2}{(2 x_2 P_2^z)^2} \frac{\zeta_2}{\zeta_1} \biggr)  \biggr]
   \nn\\*&\quad\times
   \frac{f^\TMD_\ns(x_1, \bt, \mu, \zeta_1)}{f^\TMD_\ns(x_2, \bt, \mu, \zeta_2)}
\,.\end{align}
For example, one can choose $x_1 = x_2$ and $\zeta_1 = \zeta_2$ to expose the Collins-Soper kernel $\gamma_\zeta^q$ as
\begin{align}
 \frac{\tilde f_\ns^\TMD(x, \bt, \mu, P_1^z)}{\tilde f_\ns^\TMD(x, \bt, \mu, P_2^z)}
 &= \frac{C^\TMD_\ns\bigl(\mu, x P_1^z\bigr)}{C^\TMD_\ns\bigl(\mu, x P_2^z\bigr)}
   \exp\biggl[ \gamma_\zeta^q(\mu, b_T) \ln\frac{P_1^z}{P_2^z} \biggr]
\,.\end{align}
This allows one to determine the nonperturbative $b_T$ dependence of $\gamma_\zeta^q$ from the ratio of the quasi-TMDPDFs $\tilde f^\TMD_\ns$ computed in lattice QCD, and was proposed in \mycite{Ebert:2018gzl}.%
\footnote{In the context of lattice calculations, it is also worth noting that $\gamma_\zeta^q$ depends on the quark masses in the discretized action used in the computation, however, due to its independence of the hadronic state it is independent of the valence quark masses used to construct correlation functions. We thank Phiala Shanahan and Michael Wagman for discussions on this point.}

Without this knowledge of $\gamma_\zeta^q$, one is forced to choose $x_1 P_1^z = x_2 P_2^z$ and $\zeta_1 = \zeta_2$ in \eq{qTMD_ratio} to cancel the Collins-Soper evolution kernel,
\begin{align} \label{eq:qTMD_ratio2}
 \frac{\tilde f_{\ns/h_1}^\TMD(x, \bt, \mu, P^z)}{\tilde f_{\ns/h_2}^\TMD(x, \bt, \mu, P^z)}
 &= \frac{C^\TMD_{\ns/h_1}\bigl(\mu, x P^z\bigr)}{C^\TMD_{\ns/h_2}\bigl(\mu, x P^z\bigr)}
   \frac{f^\TMD_{\ns/h_1}(x, \bt, \mu, \zeta)}{f^\TMD_{\ns/h_2}(x, \bt, \mu, \zeta)}
\,.\end{align}
Here, $h_1$ and $h_2$ for example refers to TMDPDF in different hadron states (which does not affect $C^\TMD_{\ns/h_i}$), or TMDPDFs of different spin structures (which can affect $C^\TMD_{\ns/h_i}$).
The latter requires both spin structures to be either $T$-even or $T$-odd, as the soft function is not $T$-invariant, see e.g.\ \mycite{Collins:1350496}. 
Using the results given in \sec{nlo_results}, it is easy to confirm that the infrared logarithms cancel independently at one loop on the left and right hand sides of \eq{qTMD_ratio2}, confirming that it indeed is a matching equation at this order.

In Eqs.\ \eqref{eq:qTMD_ratio}--\eqref{eq:qTMD_ratio2} we write ratios of quasi-TMDPDFs, which contains some choice for the quasi soft factor that satisfies \eq{relation_final_g}.
However, one can completely avoid the need for including a quasi soft factor in such ratios by employing \eq{qtmdpdf}, which yields
\begin{align} \label{eq:fgiveBratio}
 \frac{\tilde f_\ns^\TMD(x_1, \bt, \mu, P_1^z)}{\tilde f_\ns^\TMD(x_2, \bt, \mu, P_2^z)} &
 = \frac{\int\!\df b^z \, e^{\img b^z (x_1 P_1^z)} \tilde Z'(b^z,\mu,\tilde \mu) \tilde Z_{\rm uv}(b^z,\tilde \mu, a)
 \tilde B_{q}(b^z, \bt, a, L, P_1^z)}
        {\int\!\df b^z \, e^{\img b^z (x_2 P_2^z)} \tilde Z'(b^z,\mu,\tilde \mu) \tilde Z_{\rm uv}(b^z,\tilde \mu, a)
 \tilde B_{q}(b^z, \bt, a, L, P_2^z)}
\,.\end{align}
In this ratio, the quasi soft factor cancels because it is independent of $b^z$ and $P_{1,2}^z$. Thus \eq{fgiveBratio}
removes the necessity to calculate a quasi soft matrix element on lattice.
The leftover divergences as $L\to\infty$ present in the individual $\tilde B_q$ functions also cancel between the numerator and denominator.

Note that our analysis of ratios of quasi-TMDPDFs here differs somewhat from that of \mycites{Engelhardt:2015xja,Yoon:2016dyh,Yoon:2017qzo}. In those references Lorentz invariance is used to directly relate ratios of $b_T$-dependent spacelike TMDPDFs to the corresponding ratios of the physical TMDPDFs, analogous to our \eq{qTMD_ratio2}. However this is done by considering an adjustable spatial path for the beam functions, and taking the limit where this path approaches the light-cone. So far the focus has been on the case with $b^z\to 0$ corresponding to integrating over $x$. The relations they use do not require a non-trivial matching coefficient $C^\TMD$, but they do require a non-trivial adjustment of the path as the light-cone limit is taken. It would be interesting to consider a more detailed analysis of the difference between our approach and theirs.

\section{Conclusion}
\label{sec:outlook}

In this paper we have studied the possibility to obtain quark TMDPDFs from lattice QCD using the LaMET approach.
LaMET has been successfully applied to obtain collinear PDFs using a perturbative matching relation from quasi-PDFs, which are equal-time correlators evaluated in a highly-boosted hadron state.
The construction of quasi-TMDPDFs is severely complicated by the presence of rapidity divergences that require a dedicated regulator and the need to combine the beam function $B_q$, a collinear hadronic matrix element, with a soft factor $\Delta_S^q$ defined through a soft vacuum matrix element.

We have first discussed why the analog of rapidity divergences do not pose a problem for lattice calculations,
as they are fully regulated by the finite length $L$ of Wilson lines, both in the lightlike case of TMDPDFs and the equal-time case of quasi-TMDPDFs.

We then separately discussed constructions of the quasi beam function $\tilde B_q$ and the quasi soft factor $\tilde \Delta_S^q$.
Since beam functions only depend on the flavor of the probed parton and the direction of the struck hadron, similar to normal PDFs, there is a natural definition of the quasi beam function as an equal-time matrix element following the LaMET procedure. With this definition the bare operator matrix element can be boosted onto the bare beam function operator matrix element.
On the other hand, the soft function depends on the direction of both hadrons and thus can not be obtained from boosting a purely spatial matrix element.
 
We hence first studied a naive quasi soft function where one replaces the lightlike directions by $n^\mu\to \hat z^\mu$ and $\bn^\mu\to -\hat z^\mu$.
Both matrix elements separately suffer from rapidity divergences that are regulated by the length $L$, and we construct the naive quasi-TMDPDF $\tilde f_q^\TMD = \tilde B_q \tilde\Delta_S^q$ by demanding the cancellation of the $L$ dependence.
We also discussed implications of the Collins-Soper evolution on the matching relation between $f_q^\TMD$ and $\tilde f_q^\TMD$, which (if it exists) requires one to fix the Collins-Soper scale $\zeta$ in the TMDPDF through the proton momentum $P^z$ in the quasi-TMDPDF, and for the relation not to involve a convolution in a momentum fraction.

We have carried out a detailed one-loop calculation to study whether a perturbative matching is feasible despite the failure of the physical boost picture. For this, we require that all logarithmic dependence on the transverse separation $b_T \sim q_T^{-1} \sim \LQCD^{-1}$, assumed to be a nonperturbative scale, must agree between the lightlike functions and their quasi constructions.
We find that this consistency test fails for the quasi beam function, the naive quasi soft function and the naive quasi-TMDPDF.
For the quasi beam function, we interpret this to be due the need to regulate rapidity divergences, which necessarily breaks boost invariance and thereby invalidates the simple boost relation.
Even after combining $\tilde B_q$ and $\tilde\Delta_q^S$ into the quasi-TMDPDF $\tilde f_q^\TMD$, in which case all regularization dependence cancels, there is still a mismatch between TMDPDF and quasi-TMDPDF. 

To fix this inconsistency we were motivated to consider a ``bent'' quasi soft function which involves an equal time operator with Wilson lines on two different spatial paths. 
At one loop, one can identify the diagrams that violate the boost relation in the soft function, which motivated the precise definition of a ``bent'' soft function. This leads to a quasi soft factor $\tilde \Delta_S^q$ which gives a quasi-TMDPDF which matches the $b_T$ dependence of the TMDPDF, establishing a perturbative matching relation at least up to one loop. 

If our construction with a bent quasi soft function and the natural quasi beam function within LaMET works beyond one loop order then it, for the first time, provides a definite method to fully access the complete physical TMDPDF from lattice QCD computations. Given the importance of such a construction, it is very important to further test this relationship beyond one loop, as we have emphasized repeatedly.  

Even if this construction breaks down at higher order, one can still consider ratios of quasi-TMDPDFs, where any mismatch in the soft physics cancels. This enables one to study ratios of TMDPDFs.
For example, this allows one to nonperturbatively determine the Collins-Soper kernel, which has been presented in detail in \mycite{Ebert:2018gzl}, or to study ratios of TMDPDFs using different hadron states or spin structures as described in \sec{results}.
Such information is potentially useful to constrain TMDPDFs and thereby aid their extraction from experiment,
particularly for TMDPDF spin structures or parameter ranges where only limited data is available.

\begin{acknowledgments}
We thank W.~Detmold, M.~Diehl, M.~Engelhardt,  P.~Shanahan, and M.~Wagman for discussions.
This work was supported by the U.S.\ Department of Energy, Office of Science,
Office of Nuclear Physics, from DE-SC0011090
and within the framework of the TMD Topical Collaboration.
I.S.\ was also supported in part by the Simons Foundation through
the Investigator grant 327942.
M.E.\ was also supported by the Alexander von Humboldt Foundation
through a Feodor Lynen Research Fellowship.
M.E.\ and I.S.\ also thank KIT for hospitality while portions of this work were completed.

\end{acknowledgments}

\appendix

\section{Notation and conventions}
\label{app:conventions}

We briefly summarize the conventions used in this paper
and compare them to different conventions used in the literature.

\paragraph{Lightcone coordinates.}
We use the SCET notation based on the two reference vectors
\begin{align}
 n^\mu = (1,0,0,1) \,,\quad \bn^\mu = (1,0,0,-1)
\,,\end{align}
which satisfy $n^2 = \bn^2 = 0$ and $n \cdot \bn = 2$.
One can decompose any fourvector as
\begin{align}
 k^\mu &= k^- \frac{n^\mu}{2} + k^+ \frac{\bn^\mu}{2} + k_\perp^\mu
 \equiv (k^+, k^-, \kt)
\,,\end{align}
where the plus and minus components are defined as
\begin{align}
 k^+ = n \cdot k = k^0 - k^z \,,\quad k^- = \bn \cdot k = k^0 + k^z
\,,\end{align}
and the transverse component is orthogonal to both reference vectors, $n \cdot k_\perp = \bn \cdot k_\perp = 0$.
The Minkowski product of two fourvectors follows to be
\begin{align}
 x \cdot k = \frac{1}{2} \bigl(x^+ k^- + x^- k^+\bigr) - \vec{x}_T \cdot \kt
\,.\end{align}
In particular, one has $k^2 = k^+ k^- - k_T^2$.

Another convention often used in the literature is $k^\pm = (k^0 \pm k^z)/\sqrt{2}$, see e.g.\ \mycite{Collins:1350496}.
In that notation, the Minkowski product is $x \cdot k =\bigl(x^+ k^- + x^- k^+\bigr) - \vec{x}_T \cdot \kt$, and $k^2 = 2 k^+ k^- - k_T^2$.
To translate results from this paper to that convention, replace $k^\pm \to \sqrt{2} k^\mp$.
Often, this leaves factor $\sqrt{2}$, that can be absorbed e.g.\ in integration measures.

\paragraph{Position space.}
We often write the collinear and soft matrix elements in position space,
with the position denoted by $b^\mu = (b^0, \bt, b^z)$.
The Fourier transform to and from momentum space is defined as
\begin{align}
 f(q^\mu) = \int\frac{\df^4 b}{(2\pi)^4} e^{-\img q \cdot b} f(b^\mu)
\,,\quad
 f(b^\mu) = \int \df^4q \, e^{\img q \cdot b} f(q^\mu)
\,.\end{align}
We use the same symbol for a function in $f$ in both spaces,
as we reserve the symbol $\tilde f$ for the quasi-construction of $f$.
In practice, we only perform the Fourier transform with respect to $b^+$ to obtain the momentum fraction $x$ ($b^z$ for the quasi-TMD), while we keep the transverse dependence in position space (often called impact parameter space),
where the canonical logarithm is given by
\begin{align} \label{eq:Lb}
 L_b \equiv \ln\frac{b_T^2 \mu^2}{b_0^2} \,, \quad
 b_0 = 2 e^{-\gamma_E} \approx 1.12291\dots
\,.\end{align}

\paragraph{Renormalization scheme.}
Our results are expressed in the $\MS$ scheme.
The associated renormalization scale $\mu \equiv \mu_{\MS}$ is related to the MS scale $\mu_0$ by
\begin{align}
 \mu^2 \equiv \mu^2_{\overline{\rm MS}} = \frac{4\pi}{e^{\gamma_E}} \mu_0^2
\,.\end{align}
Note that this differs from the convention used by Collins \cite{Collins:1350496},
where $\mu^2 = 4\pi \mu_0^2 / \Gamma(1-\eps)$, by terms of $\cO(\eps^2)$.

\section{Comparison of different schemes for TMD definitions}
\label{app:overview_tmdpdfs}

Here, we give more details on the various rapidity regulators employed in the literature that are compatible with the generic notation used in \sec{tmd_review}, where the TMDPDF was defined in \eq{tmdpdf1b} as
\begin{align} \label{eq:tmdpdf_app}
 f^\TMD_q(x, \bt, \mu , \zeta) &
 = \lim_{ \substack{\epsilon\to 0 \\ \tau\to 0}}\, Z_{\rm uv}(\mu,\zeta,\eps) \, B_{q}(x, \bt, \eps, \tau, \zeta)\, \Delta_S^q(b_T,\eps,\tau)
\nn\\&
 \equiv \lim_{ \substack{\epsilon\to 0 \\ \tau\to 0}}\, Z_{\rm uv}(\mu,\zeta,\eps) f_q^{\TMD}(x, \bt, \epsilon, \zeta)
\,,\end{align}
and the Collins-Soper scale was given by \eq{zeta},
\begin{align} \label{eq:zetadefn_app}
 \zeta &= (x P^- e^{- y_n})^2  = (x m_P e^{y_P - y_n})^2
\,.\end{align}
We show how this formulation for the TMDPDF arises in the various regulators, 
including Wilson lines off the light-cone, the $\delta$ regulator, the $\eta$ regulator, and the exponential regulator from \mycites{Collins:1350496,GarciaEchevarria:2011rb,Chiu:2012ir,Li:2016axz} respectively.
We also explicitly give one-loop results for the individual matrix element ingredients in the TMDPDF where available. 
In all cases, while the ingredients differ, the same universal result is obtained for $f_q^{\TMD}$.  We also give the correspondence to  results for the analytic regulator of \mycite{Becher:2010tm}, where only the product of two TMDs are defined, but which also agree with this $f_q^\TMD$.

At one-loop, the common result when evaluating the TMDPDF in an on-shell quark state of momentum $P$ using pure dimensional regularization is
\begin{align} \label{eq:tmdpdf_nlo}
 f_q^{\TMD\,(1)}(x, \bt, \epsilon, \zeta) &
 = \frac{\as C_F}{2\pi} \biggl[ - \biggl(\frac{1}{\eps_\IR} + L_b\biggr) P_{qq}(x) +  (1-x) \biggr]_+^1\Theta(1-x)\Theta(x)
 \\\nn&
  + \frac{\as C_F}{2\pi} \delta(1-x) \biggl[ \frac{1}{\eps^2} - \frac{1}{2} L_b^2
   + \biggl(\frac{1}{\eps} + L_b\biggr) \biggl(\frac{3}{2} + \ln\frac{\mu^2}{\zeta} \biggr)
   + \frac{1}{2} - \frac{\pi^2}{12} \biggr]
\,.\end{align}
Here, $L_b$ is defined in \eq{Lb} and $P_{qq}(x) = (1+x^2)/(1-x)$ is the regular part of the quark-quark splitting function.
Absorbing the UV divergences in a multiplicative counter term
\begin{align}
 Z_{\rm uv}(\mu,\zeta,\eps) = 1 - \frac{\as C_F}{2\pi} \biggl[ \frac{1}{\eps^2}
   + \frac{1}{\eps} \biggl(\frac{3}{2} + \ln\frac{\mu^2}{\zeta} \biggr) \biggr] + \cO(\as^2)
\,,\end{align}
one obtains the renormalized TMDPDF as
\begin{align} \label{eq:tmdpdf_nlo_renom}
 f_q^{\TMD\,(1)}(x, \bt, \mu, \zeta) &
 = \frac{\as C_F}{2\pi} \biggl[ - \biggl(\frac{1}{\eps_\IR} + L_b\biggr) P_{qq}(x) +  (1-x) \biggr]_+^1\Theta(1-x)\Theta(x)
 \\\nn&
  + \frac{\as C_F}{2\pi} \delta(1-x) \biggl[ - \frac{1}{2} L_b^2 + L_b \biggl(\frac{3}{2} + \ln\frac{\mu^2}{\zeta} \biggr) + \frac{1}{2} - \frac{\pi^2}{12} \biggr]
\,.\end{align}
The remaining $1/\eps_\IR$ pole here is of infrared origin and is the same collinear divergence that is present for the PDF. This correspondence between infrared divergences enables the TMDPDF to be matched on to the PDF for perturbative $b_T$.

\subsection{Wilson lines off the light-cone}
In the modern definition by Collins \cite{Collins:1350496},
the lightlike Wilson lines are replaced by spacelike Wilson lines,
\begin{align} \label{eq:Collins_rap}
n^\mu &\quad\to\quad n_{y_A}^\mu \equiv n^\mu - e^{-2 {y_A}} \bn^\mu
\,,\nn\\
\bn^\mu &\quad\to\quad \bn_{y_B}^\mu \equiv \bn^\mu - e^{+2 {y_B}} n^\mu
\,.\end{align}
This affects both the beam function as well as the soft factor, which is now a combination of soft matrix elements.
From Eq.~(13.106) in \mycite{Collins:1350496} we have for the $n$-collinear TMDPDF
\begin{align} \label{eq:tmdpdf_collins}
&f_{q}^\TMD(x, \bt, \mu, \zeta)
\nn\\&
= \lim_{\substack{y_A\to+\infty\\y_B\to-\infty}} Z_{\rm uv}\:
B_q^C(x, \bt, \eps, y_P- y_B) \sqrt{\frac{S^q_C(b_T,\eps, y_A-y_n)}{S^q_C(b_T,\eps,y_A-y_B) S^q_C(b_T,\eps,y_n-y_B)}}
\nn\\&
= \lim_{\substack{y_B\to-\infty}} Z_{\rm uv}\:
\frac{B_q^C(x, \bt, \eps, y_P- y_B)}{\sqrt{S^q_C(b_T,\eps,2y_n-2y_B)}}
\,,\end{align}
where the result in the last line was derived in \mycite{Buffing:2017mqm}.
Here, $y_P$ is the rapidity of the hadron (not the Wilson line direction), $y_B$ is the direction of the Wilson line as in \eq{Collins_rap} which acts as a rapidity regulator, and $y_n$ is a rapidity parameter that controls the split of soft radiation into the two TMDPDFs.
The $\zeta$ scale is defined as in \eq{zetadefn_app}.
To relate \eq{tmdpdf_collins} to our notation in \eq{Collins_rap} one identifies $\tau = 1/(y_B - y_n)$ and notes that
\begin{align} \label{eq:tau_Collins}
 1/\tau - \ln\sqrt{\zeta} = y_B-y_n -(y_P-y_n)-\ln (x\, m_P) = y_B-y_P-\ln(x\, m_P)
\,,\end{align}
so that the translation of functional forms is
\begin{align} \label{eq:BqDeltaC}
 B_q(x,\bt,\epsilon,\tau,\zeta) &= B_q^C(x,\bt,\epsilon,y_P-y_B)
\,,\nn\\
 \Delta_S^q(b_T,\epsilon,\tau) &= \frac{1}{\sqrt{S^q_C(b_T,\eps,2y_n-2y_B)}}
\,,\end{align}
where $B_q^C \equiv \tilde f^{\rm unsub}_q$ is the collinear matrix element as defined in Eq.\ (13.108) in \mycite{Collins:1350496} and $S^q_C \equiv \tilde S_{(0)}$ is the soft function as defined in Eq.\ (13.39) therein.
As usual, the dependence on $m_P$ and $\LQCD$ is implicit.
Note that the two $B_q$'s have the same number of arguments after using the fact that only the combination $1/\tau-\ln\sqrt{\zeta}$ appears in $B_q$.

Since $\zeta = (x m_p e^{y_P-y_n})^2$, one can obtain the Collins-Soper kernel as
\begin{align}
 \gamma_\zeta^q(\mu,b_T) = 2 \frac{\df\ln f^\TMD_q}{\df\ln\zeta} = \frac{\df\ln B_q^C}{\df y_P} = -\frac{\df\ln \Delta_q^S}{\df y_n}
\,,\end{align}
which is the analog of \eq{gamma_zeta}.
This makes clear that the Collins-Soper kernel can be obtained as a differential equation involving $f^\TMD$, or $B_q$, or $\Delta_S^q$ independently. This fact is commonly exploited in the perturbative TMDPDF literature.

\paragraph{One-loop results.}
We are not aware of explicit individual one-loop results for $B_q^C$ and $S_C^q$ in the literature.
Instead, in \mycites{Aybat:2011zv,Collins:1350496} the two functions are combined at the integrand level before integrating over the momentum of either real or virtual emissions.
For example, in \mycite{Collins:1350496}, where the soft factor is calculated in the context of the TMD fragmentation function, the limit $y_{A,B} \to \pm\infty$ is taken right away. The resulting divergent integral is then directly canceled against a divergent integral in the beam function, which for the TMDPDF then yields \eq{tmdpdf_nlo_renom}.
The same order of operations was used for the calculation of the TMDPDF for double-parton scattering in~\mycite{Buffing:2017mqm}.

\begin{figure*}
 \centering
 \begin{subfigure}{0.3\textwidth}
 \includegraphics[width=\textwidth]{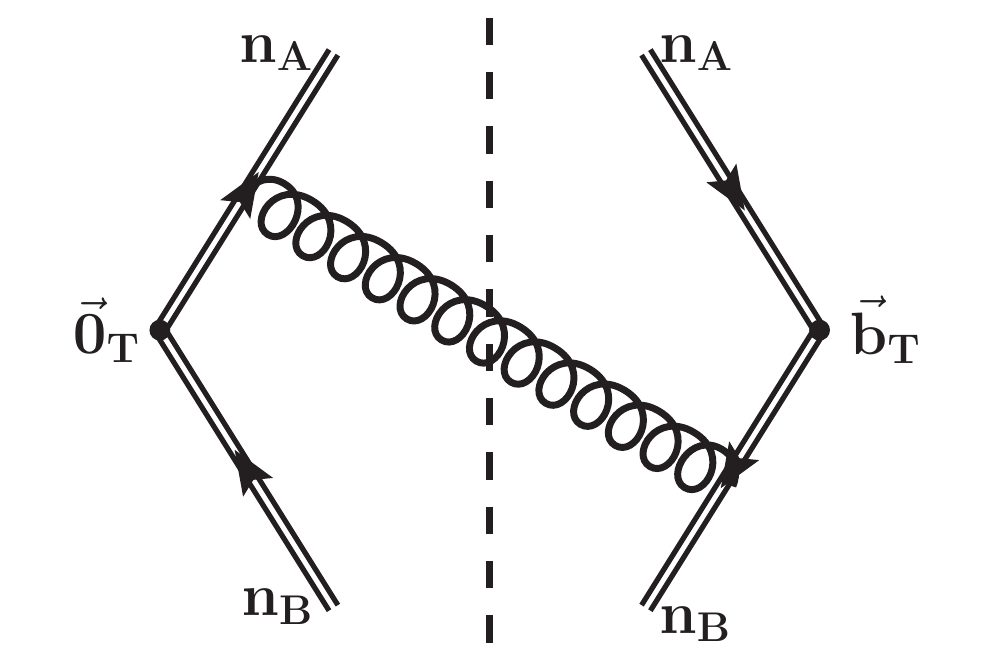}
 \caption{}
 \label{fig:softfunc_collins_a}
 \end{subfigure}
 \quad
 \begin{subfigure}{0.3\textwidth}
 \includegraphics[width=\textwidth]{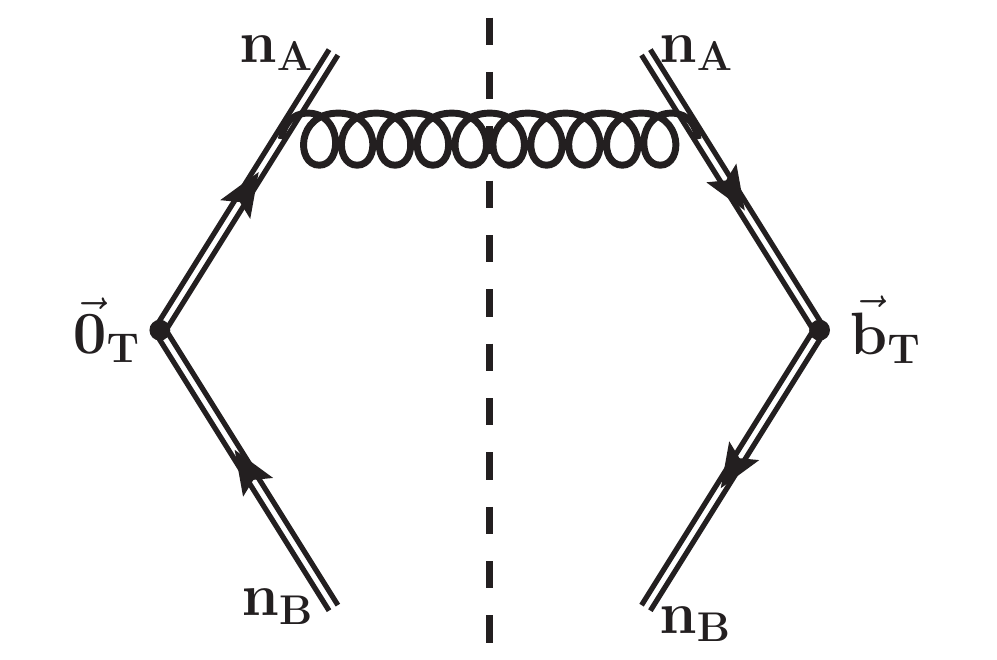}
 \caption{}
 \label{fig:softfunc_collins_b}
 \end{subfigure}
 \caption{Real emission diagrams contributing to the soft function with two spacelike Wilson lines at one loop, up to mirror diagrams. The dashed line indicates the cut gluon.}
 \label{fig:softfunc_collins}
\end{figure*}

We can calculate the soft function in the Collins scheme with Wilson lines along two spacelike vectors $n_A$ and $n_B$.
In Feynman gauge, purely virtual diagrams are scaleless and vanish, so we only need to take real emissions into account, as shown in \fig{softfunc_collins}.
The two diagrams are given by
\begin{align} \label{eq:soft_collins_a}
 S_{C,a}^{q\,(1)}(b_T,\eps,y_A-y_B) &
 = g^2 C_F (n_A{\cdot}n_B) \mu_0^{2\eps} \int\frac{\df^dk}{(2\pi)^d} (2\pi) \delta_+(k^2)
   \frac{1}{n_A \cdot k - \img0} \frac{e^{-\img \bt \cdot \kt}}{n_B \cdot k + \img0}
\nn\\&
 = \frac{\as C_F}{2\pi} (y_A - y_B) \frac{1 + e^{2(y_B-y_A)}}{1 - e^{2(y_B-y_A)}}
   \biggl( - \frac{1}{\eps} - \ln\frac{b_T^2\mu^2}{b_0^2} \biggr)
\,,\\ \label{eq:soft_collins_b}
 S_{C,b}^{q\,(1)}(b_T,\eps,y_A) &
 = - g^2 C_F n_A^2 \mu_0^{2\eps} \int\frac{\df^dk}{(2\pi)^d} (2\pi) \delta_+(k^2)
   \frac{1}{n_A \cdot k + \img0} \frac{e^{-\img \bt \cdot \kt}}{n_A \cdot k - \img0}
 \nn\\&
 = \frac{\as C_F}{2\pi} \biggl( \frac{1}{\eps} + \ln\frac{b_T^2\mu^2}{b_0^2} \biggr)
\,.\end{align}
Note that integral in in \eq{soft_collins_b} is ill-defined due to a pinched pole singularity arising from the opposite signs of the $\img0$ prescription in the two eikonal vertices, and we have evaluated it using the principal-value prescription.
This is similar to the calculation in \mycite{Bacchetta:2008xw}, where the soft function as defined in \mycite{Collins:1981uk} was calculated in an axial gauge with $v \cdot \cA = 0$ for a spacelike $v$.
However, there the principal-value prescription is part of the axial gauge, while we employ it because a corresponding regularization in terms of Wilson lines for Feynman gauge is currently unknown, see also the discussion in appendix A of \mycite{Bacchetta:2008xw}.
In the calculation of \mycite{Ji:2004wu}, the pinched pole singularities do not pose a problem due to employing timelike reference vectors $n_A$ and $n_B$.

Taking the mirror diagrams into account and taking the limit $|y_A-y_B| \gg 1$, we obtain
\begin{align}
 S_C^q(b_T,\eps,y_A-y_B) &= 1 + \frac{\as C_F}{2\pi} \biggl( \frac{1}{\eps} + \ln\frac{b_T^2\mu^2}{b_0^2} \biggr) (2 - 2 |y_A-y_B|) + \cO(\as^2)
\,,\end{align}
which after renormalization agrees with the result for timelike Wilson lines in \mycite{Ji:2004wu}.
Thus at one loop using \eq{BqDeltaC} with $\tau=1/(y_B-y_n)$ we find
\begin{align}
 \Delta_S^q(b_T,\eps,\tau) &
 =  1 + \frac{\as C_F}{2\pi} \biggl( \frac{1}{\eps} + \ln\frac{b_T^2\mu^2}{b_0^2} \biggr) \biggl( -\frac{2}{\tau} - 1 \biggr) + \cO(\as^2)
\,.\end{align}

The one-loop result for $f_q^\TMD$ was calculated in \mycite{Aybat:2011zv} and agrees with \eq{tmdpdf_nlo_renom} after adjusting for the difference in definitions of the $\MS$ scheme, see \app{conventions}.
Hence we can also deduce the bare beam function in this scheme, using
\begin{align} \label{eq:beam_Collins}
 B_q^{C(1)}(x, \bt, &\epsilon, y_P-y_B) 
 = f_q^{\TMD\,(1)}(x, \bt, \epsilon, \zeta) - \delta(1-x) \Delta_S^{q\,(1)}(b_T, \eps, \tau)
\\\nn&
 = \frac{\as C_F}{2\pi} \biggl[ - \biggl(\frac{1}{\eps_\IR} + L_b\biggr) P_{qq}(x) +  (1-x) \biggr]_+^1\Theta(1-x)\Theta(x)
 \\\nn&\ \ 
  + \frac{\as C_F}{2\pi} \delta(1-x) \biggl[ \frac{1}{\eps^2} - \frac{1}{2} L_b^2
   + \biggl(\frac{1}{\eps} + L_b\biggr) \biggl(\frac{5}{2} + \ln\frac{\mu^2}{\zeta} + \frac{2}{\tau} \biggr)
   + \frac{1}{2} - \frac{\pi^2}{12} \biggr]
\,.\end{align}
Using \eq{tau_Collins}, one can see that this result only depends on $y_P - y_B$.

\subsection[\texorpdfstring{$\delta$ regulator}{delta regulator}]
           {\texorpdfstring{\boldmath $\delta$}{Delta} regulator}
The $\delta$ regulator was originally introduced in \mycite{GarciaEchevarria:2011rb}
and then modified in \mycites{Echevarria:2015usa,Echevarria:2015byo} to be applicable for NNLO calculations.
Note that their notation for the light-cone momenta $x^\pm$ are related to our notation by $x^\pm \leftrightarrow x^\mp$.
For consistency of this paper, we convert results for the $\delta$ regulator in the literature to our notation.

Here, we follow the definition of the TMDPDF as presented in \mycite{Echevarria:2012js},
which generalizes the definition in \mycite{GarciaEchevarria:2011rb}.
At one loop, the $\delta$ regulator amounts to modifying propagators in the full theory as
\begin{align}
 \frac{\img}{(p+k)^2+\img0} \to \frac{\img}{(p+k)^2+\img \Delta^+}
\,,\qquad
 \frac{\img}{(\bar p+k)^2+\img0} \to \frac{\img}{(\bar p+k)^2+\img \Delta^-}
\,,\end{align}
where
\begin{align}
 p^\mu = p^- \frac{n^\mu}{2} = Q e^Y \frac{n^\mu}{2}
\,,\quad
 \bar p^\mu = \bar p^+ \frac{\bar n^\mu}{2} = Q e^{-Y} \frac{\bar n^\mu}{2}
\end{align}
are the momenta of the struck quark in the $n$-collinear and $\bn$-collinear proton, respectively.
One can equivalently modify the eikonal propagators as
\begin{align}
 \frac{\img}{k^++\img0} \to \frac{\img}{k^++\img \delta^+}
\,,\qquad
 \frac{\img}{k^-+\img0} \to \frac{\img}{k^-+\img \delta^-}
\,.\end{align}
(Additional integer $m>0$ multiplicative factors, $k^\pm + \img\,m\,\delta^\pm$, appear at higher loop orders).
The regulators are related by
\begin{align}  \label{eq:deltadefn}
 \delta^+  = \frac{\Delta^+}{p^-} \,,\quad \delta^- = \frac{\Delta^-}{\bar p^+}
\,.\end{align}
When working with $\Delta^\pm$ as arising in the full theory, the soft function can be written as \cite{Echevarria:2012js}
\begin{align}
 \ln S^q_{\rm EIS} = R_{s1}(\as, L_b) + R_{s2}(\as, L_b) \ln\frac{\Delta^+ \Delta^-}{Q^2 \mu^2}
\,.\end{align}
However, the $Q^2$ dependence appearing in $S$ here is artificial, as it is the $\delta^\pm$ regulators that are the fundamental Wilson line regulators.
Using \eq{deltadefn} together with $\bar p^+ p^- = Q^2$, one can equivalently write this as
\begin{align} \label{eq:S_delta_tmp}
 \ln S^q_{\rm EIS}\bigl(b_T, \eps, \sqrt{\delta^+ \delta^-}\bigr) &
 = R_{s1}(\as, L_b) + R_{s2}(\as, L_b) \ln\frac{\delta^+ \delta^-}{\mu^2}
\\*\nn&
 = R_{s1}(\as, L_b) + R_{s2}(\as, L_b) \ln\frac{\delta^- e^{-y_n}}{\mu}
                    + R_{s2}(\as, L_b) \ln\frac{\delta^+ e^{+y_n}}{\mu}
\,,\end{align}
where $y_n$ is an arbitrary parameter that cancels between the two logarithms.
\eq{S_delta_tmp} allows one to split the soft function into a ``left-moving'' and ``right-moving'' component,%
\footnote
{
   \eq{S_delta_split} might seem to differ from Eq.~(16) in \mycite{Echevarria:2012js}, where the soft function is expressed as
   \begin{align} \label{eq:Sscimemi}
   S\biggl(b_T, \frac{\Delta^-}{p^+},  \frac{\Delta^+}{\bar p^-}\biggr)
   =  \sqrt{S\biggl(b_T, \frac{\Delta^-}{\bar p^+}, \alpha \frac{\Delta^-}{ p^-}\biggr)}
      \sqrt{S\biggl(b_T, \frac{1}{\alpha} \frac{\Delta^+}{\bar p^+}, \frac{\Delta^+}{ p^-}\biggr)}
   \,.\end{align}
   Since $S$ only depends on the product of the last two arguments, one can identify $\alpha= e^{2 Y-2y_n}$ and use that $\alpha (\Delta^-)^2/Q^2 = \alpha (\delta^- e^{-Y})^2= (\delta^- e^{-y_n})^2$ and $(\Delta^+)^2/(\alpha Q^2) = (\delta^+ e^Y)^2/\alpha = (\delta^+ e^{y_n})^2$ to show that \eq{Sscimemi} is equivalent to our \eq{S_delta_split}.
}
\begin{align} \label{eq:S_delta_split}
 S^q_{\rm EIS}\bigl(b_T, \eps, \sqrt{\delta^+ \delta^-}\bigr) &
 = \sqrt{ S^q_{\rm EIS}(b_T,\eps, \delta^- e^{-y_n})} \sqrt{ S^q_{\rm EIS}(b_T,\eps, \delta^+ e^{+y_n})}
\,,\end{align}
and it is this form of $S^q_{\rm EIS}$ that was used in NNLO calculations in \mycites{Echevarria:2015usa,Echevarria:2015byo,Echevarria:2016scs}.
The two factors in \eq{S_delta_split} are then combined with the $n$-collinear and $\bn$-collinear matrix elements, respectively.
Since with the $\delta$ regulator the soft zero-bin subtractions on the two beam functions are identical to dividing by the original soft function, the soft subtraction amounts to dividing by $\sqrt{S}$.
For the $n$-collinear case, one identifies
\begin{align} \label{eq:TMDPDF_EIS}
 1/\tau &= \ln( \delta^- e^{-y_n} )
\,,\nn\\
 B_q(x,\bt,\epsilon,\tau,\zeta) &= B_q^{\rm EIS}\bigl(x, \bt, \eps, \Delta^-/Q^2\bigr) =  B_q^{\rm EIS}(x, \bt, \eps, \delta^-/(x P^-)\bigr)
\,,\nn\\
 \Delta_S^q(b_T,\epsilon,\tau) &= \frac{1}{\sqrt{S^q_{\rm EIS}\bigl(b_T,\eps, \delta^- e^{-y_n}\bigr)} }
\,,\end{align}
where $B_q^{\rm EIS} \equiv J_n$ appears in Eq.~(12) in \mycite{Echevarria:2012js}.
Once again, only the combination
\begin{align}
 1/\tau - \ln\sqrt{\zeta} = \ln( \delta^-\, e^{-y_n} ) - \ln(x P^- e^{-y_n}) = \ln\bigl[\delta^-/(xP^-)\bigr]
\end{align}
arises in $B_q$, such that the functional dependence of $B_q$ and $J_n$ in \eq{TMDPDF_EIS} agree.

\paragraph{One-loop results.}
The unsubtracted beam function has first been calculated in \mycite{GarciaEchevarria:2011rb},
where the $\delta$ regulator also acts as an IR regulator.
The corresponding result with dimensional regularization as the IR regulator can be extracted from \mycite{Echevarria:2016scs},
\begin{align} \label{eq:beam_naive_nlo_delta}
 J_n^{(1)}(x,\bt,\eps,\delta^-/p^-) &
 = \frac{\as C_F}{2\pi} \biggl[ -\left(\frac{1}{\eps} + L_b\right) P_{qq}(x) + (1-x)\biggr]_+^1\Theta(1-x)\Theta(x)
 \nn\\&
 + \frac{\as C_F}{2\pi} \delta(1-x) \biggl[
   \biggl(\frac{1}{\eps} + L_b\biggr)\biggl(\frac{3}{2} + 2 \ln\frac{\delta^-}{p^-}\biggr) + \frac{1}{2} \biggr]
\,.\end{align}
The one-loop soft function is given by \cite{Echevarria:2012js}
\begin{align} \label{eq:soft_nlo_delta}
 S_{\rm EIS}^{q\,(1)}\bigl(b_T,\eps, \delta^- e^{-y_n}\bigr) &= \frac{\as C_F}{2\pi} \biggl[
  - \frac{2}{\eps^2} - 2 \left(\frac{1}{\eps} + L_b\right) \ln\frac{\mu^2}{(\delta^- e^{-y_n})^2}
  + L_b^2 + \frac{\pi^2}{6}  \biggr]
\,.\end{align}
According to \eqs{TMDPDF_EIS}{tmdpdf_app}, the TMDPDF at NLO follows as
\begin{align}
 f^{\TMD\,(1)}_q(x,\bt,\mu,\zeta) = J_n^{(1)}(x,\bt,\eps,\delta^-/p^-) - \frac{1}{2} S_{\rm EIS}^{q\,(1)}\bigl(b_T,\eps, \delta^- e^{-y_n}\bigr)
\,,\end{align}
which reproduces the result in \eq{tmdpdf_nlo} with $\zeta$ as given in \eq{zetadefn_app}.

\subsection[\texorpdfstring{$\eta$ regulator}{eta regulator}]
           {\texorpdfstring{\boldmath $\eta$}{Eta} regulator}
\label{app:eta_regulator}

The $\eta$ regulator \cite{Chiu:2011qc,Chiu:2012ir} modifies collinear and soft Wilson lines
in momentum space as
\begin{align} \label{eq:eta_regulator}
W &= \sum_\text{perms} \exp\biggl[ - g_s w^2 \frac{|\bn \cdot \cP_g|^{-\eta}}{\nu^{-\eta}} \frac{\bn \cdot A_n}{\bn \cdot \cP} \biggr]
\,,\\
S_n &= \sum_\text{perms} \exp\biggl[ - g_s w \frac{|2 \cP_{g3}|^{-\eta/2}}{\nu^{-\eta/2}} \frac{n \cdot A_s}{n \cdot \cP} \biggr]
\,,\end{align}
where the momentum operator $\cP$ picks up the momentum of the gluon fields $A$, and $\eta$ is the rapidity regulator with an associated rapidity scale $\nu$.
While the authors of \mycite{Chiu:2012ir} separately renormalize the beam and soft functions in $\eta$, one can trivially define a TMDPDF in the form of \eq{tmdpdf_app} by combining the bare beam and soft functions, in which case the dependence on $\eta$ and $\nu$ cancels,
\begin{align} \label{eq:tmdpdf_eta}
f^\TMD_q(x, \bt, \mu, x P^-) &=  \lim_{\substack{\eps\to0 \\ \eta\to0}} \,
Z_{\rm uv}(\mu,x P^-,\eps) \, B_{q}(x, \bt, \eps, \eta, x P^-) \sqrt{S^q(b_T,\eps,\eta)}
\,,\end{align}
where $B_q \equiv B_q^{\rm CJNR}$ and $S^q \equiv S^q_{\rm CJNR}$ are the beam and soft function as defined in \mycite{Chiu:2012ir}.
For this regulator, the soft zero-bin subtractions vanish, so $\Delta_S^q=\sqrt{S^q_{\rm CJNR}}$ only involves a $\sqrt{S^q}$ in the numerator.
For this regulator the correspondence with our notation is
\begin{align}
\eta = \tau \,,\qquad \zeta = (x P^-)^2 \,, \quad y_n = 0
\,.\end{align}
The choice of fixing $y_n=0$ arises because of the symmetric treatment of the two beam functions, but can be relaxed as in the other definitions if so desired.

In \eq{tmdpdf_eta}, we combined bare beam and soft functions as usual for TMDPDFs.
However, in practice the $\eta$ regulator is often used to define rapidity-renormalized beam and soft functions
by first expanding them individually about $\eta\to 0$ and $\epsilon \to 0$ and then absorbing poles in $\eta$ and $\eps$ through separate $\MS$ counterterms~\cite{Chiu:2012ir},
\begin{align} \label{eq:B_S_renorm}
 B_q^{\rm CJNR}(x, \bt, \mu, x P^-/\nu) &= Z_B^{\rm CJNR}(b_T, \mu, \nu, \eps, \eta, x P^-) \, B_q^{\rm CJNR}(x, \bt, \eps, \eta, x P^-)
\,,\nn\\
 S^q_{\rm CJNR}(b_T,\mu,\nu) &= Z_S^{\rm CJNR}(b_T, \mu, \nu, \eps, \eta) \, S^q_{\rm CJNR}(b_T,\eps,\eta)
\,.\end{align}
When combined these counterterms give
\begin{align} \label{eq:Z_uv_eta_2}
 Z_B^{\rm CJNR}(b_T, \mu, \nu, \eps, \eta, x P^-) \sqrt{Z_S^{\rm CJNR}(b_T, \mu, \nu, \eps, \eta)} = Z_{\rm uv}(\mu, x P^-, \eps)
\,.\end{align}
Using the fact the bare $B_q^{\rm CJNR}$ and $S^q_{\rm CJNR}$ are $\mu$ and $\nu$ independent yields the $\mu$ RGEs
\begin{align}
 \mu \frac{\df}{\df\mu} B_q^{\rm CJNR}(x, \bt, \mu, x P^-/\nu) &= \gamma_B^q(\mu, x P^-) B_q^{\rm CJNR}(x, \bt, \mu, x P^-/\nu)
\,,\nn\\
 \mu \frac{\df}{\df\mu} S^q_{\rm CJNR}(b_T,\mu,\nu) &= \gamma_S^q(\mu, \nu) S^q_{\rm CJNR}(b_T,\mu,\nu)
 \,,
\end{align}
and the rapidity RGEs
\begin{align}
 \nu \frac{\df}{\df\nu} B_q^{\rm CJNR}(x, \bt, \mu, x P^-/\nu) &= -\frac{1}{2} \gamma_\nu^q(\mu, b_T) B_q^{\rm CJNR}(x, \bt, \mu, x P^-/\nu)
\,,\nn\\
 \nu \frac{\df}{\df\nu} S^q_{\rm CJNR}(b_T,\mu,\nu) &= \gamma_\nu^q(\mu, b_T) S^q_{\rm CJNR}(b_T,\mu,\nu)
\,.\end{align}
Using Eqs.~\eqref{eq:tmdpdf_eta}, \eqref{eq:B_S_renorm} and \eqref{eq:Z_uv_eta_2} gives the renormalized TMDPDF in terms of the renormalized beam and soft functions as
\begin{align} \label{eq:tmdpdf_eta_ren}
f^\TMD_q(x, \bt, \mu, \zeta) &=  B_q^{\rm CJNR}\bigl(x, \bt, \mu, \sqrt{\zeta}/\nu\bigr) \sqrt{S^q_{\rm CJNR}(b_T,\mu,\nu)}
\,,\end{align}
where the $\nu$ independence cancels on the right hand side, and we remind the reader that here $\zeta = (x P^-)^2$.
Since $B_q^{\rm CJNR}$ only depends on the ratio $x P^-/\nu$, the Collins-Soper kernel can be extracted as
\begin{align}
 \gamma_\zeta^q = 2 \frac{\df\ln f_q^\TMD}{\df\ln \zeta} = 2 \frac{\df\ln B_q^{\rm CJNR}}{\df\ln \zeta}
 = - \frac{\df\ln B_q^{\rm CJNR}}{\df\ln\nu} = \frac{1}{2} \frac{\df\ln S^q_{\rm CJNR}}{\df\ln\nu}
\,.\end{align}
This also shows that the rapidity anomalous dimension of Ref.~\cite{Chiu:2012ir} is related to the Collins-Soper kernel by $\gamma_\nu^q(\mu,b_T) = 2 \gamma_\zeta^q(\mu,b_T)$.

\paragraph{One-loop results.}
The bare quark-quark matrix element for the beam function is not explicitly given in the literature, but can easily be calculated analogous to the gluon beam function in \mycite{Chiu:2012ir}. We find
\begin{align} \label{eq:beam_nlo_eta}
 B_{q}^{{\rm CJNR}\,(1)}(x, \bt, \eps,\eta)
 &= \frac{\as C_F}{2\pi} \biggl[ - \left(\frac{1}{\eps} + L_b\right) P_{qq}(x) + (1-x) \biggr]_+^1\Theta(1-x)\Theta(x)
 \nn\\*&
  + \frac{\as C_F}{2\pi} \delta(1-x) \biggl[ \left(\frac{1}{\eps} + L_b\right)
   \left( \frac{2}{\eta} + \frac{3}{2} - 2 \ln\frac{x P^-}{\nu}\right) + \frac{1}{2} \biggr]
\,.\end{align}
The NLO soft function using this regulator can be found in \mycites{Chiu:2012ir,Luebbert:2016itl},
\begin{align} \label{eq:soft_nlo_eta}
 S^{q\,(1)}_{\rm CJNR}(b_T,\epsilon,\eta) = \frac{\as C_F}{2 \pi} \biggl[
    \frac{2}{\eps^2} + 4 \Bigl(\frac{1}{\eps} + L_b\Bigr)\Bigl( - \frac{1}{\eta} + \ln\frac{\mu}{\nu}\Bigr)
    - L_b^2 - \frac{\pi^2}{6} \biggr]
\,.\end{align}
Combining the bare functions as in \eq{tmdpdf_eta} gives the NLO TMDPDF as
\begin{align}
 f^{\TMD\,(1)}_q(x,\bt,\mu,\zeta) = B_{q}^{{\rm CJNR}\,(1)}(x,\bt,\eps,\eta) + \frac{1}{2} S_q^{{\rm CJNR}\,(1)}(b_T,\eps,\eta)
\,,\end{align}
which yields \eq{tmdpdf_nlo} with $\zeta = (x P^-)^2$.
Here, the poles in $\eta$ manifestly cancel.

Alternatively, one can separately UV and rapidity renormalize beam and soft functions by absorbing the poles in $\eta$ and $\eps$ in separate counterterms.
The renormalized functions are given by
\begin{align} \label{eq:beam_nlo_eta_ren}
 B_{q}^{{\rm CJNR}\,(1)}\bigl(x, \bt, \mu, x P^-/\nu\bigr)
 &= \frac{\as C_F}{2\pi} \biggl[ - \left(\frac{1}{\eps} + L_b\right) P_{qq}(x) + (1-x) \biggr]_+^1\Theta(1-x)\Theta(x)
 \nn\\*&
  + \frac{\as C_F}{2\pi} \delta(1-x) \biggl[ L_b \left(\frac{3}{2} - 2 \ln\frac{x P^-}{\nu}\right) + \frac{1}{2} \biggr]
\,,\\
 \label{eq:soft_nlo_eta_ren}
 S^{q\,(1)}_{\rm CJNR}(b_T,\mu,\nu) &= \frac{\as C_F}{2 \pi} \biggl[- L_b^2  + 4  L_b \ln\frac{\mu}{\nu}  - \frac{\pi^2}{6} \biggr]
\,,\end{align}
where the remaining $1/\epsilon$ pole in \eq{beam_nlo_eta_ren} is of infrared origin.
Combining these as in \eq{tmdpdf_eta_ren} yields the UV-renormalized TMDPDF \eq{tmdpdf_nlo_renom} with $\zeta = (x P^-)^2$.

\subsection{Exponential regulator}

The exponential regulator \cite{Li:2016axz} has been designed to connect the TMD soft function to the
threshold soft function, which enabled it to be calculated up to ${\cal O}(\alpha_s^3)$~\cite{Li:2016ctv}.
The regulator introduces a factor
\begin{equation}
 \exp\Bigl[ - k^0 \tau e^{-\gamma_E} \Bigr]
\end{equation}
into the phase space of real emissions.
One then takes the $\tau\to0$ limit, keeping only divergent terms.
Identifying $\tau = \nu^{-1}$, one can connect this to the $\eta$ regulator shown above.
Here, we will keep $\tau$ for clarity.

Similar to the $\eta$ regulator, in the exponential regulator one typically combines UV- and rapidity-renormalized beam and soft functions. A key difference is that the exponential regulator requires a zero-bin subtraction, which is equivalent to dividing by the soft function. Thus identifying
\begin{align}
 \Delta_S^q(b_T,\eps,\tau) = \frac{1}{\sqrt{S^q_{\rm LNZ}(b_T,\eps,\tau)}}
\,,\end{align}
one obtains the TMDPDF. Similar to our discussion for the $\eta$ regulator this can be done either by first combining and then renormalizing the bare matrix elements,
or by combining renormalized beam and soft functions.

\paragraph{One-loop results.}
The beam function is only given after UV renormalization in \mycite{HuaXing},
\begin{align} \label{eq:beam_unsub_nlo_exp}
 B_{q}^{{\rm LNZ}\,(1)}(x, \bt, \mu, \nu Q)
  &= \frac{\as C_F}{2\pi} \biggl[ - \left(\frac{1}{\eps} + L_b\right) P_{qq}(x) + (1-x) \biggr]_+^1\Theta(1-x)\Theta(x)
 \nn\\&
  + \frac{\as C_F}{2\pi} \delta(1-x) \biggl[ - L_b^2 + L_b \left(\frac{3}{2} - \ln\frac{Q^2 \nu^2}{\mu^4} \right) + \frac{1}{2}  - \frac{\pi^2}{6}\biggr]
\,.\end{align}
The renormalized NLO soft function in this scheme is given by \cite{Li:2016axz}
\begin{align} \label{eq:soft_nlo_exp}
 S^{(1)}_{\rm LNZ}(b_T,\mu,\nu) &
 = \frac{\as C_F}{2\pi} \biggl[ - L_b^2 + 2 L_b \ln\frac{\mu^2}{\nu^2} - \frac{\pi^2}{6} \biggr]
\,.\end{align}
Combining the renormalized functions gives the NLO TMDPDF as
\begin{align}
 f^{\TMD\,(1)}_q(x,\bt,\mu,\zeta) = B_{q}^{{\rm LNZ}\,(1)}(x, \bt, \mu, \nu Q) - \frac{1}{2} S^{(1)}_{\rm LNZ}(b_T,\mu,\nu)
\,,\end{align}
which reproduces \eq{tmdpdf_nlo_renom} with $\zeta = Q^2$.

\subsection{Analytic regulator}
The analytic regulator as introduced in \mycite{Becher:2010tm} for TMDs
raises propagators including $n$ and $\bn$-collinear momenta to a power,%
\footnote{In practice, one can also introduce a factor $(\nu/n\cdot k)^\alpha$
for each unresolved final-state parton with momentum $k$, as suggested in \mycite{Becher:2011dz}
and employed in the NNLO calculation in \mycite{Gehrmann:2014yya}.}
\begin{align}
  {n-}\text{collinear}:\quad \frac{1}{-(p-k)^2} \to \frac{\nu_1^{2\alpha}}{[-(p - k)^2]^{1+\alpha}}
\,,\\
{\bn-}\text{collinear}:\quad \frac{1}{-(p-k)^2} \to \frac{\nu_2^{2\beta}}{[-(p - k)^2]^{1+\beta}}
\,,\end{align}
where $\alpha$ and $\beta$ are distinct parameters.
One then has to consistently expand first in $\beta\to0$ and then in $\alpha\to0$, or vice versa.
Since this breaks the $n \leftrightarrow \bn$ symmetry, it regulates all rapidity divergences.
However, the regulator does not act symmetrically in the $n$ and $\bn$-collinear beam function and thus yields two distinct bare quantities $\cB_{q\,n}^{\rm BN}$ and $\cB_{q\,\bn}^{\rm BN}$.
Following \mycite{Becher:2010tm}, we give all results for expanding first in $\beta\to0$ and then in $\alpha\to0$, such that rapidity divergences are regulated by $\alpha$ alone.

A particular feature of this regulator is that the soft function is unity all orders,
\begin{equation}
 S_q^{\rm BN}(b_T,\eps,\alpha) \equiv 1
\,,\end{equation}
as all loop corrections are scaleless and vanish.
This implies that rapidity divergences, i.e.\ poles in $1/\alpha$, cannot cancel between  a single beam function and the soft factor, and hence one cannot define a rapidity-finite TMDPDF as in \eq{tmdpdf_app}.
Instead, divergences in $\alpha$ only cancel in the product, so one has to define the TMDPDFs through the limit
\begin{align} \label{eq:BBSBN}
 \lim_{\substack{\eps\to0\\\alpha\to0}} & \Bigl[ \cB_{q\,n}^{\rm BN}(x_1,\bt,\eps,\alpha) \cB_{q\,\bn}^{\rm BN}(x_2,\bt,\eps,\alpha) S_q^{\rm BN}(b_T,\eps,\alpha)\Bigr]
 \nn\\
 & = \biggl(\frac{b_T^2 Q^2}{b_0^2}\biggr)^{-F_{q\bar q}(\mu,b_T)} B_{q\,n}^{\rm BN}(x_1, \bt, \mu) B_{q\,\bn}^{\rm BN}(x_2, \bt, \mu)
\,.\end{align}
Here, $F_{q\bar q}(\mu,b_T) = - \gamma^q_\zeta(\mu,b_T)$ is the so-called collinear anomaly coefficient \cite{Becher:2010tm}, which exponentiates the rapidity logarithms.
Note that the TMDPDFs on the right hand side of \eq{BBSBN} do not have an explicit $\zeta$ dependence, as $\zeta_a = \zeta_b = Q$ has already been chosen to fully exponentiate all rapidity logarithms.

\paragraph{One-loop results.}
The one-loop beam functions have been calculated in \cite{Becher:2010tm}.
As discussed, the $n$ and $\bn$-collinear functions take different forms
due to the regulator breaking the $n\leftrightarrow\bn$ symmetry.
The one-loop result for the matrix element in an on-shell quark state can be extracted from \mycite{Becher:2010tm} in the form
\begin{align}
 \cB^{{\rm BN}\,(1)}_{q\,n}(x_1, \bt, \eps, \alpha) &
 = \frac{\as C_F}{2\pi} \biggl[ - \left(\frac{1}{\eps} + L_b\right) P_{qq}(x_1) + (1-x_1) \biggr]_+^1\Theta(1-x_1)\Theta(x_1)
 \\\nn&
 + \frac{\as C_F}{2\pi}\delta(1-x_1) \biggl[ \frac{2}{\eps^2}
   - \left(\frac{1}{\eps} + L_b\right) \left(-\frac{3}{2} + \frac{2}{\alpha} - 2 \ln\frac{\mu^2}{\nu^2} \right)
 - L_b^2 - \frac{\pi^2}{6} + \frac{1}{2} \biggr]
\,,\\
 \cB^{{\rm BN}\,(1)}_{q\,\bn}(x_2, \bt, \eps, \alpha) &
 = \frac{\as C_F}{2\pi} \biggl[ - \left(\frac{1}{\eps} + L_b\right) P_{qq}(x_2) + (1-x_2)\biggr]_+^1\Theta(1-x_2)\Theta(x_2)
 \nn\\&
 + \frac{\as C_F}{2\pi} \delta(1-x_2) \biggl[ - \left(\frac{1}{\eps} + L_b\right)  \left( -\frac{3}{2} - \frac{2}{\alpha} + 2 \ln\frac{Q^2}{\nu^2} \right)
  + \frac{1}{2} \biggr]
\,.\end{align}
Only the product of the two beam functions is meaningful, as the poles in $\alpha$ and the $\nu$ dependence cancel,
and one obtains the product of the two TMDPDFs,
\begin{align}
 &\Bigl[\cB^{{\rm BN}\,(1)}_{q\,n}(x_1, \bt, \eps, \alpha) \cB^{{\rm BN}\,(1)}_{q\,\bn}(x_2, \bt, \eps, \alpha)\Bigr]_{\cO(\as)}
 \nn\\&
 = \frac{\as C_F}{2\pi} \biggl\{ \delta(1-x_2) \biggl[ - \left(\frac{1}{\eps} + L_b\right) P_{qq}(x_1) + (1-x_1)  \biggr]_+^1\Theta(1-x_1)\Theta(x_1)
  + (x_1 \leftrightarrow x_2) \biggr\}
  \nn\\&
  + \frac{\as C_F}{2\pi} \delta(1-x_1) \delta(1-x_2) \biggl[
   \frac{2}{\eps^2} + \left(\frac{1}{\eps} + L_b\right) \left( 3 + 2 \ln\frac{\mu^2}{Q^2} \right)
   - L_b^2 + 1  - \frac{\pi^2}{6}\biggr]
\,.\end{align}
One can easily check that the same is obtained for $f_q^\TMD(x_1,\bt,\eps,\zeta_a) f_q^\TMD(x_2,\bt,\eps,\zeta_b)$ using the result in \eq{tmdpdf_nlo} together with $\zeta_a \zeta_b = Q^4$.

\section{Calculation of the quasi beam function}
\label{app:qbeamfunc}

\begin{figure}[pt]
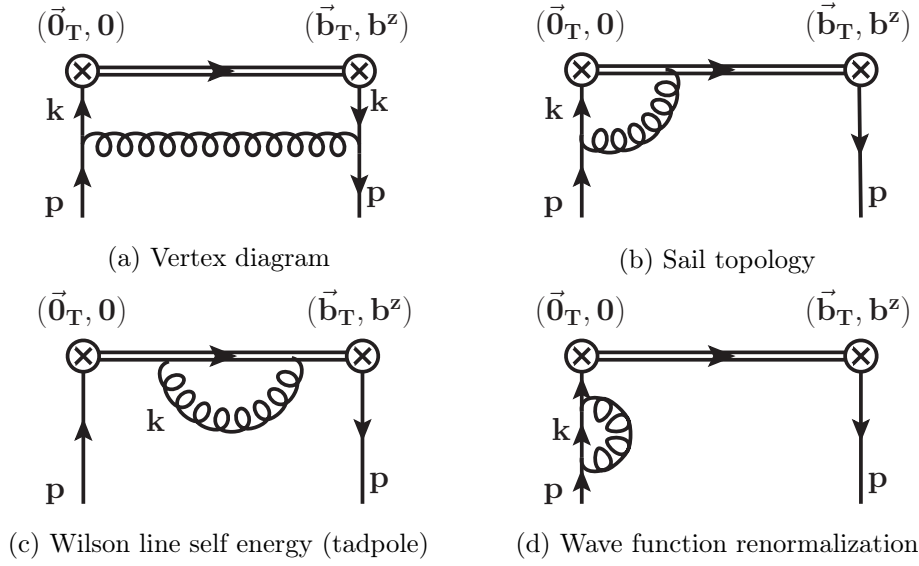

 \centering
 \begin{subfigure}{0.4\textwidth}
  \includegraphics[width=\textwidth]{fig/qbeamfunc_a.pdf}
  \caption{Vertex diagram}
  \label{fig:TMD a}
 \end{subfigure}
 \quad
 \begin{subfigure}{0.4\textwidth}
  \includegraphics[width=\textwidth]{fig/qbeamfunc_b1.pdf}
  \caption{Sail topology}
  \label{fig:TMD b}
 \end{subfigure}
\\
 \begin{subfigure}{0.4\textwidth}
  \includegraphics[width=\textwidth]{fig/qbeamfunc_c.pdf}
  \caption{Wilson line self energy (tadpole)}
  \label{fig:TMD c}
 \end{subfigure}
 \quad
 \begin{subfigure}{0.4\textwidth}
  \includegraphics[width=\textwidth]{fig/qbeamfunc_d.pdf}
  \caption{Wave function renormalization}
  \label{fig:TMD d}
 \end{subfigure}
 \caption{One-loop diagrams contributing to the quasi TMD beam function
          in Feynman gauge, up to mirror diagrams.
          The double line represents the Wilson line.}
 \label{fig:qbeamfunc_nlo}
\end{figure}

We calculate the spin-averaged quasi beam function by evaluating \eq{qbeam}
with an on-shell quark with momentum $P^\mu = (P^z, 0, 0, P^z)$,
and carry out the Fourier transform to momentum space only for the $z$ momentum,
while the transverse component is kept in position space,
\begin{align} \label{eq:qtmd matrix element}
 \tilde q_n(x,\bt,P^z) &= \int\!\frac{\df b^z}{2\pi} e^{\img (x P^z) b^z} \frac{1}{2} \sum_\mathrm{spins} \Bigl<q(P) \Big|
  \bar q(b^\mu) \mathrm{P} \exp\Bigl[\img g \int_\gamma \df \vec s \cdot \cA(\vec s\,)\Bigr]
  \frac{\gamma^\lambda}{2} q(0)
  \Big| q(P) \Bigr>
\,,\end{align}
where for compactness we suppress the dependence of $\tilde q_n$ on $\eps$ and $L$.
Here $\lambda = 0$ or $\lambda=3$, and the path ordered exponential represents the Wilson
line stretching along the path $\gamma$, as illustrated in \fig{qbeam}.
The explicit parameterization is given by three segments
\begin{align} \label{eq:qtmd_path}
 \gamma_1(s) &= \left(\begin{matrix} 0 \\ \vec 0_T \\ L s \end{matrix}\right)
\,,\quad
 \gamma_2(s) = \left(\begin{matrix} 0 \\ s \bt  \\ L \end{matrix}\right)
\,,\quad
 \gamma_3(s) = \left(\begin{matrix} 0 \\ \bt \\ L  + s (b^z - L) \end{matrix}\right)
\,,\quad s \in [0,1]
\,.\end{align}
In Feynman gauge, there are four topologies contributing to the one-loop beam function,
shown in \fig{qbeamfunc_nlo}. The sail and wave function diagram each have a mirror diagram.
Since we work in pure dimensional regularization with on-shell quarks,
the wave function renormalization diagram is scaleless and vanishes.
In practice, it converts IR poles into UV poles, which is crucial for the renormalization of
the beam function.

The general strategy of the NLO calculation is as follows.
First, we write down the integral fully in momentum space and introduce Feynman parameters.
The resulting integral over the loop momentum $q$ can then be reduced
to a set of master integrals, defined as
\begin{alignat}{3}
 \label{eq:I0}
 I_0(l) &= \int \frac{\df^d q}{(2\pi)^d} \frac{e^{-\img \qt \cdot \bt}}{(q^2 + \img0)^2} \delta(q^z - l)
 &&= \frac{\img}{16 \pi^3} (2\pi)^{\frac{1}{2}+\eps} K_{\frac{1}{2}+\eps}(b_T |l|) \biggl(\!\frac{b_T}{|l|}\!\biggr)^{\frac{1}{2}+\eps}
\,,\\
 \label{eq:I1}
 I_1(l) &= \int \frac{\df^d q}{(2\pi)^d} \frac{\delta(q^z - l)}{(q^2 + \img0)^2}
 &&= \frac{\img}{16 \pi^2} \frac{(4\pi)^\eps}{\sqrt{\pi}} \Gamma\left(\frac{1}{2}+\eps\right) |l|^{-1-2\eps}
\,,\\
 \label{eq:I2}
 I_2(l) &= \int \frac{\df^d q}{(2\pi)^d} \frac{e^{-\img \qt \cdot \bt}}{(q^2 + \img0)^2}
 &&= \frac{\img}{16 \pi^2} (b_T^2 \pi)^\eps \Gamma(-\eps)
\,.\end{alignat}
Here, $K_n(x)$ is the modified Bessel function of the second kind.
We need only the following integrals over the Feynman parameter $y$,
\allowdisplaybreaks
\begin{align}
 \label{eq:I0a}
 \mu_0^{2\eps} I_0^a &= \int_0^1 \df y \, I_0[(x-y) P^z]
 \nn\\*&
 = \frac{-\img}{16\pi^2} \frac{1}{P^z} \biggl[
   \left(\frac{1}{\eps} + \Lb{}\right)\Theta(1-x)\Theta(x)
  \nn\\*&\hspace{2cm}
  + \sgn{x} \Gamma\bigl(0, b_T P^z |x|\bigr) + \sgn{1-x} \Gamma\bigl(0, b_T P^z |1-x|\bigr) \biggr]
\,,\\
 \label{eq:I0b}
 \mu_0^{2\eps} I_0^b &= \int_0^1 \df y \, (x-y) I_0[(x-y) P^z]
 = \frac{-\img}{16 \pi^2} \frac{1}{P^z} \frac{e^{-b_T P^z |x|} - e^{-b_T P^z |1-x|}}{b_T P^z}
\,,\\
 \label{eq:I1a}
 \mu_0^{2\eps} I_1^a &= \int_0^1 \df y \, I_1[(x-y) P^z]
 \nn\\*&
 = \frac{-\img}{16\pi^2} \frac{1}{P^z} \biggl\{
     \left( \frac{1}{\eps} + \Lb{} \right)\Theta(1-x)\Theta(x)
    - \sgn{x} \Bigl[\gamma_E + \ln(b_T P^z |x|)\Bigr]
    \nn\\*&\hspace{2.cm}
    - \sgn{1-x} \Bigl[\gamma_E + \ln(b_T P^z |1-x|)\Bigr] \biggr\}
\,,\\
 \label{eq:I1b}
 \mu_0^{2\eps} I_1^b &= \int_0^1 \df y \, (x-y) I_1[(x-y) P^z]
 = \frac{-\img}{16 \pi^2} \frac{1}{P^z} \bigl( |1-x|-|x| \bigr)
\,,\end{align}
where $\mu_0$ is the MS scale and $\mu$ is the $\MS$ scale, and $x$ is the momentum fraction carried by the parton, not a Feynman parameter.
Finally, we introduce plus distributions taking into account that prior to taking the physical limit,
the quasi beam function has support $x \in [-\infty,\infty]$, rather than the physical support $x \in [0,1]$.
For the calculation, it is most convenient to define the plus distribution such that the integral over all $x$ vanishes,
\begin{align} \label{eq:plusinf}
 \Bigl[ h(x) \Bigr]\plusinf &= h(x) \,,\quad x \ne 1
\,,\nn\\
 \int_{-\infty}^\infty \df x\, \Bigl[ h(x) \Bigr]\plusinf &= 0
\,.\end{align}
For functions with physical support $x \in [0,1]$, this naturally reduces to the standard plus distribution.
Any integrable expression $h(x)$ can be written as a distribution using
\begin{align} \label{eq:into_plus}
 h(x) = \Bigl[ h(x) \Bigr]\plusinf + \delta(1-x) \int_{-\infty}^\infty \df x' \, h(x')
\,.\end{align}

\subsection{Vertex correction}
The vertex correction \fig{TMD a} is given by
\begin{align} \label{eq:qtmd_a_1}
 \tilde q^{(a)}_n(b^\mu) &
 = -\img g^2 C_F \mu_0^{2\eps} \frac{1}{2} \sum_S
  \bar u_S(P) \left[\int\frac{\df^dk}{(2\pi)^d}
  \frac{\gamma^\mu \slashed{k} \gamma^\lambda \slashed{k} \gamma_\mu}{2 k^4 (P-k)^2}
  e^{-\img \vec k \cdot \vec b} \right] u_S(P)
\nn\\&
 = -\img g^2 C_F \mu_0^{2\eps} (d-2) \int\frac{\df^dk}{(2\pi)^d}
   \left[ \frac{P^\lambda - k^\lambda}{k^2 (P-k)^2} + \frac{k^\lambda}{k^4} \right]
  e^{-\img \vec k \cdot \vec b}
\,.\end{align}
Note that the second term in square brackets vanishes for $\lambda = 0$, as the integrand is odd in $k^0$.
The Fourier transform w.r.t.\ $b^z$ can be carried out trivially,
\begin{align}
 \tilde q^{(a)}_n(x, \bt, P^z) &
 = -\img g^2 C_F \mu_0^{2\eps} (d-2) \int\frac{\df^dk}{(2\pi)^d}
  \left[ \frac{P^\lambda - k^\lambda}{k^2 (P-k)^2} + \frac{k^\lambda}{k^4} \right]
  e^{-\img \kt \cdot \bt} \delta(x P^z - k^z)
\,.\end{align}
Introducing Feynman parameters, we obtain
\begin{align} \label{eq:qtmd_a_2}
 \tilde q^{(a)}_n(x, \bt, P^z) &
 = -\img g^2 C_F \mu_0^{2\eps} (d-2) \int_0^1 \df y
  \int\frac{\df^d q}{(2\pi)^d} \frac{e^{-\img \qt \cdot \bt}}{q^4}
  \Bigl\{ (1-y) P^\lambda \delta\bigl[(x-y)P^z - q^z\bigr]
  \nn\\*&\hspace{3cm}
  + \delta^{\lambda z} q^z \Bigr[ \delta(x P^z - q^z) - \delta\bigl[(x-y)P^z - q^z\bigr] \Bigr]
  \Bigr\}
\,.\end{align}
Using \eqs{I0a}{I0b}, this can be expressed in terms of the master integrals,
\begin{align} \label{eq:qtmd_a_3}
 \tilde q^{(a)}_n(x, \bt, P^z) &
 = -\img g^2 C_F \mu_0^{2\eps} (d-2) P^z
  \left[ (1-x) I_0^a + (1 - \delta^{\lambda z}) I_0^b
  + \delta^{\lambda z} \, x I_0(x P^z) \right]
\,.\end{align}
Next we use \eq{into_plus} to rewrite \eq{qtmd_a_3} in terms of a plus distribution.
Note that the integral of the second line in \eq{qtmd_a_2} vanishes when integrated over all $x$
and thus does not yield a boundary term. We obtain
\begin{align} \label{eq:qtmd_a_4}
 \tilde q^{(a)}_n(x, \bt, P^z)
 =&~ -\img g^2 C_F \mu_0^{2\eps} (d-2) P^z
  \left[ (1-x) I_0^a + (1 - \delta^{\lambda z}) I_0^b
  + \delta^{\lambda z} \, x I_0(x P^z) \right]\plusinf
 \nn\\&~
 -\img g^2 C_F \mu_0^{2\eps} (d-2) \delta(1-x) \frac{I_2}{2}
\,.\end{align}
Plugging in the master formulas and expanding in $\eps$ gives the final result
\begin{align} \label{eq:qtmd_a}
 \tilde q^{(a)}_n(x, \bt, P^z) &
 = \frac{\as C_F}{4\pi} \biggl(\frac{1}{\eps} + \Lb{} - 1 \biggr)
   \Bigl\{ \Bigl[ 2 (x-1) \Theta(1-x)\Theta(x) \Bigr]_+^1 - \delta(1-x) \Bigr\}
\nn\\&\quad
 + \Delta\tilde q^{(a)}_n(x, \bt, P^z)
\,,\end{align}
where we singled out the terms with physical support $x \in [0,1]$.
These will be crucial to recover the collinear singularity of the standard PDF.
The contribution with unphysical support $x \in [-\infty,\infty]$ is given by
\begin{align} \label{eq:qtmd_a_suppressed}
 \Delta\tilde q^{(a)}_n(x, \bt, P^z) &=
 - \frac{\as C_F}{2\pi} \biggl[
    (1 - \delta^{\lambda z}) \frac{e^{-b_T P^z |x|} - e^{-b_T P^z |1-x|}}{b_T P^z}
    \nn\\*&\hspace{1.75cm}
    + |1-x| \Gamma\bigl(0, b_T P^z |1-x|\bigr) - |x| \Gamma\bigl(0, b_T P^z |x|\bigr)
    \nn\\*&\hspace{1.75cm}
    + \sgn{x} \Gamma\bigl(0, b_T P^z |x|\bigr)
    - \delta^{\lambda z} \sgn{x} e^{- b_T P^z |x|} \biggr]\plusinf
\,.\end{align}
The first line in \eq{qtmd_a_suppressed} is clearly suppressed for $b_T P^z \gg 1$.
The second line in \eq{qtmd_a_suppressed} also vanishes in this limit, because the incomplete Gamma function behaves as
\begin{equation}
 \Gamma(0, z) \stackrel{z \to 0}{\approx} -\ln(z) - \gamma_E
\,,\qquad
 \Gamma(0, z) \stackrel{z \to \infty}{\approx} \frac{1}{z} e^{-z}
\,,\end{equation}
and the logarithmic enhancement for small arguments is compensated by the prefactor.
The third line of \eq{qtmd_a_suppressed} yields a power-suppressed contribution when convoluted with the TMDPDF, see \app{pwr}.

\subsection{Sail diagrams}

\begin{figure}[ht]
 \centering
 \begin{subfigure}{0.4\textwidth}
  \includegraphics[width=\textwidth]{fig/qbeamfunc_b1.pdf}
 \end{subfigure}
 \quad
 \begin{subfigure}{0.4\textwidth}
  \includegraphics[width=\textwidth]{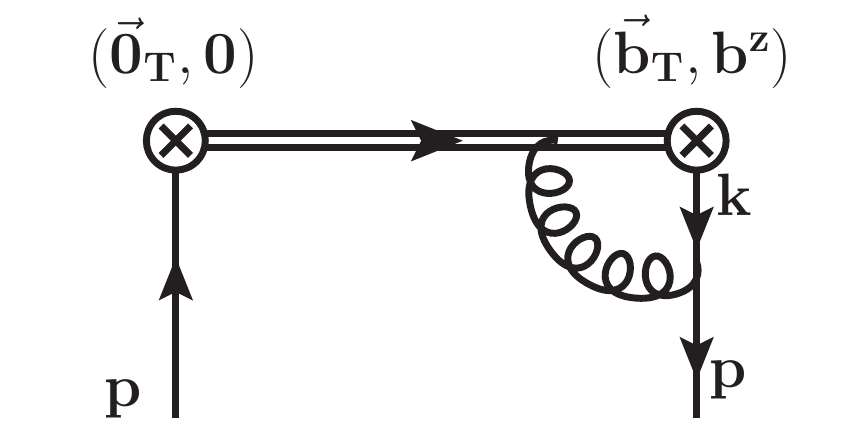}
 \end{subfigure}
 \caption{Sail diagrams contributing to the quasi beam function at NLO.}
 \label{fig:qtmd sail}
\end{figure}

The two sail diagrams are shown in \fig{qtmd sail}.
The Feynman rule for the Wilson line following from \eq{qtmd matrix element} is given by
\begin{align}
 \img g t^a \int_0^1 \df s\, \gamma_\nu'(s) e^{-\img k \cdot \gamma(s)}
\,,\end{align}
where $k$ is the momentum flowing into the Wilson line
and $\gamma$ the Wilson line path, given in \eq{qtmd_path}.
The two diagrams prior to Fourier transforming are given by
\begin{align} \label{eq:qtmd_b_1}
 \tilde q^{(b)}_n (b^\mu)
=&~-g^2 C_F \mu_0^{2\eps} \frac{1}{2} \sum_S
  \int_0^1 \df s \, \gamma'(s)_\mu \int\frac{\df^d k}{(2\pi)^d}
  \bar u_S(P) \frac{\gamma^\lambda \slashed{k} \gamma^\mu}{2k^2(P-k)^2} u_S(P)
  \, e^{\img P \cdot b -\img (P-k) \cdot \gamma(s)}
\nn\\*&~
  - g^2 C_F \mu_0^{2\eps} \frac{1}{2} \sum_S
  \int_0^1 \df s \, \gamma'(s)_\mu \int\frac{\df^d k}{(2\pi)^d}
  \bar u_S(P) \frac{\gamma^\mu \slashed{k} \gamma^\lambda }{2 k^2 (P-k)^2} u_S(P)
  \,e^{\img k \cdot b + \img (P-k)\cdot\gamma(s)}
\,.\end{align}
After averaging over spin, they can be combined into
\begin{align} \label{eq:qtmd_b_2}
 \tilde q^{(b)}_n (b^\mu)
= -g^2 C_F &\mu_0^{2\eps} \int\frac{\df^d k}{(2\pi)^d}
  \frac{P^\lambda k^\mu - (P\cdot k) g^{\lambda\mu} + P^\mu k^\lambda}{k^2(P-k)^2}
  \nn\\&\times
  \int_0^1 \df s \, \gamma'(s)_\mu
  \left[ e^{\img P \cdot b -\img (P-k) \cdot \gamma(s)} + e^{\img k \cdot b + \img (P-k)\cdot\gamma(s)}\right]
\,.\end{align}
For both $\lambda = 0$ and $\lambda = z$, the numerator can be simplified to
\begin{align}
 P^\lambda k^\mu - (P\cdot k) g^{\lambda\mu} + P^\mu k^\lambda
 = P^z \bigl[ k^\mu + k^z g^{\mu0} - k^0 g^{\mu z} \bigr]
 \quad\to\quad P^z (k^\mu - k^0 g^{\mu z})
\,,\end{align}
where the last step employs that the $g^{\mu0}$ piece vanishes
when contracted with the spatial path $\gamma'(s)_\mu$.
Introducing Feynman parameters and dropping the terms linear in $q^0$ gives
\begin{align} \label{eq:qtmd_b_3}
 \tilde q^{(b)}_n (b^\mu)
= -g^2 C_F \mu_0^{2\eps} P^z
  &\int_0^1 \df y \, \int\frac{\df^d q}{(2\pi)^d}
  \frac{q^\mu -2 y P^z g^{\mu z}}{q^4}
  \int_0^1 \df s \, \gamma'(s)_\mu
  \nn\\&\times
  \left[ e^{\img P \cdot b} e^{\img [q - P(1-y)] \cdot \gamma(s)}
      + e^{\img (q + y P) \cdot b} e^{-\img [q - P(1-y)] \cdot\gamma(s)}\right]
\,.\end{align}
Next, we use the relation
\begin{align}
 [q \cdot \gamma'(s)] \, e^{\pm \img [q - P(1-y)] \cdot \gamma(s)}
 &= \left( \mp \img \frac{\df}{\df s} + (1-y) [P \cdot \gamma'(s)] \right) e^{\pm \img [q - P(1-y)] \cdot \gamma(s)}
\end{align}
to replace the $q^\mu$ piece in the numerator,
\begin{align} \label{eq:qtmd_b_4}
 \tilde q^{(b)}_n (b^\mu) &
 = \img g^2 C_F \mu_0^{2\eps} P^z
  \int_0^1 \df y \, \int\frac{\df^d q}{(2\pi)^d} \frac{1}{q^4}
  \Bigl[e^{\img P \cdot b} e^{\img [q - P(1-y)] \cdot \gamma(s)}
  - e^{\img (q + y P) \cdot b} e^{-\img [q - P(1-y)] \cdot\gamma(s)} \Bigr]_{s=0}^{s=1}
 \nn\\&\quad
  + g^2 C_F \mu_0^{2\eps} (P^z)^2
  \int_0^1 \df y \, \int\frac{\df^d q}{(2\pi)^d}
  \frac{1+y}{q^4} \int_0^1 \df s \, \gamma'(s)^z
  \nn\\&\qquad\times
  \left[ e^{\img P \cdot b} e^{\img [q - P(1-y)] \cdot \gamma(s)}
      + e^{\img (q + y P) \cdot b} e^{-\img [q - P(1-y)] \cdot\gamma(s)}\right]
 \nn\\&
 \equiv \tilde q^{(b,1)}_n (b^\mu) + \tilde q^{(b,2)}_n (b^\mu)
\,.\end{align}
Since the first line in \eq{qtmd_b_4} is much simpler than the second line,
we have split $\tilde q_n^{(b)}$ accordingly.
Note that the second line in \eq{qtmd_b_4} only involves $\gamma'(s)^z$.
All contributions from the transverse gauge link are thus
fully captured in the first line, which becomes
\begin{align} \label{eq:qtmd_b1_1}
 \tilde q^{(b,1)}_n (b^\mu) &
 = 2 \img g^2 C_F \mu_0^{2\eps} P^z
  \int_0^1 \df y \int\frac{\df^d q}{(2\pi)^d} \frac{1}{q^4}
  e^{- \img \qt \cdot \bt - \img (q^z + y P^z) b^z}
\,.\end{align}
Taking the Fourier transform of this and pulling it inside a plus distribution yields
\begin{align} \label{eq:qtmd_b1}
 \tilde q^{(b,1)}_n(x, \bt, P^z) &
 = 2 \img g^2 C_F \mu_0^{2\eps} P^z \Bigl[I_0^a\Bigr]\plusinf
 + 2 \img g^2 C_F \mu_0^{2\eps} \delta(1-x) I_2
\,.\end{align}
The second contribution to \eq{qtmd_b_4} is more involved.
We first simplify the exponentials by letting $y \to 1-y$ and then $q \to q + y P$,
\begin{align} \label{eq:qtmd_b2_1}
 \tilde q^{(b,2)}_n (b^\mu) &
 = g^2 C_F \mu_0^{2\eps} (P^z)^2
  \int_0^1 \df y \, \int\frac{\df^d q}{(2\pi)^d}
  \frac{2-y}{(q + y P)^4}
  \int_0^1 \df s \, \gamma'(s)^z
  e^{\img P \cdot b} \left[  e^{\img q \cdot \gamma(s)}
   + e^{\img q \cdot b} e^{-\img q \cdot\gamma(s)}\right]
\,.\end{align}
To proceed, we need the explicit parameterizations for the Wilson line path $\gamma$,
\eq{qtmd_path}.
Evaluating all line integrals and taking the Fourier transform w.r.t.\ $b^z$,
\eq{qtmd_b2_1} yields
\begin{align} \label{eq:qtmd_b2_2}
 \tilde q^{(b,2)}_n(x, \bt, P^z) &
 = -\img g^2 C_F \mu_0^{2\eps} (P^z)^2
  \int_0^1 \df y \, \int\frac{\df^d q}{(2\pi)^d} \frac{2-y}{(q + y P)^4}
  \nn\\&\quad\times
  \biggl[ \delta[P^z (1-x)] \frac{2 - e^{-\img q^z L}(1 - e^{-\img \qt \cdot \bt})}{q^z}
  \nn\\&\qquad
  - \delta[q^z + P^z (1-x)] \frac{2 - e^{\img q^z L}(1 - e^{\img \qt \cdot \bt})}{q^z} e^{-\img \qt \cdot \bt} \biggr]
\,.\end{align}
The first term in brackets has a singularity at $q^z = 0$.
However, when integrating over $x$, the singularity cancels with the second line,
so it can be regulated by introducing the plus distribution prior to evaluating the $q$ integration,
\begin{align} \label{eq:qtmd_b2_2.5}
 \tilde q^{(b,2)}_n(x, \bt, P^z)
 & = \Bigl[ \tilde q^{(b,2)}_n(x \ne 1, \bt, P^z) \Bigr]\plusinf
  + \delta(1-x) \int \df x' \, \tilde q^{(b,2)}_n(x', \bt, P^z)
\,.\end{align}
For $x \ne 1$, this can be expressed using our master integrals,
\begin{align} \label{eq:qtmd_b2_3}
 \tilde q^{(b,2)}_n(x \ne 1, \bt, P^z) &
 = -2 \img g^2 C_F \mu_0^{2\eps} P^z
  \left[ \frac{1+x}{1-x} I_0^a - \frac{I_0^b}{1-x} \right]
  \nn\\&\quad
  +\img g^2 C_F \mu_0^{2\eps} P^z \frac{e^{-\img P^zL (1-x) }}{1-x}
  \left[ (1+x) (I_0^a - I_1^a) - (I_0^b - I_1^b) \right]
\,.\end{align}
The $\delta(1-x)$ term is fixed by integrating $\tilde q^{(b,2)}_n(x, \bt, P^z)$ over $x$,
\begin{align}
 \int \df x \, \tilde q^{(b,2)}_n(x, \bt, P^z)
 = -\img g^2 C_F \mu_0^{2\eps} P^z
  \int_0^1 \df y \, \int\frac{\df^d q}{(2\pi)^d} &\frac{2-y}{(q + y P)^4} (1 - e^{-\img \qt \cdot \bt})
  \nn\\*&\times
  \frac{2 - e^{-\img q^z L} - e^{\img q^z L}}{q^z}
\,.\end{align}
Here, the finite $L$ terms are crucial to regulate the integral at $q^z = 0$,
which makes the integral IR finite.
Note that due to the factor $1/q^z$, the above integral is UV finite,
so we can evaluate the integral in $d=4$ dimensions,
which after some manipulations yields
\begin{align} \label{eq:qtmd_b2_4}
 \int \df x \, \tilde q^{(b,2)}_n(x, \bt, P^z) &
 = \frac{g^2 C_F}{8 \pi^2}
  \int_0^{L P^z} \!\!\!\df v \, \ln\left(1 + \frac{b_T^2 (P^z)^2}{v^2}\right)
  \frac{\sin(v) + v [\cos(v) - 2]}{v^2}
\,.\end{align}
We have not been able to obtain a closed form for this integral.
However, it is clearly convergent for $v\to0$ and $v\to\infty$,
so we can take the $L P^z \to\infty$ limit first.
To cleanly extract the logarithmic dependence on $P^z$, take
\begin{align} \label{eq:sail delta approx}
 &\int \df x \, \tilde q^{(b,2)}_n(x, \bt, P^z)
 = \frac{g^2 C_F}{8 \pi^2}
  \int_0^{L P^z} \df v \int_0^1 \df y \,(y-2)
  \ln\left(1 + \frac{b_T^2 P^2}{v^2}\right) \sin(v y)
\nn\\&
 \approx
 g^2 C_F \frac{b_T P^z}{16 \pi^{3/2}} \int_0^1 \df y \, (y-2)
 G_{2,4}^{2,2} \left(
 \frac{1}{4} b_T^2 P^z y^2 \Bigg|
 \begin{matrix}\frac{1}{2} , \frac{1}{2} \\
 \frac{1}{2}, \frac{1}{2}, -\frac{1}{2}, 0
 \end{matrix}\right)
\nn\\&
 = g^2 C_F \frac{b_T P^z}{16 \pi^{3/2}} \left[
  \frac{2}{b_T^2 (P^z)^2}
  G_{2,4}^{2,2} \left(
  \frac{b_T^2 P_z^2}{4} \Bigg|
  \begin{matrix}\frac{3}{2} , \frac{3}{2} \\
  \frac{3}{2}, \frac{3}{2}, 0, \frac{1}{2}
  \end{matrix}\right)
 -
  G_{3,5}^{2,3} \left(
  \frac{b_T^2 P_z^2}{4} \Bigg|
  \begin{matrix}\frac{1}{2} , \frac{1}{2} , \frac{1}{2} \\
  \frac{1}{2}, \frac{1}{2}, -\frac{1}{2}, -\frac{1}{2} , 0
  \end{matrix}\right) \right]
\nn\\&
 = \frac{-g^2 C_F}{16\pi^2} \biggl[ \biggl(\Lb{} + \ln\frac{(2P^z)^2}{\mu^2}\biggr)^2
   - 2 \Lb{} -2 \ln\frac{(2P^z)^2}{\mu^2} + 4 \biggr]
\,.\end{align}
The full expression for the sail diagram, \eq{qtmd_b_4}, is obtained by adding \eqs{qtmd_b1}{qtmd_b2_2.5},
\begin{align} \label{eq:qtmd_b2_5}
 \tilde q^{(b)}_n(x, \bt, P^z) &
 = \img g^2 C_F \mu_0^{2\eps} P^z \biggl [ \frac{-4x I_0^a}{1-x} + \frac{2 I_0^b}{1-x}
  + \frac{e^{-\img P^zL (1-x) }}{1-x} \left[ (1+x) (I_0^a - I_1^a) - (I_0^b - I_1^b) \right]
  \biggr]\plusinf
 \nn\\&\quad
 + 2 \img g^2 C_F \mu_0^{2\eps} \delta(1-x) I_2
  + \delta(1-x) \int \df x \, \tilde q^{(b,2)}_n(x, \bt, P^z)
\,.\end{align}
Using the approximate result \eq{sail delta approx}, but keeping all master integrals exact,
we obtain
\begin{align} \label{eq:qtmd_b}
 \tilde q^{(b)}_n(x, \bt, P^z) &
 = \frac{\as C_F}{2\pi}  \biggl [ \frac{-2x}{1-x}  \biggl(\frac{1}{\eps} + \Lb{}\biggr)\Theta(1-x)\Theta(x) \biggr]_+^1
   \nn\\*&\quad
   + \frac{\as C_F}{2\pi} \delta(1-x) \biggl[\frac{1}{\eps} - \frac{1}{2} \Lb{2}
     - \Lb{} \LPz{}  + 2\Lb{}
     \nn\\*&\hspace{3.5cm}
     - \frac{1}{2} \LPz{2} + \LPz{} - 2 \biggr]
   \nn\\*&\quad
   + \Delta \tilde q^{(b)}_n(x, \bt, P^z)
\,,\end{align}
where we have singled out the terms with unphysical support $x \in [-\infty,\infty]$,
\begin{align} \label{eq:qtmd_b_suppressed}
 \Delta\tilde q^{(b)}_n(x, \bt, P^z) &=
 - \frac{\as C_F}{2\pi}  \biggl [ \frac{2|x|}{1-x} \Gamma\bigl(0, b_T P^z |x|\bigr)
   + \frac{2x}{|1-x|} \Gamma\bigl(0, b_T P^z |1-x|\bigr) \biggr]\plusinf
\nn\\*&
 + \frac{\as C_F}{2\pi} \biggl [ \frac{e^{-b_T P^z |x|} - e^{-b_T P^z |1-x|}}{b_T P^z (1-x)}
   \biggl(1 -\frac{1}{2} e^{-\img P^zL (1-x)} \biggr) \biggr]\plusinf
\nn\\*&
 + \frac{\as C_F}{4\pi} \biggl [
  e^{-\img P^zL (1-x)} \frac{1+x}{1-x} \sgn{x} \Bigl(\Gamma(0, b_T P^z |x|) + \gamma_E + \ln(P^z b_T |x|) \Bigr)
  \biggr]\plusinf
\nn\\*&
 + \frac{\as C_F}{4\pi} \biggl [
  e^{-\img P^zL (1-x)} \frac{1+x}{|1-x|} \Bigl(\Gamma(0, b_T P^z |1-x|) + \gamma_E + \ln(P^z b_T |1-x|) \Bigr)
  \biggr]\plusinf
\nn\\*&
 - \frac{\as C_F}{4\pi}  \biggl [
  \frac{e^{-\img P^zL (1-x) }}{1-x} \bigl( |x| - |1-x| \bigr)
  \biggr]\plusinf
\,.\end{align}

\subsection{Wilson line self energy}

\begin{figure}[h]
 \centering
 \begin{subfigure}{0.4\textwidth}
  \includegraphics[width=\textwidth]{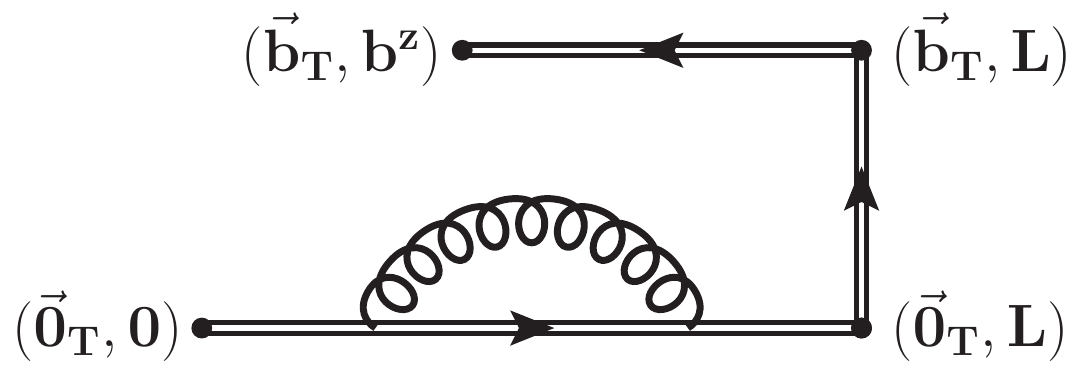}
  \caption{}
 \end{subfigure}
 \quad
 \begin{subfigure}{0.4\textwidth}
  \includegraphics[width=\textwidth]{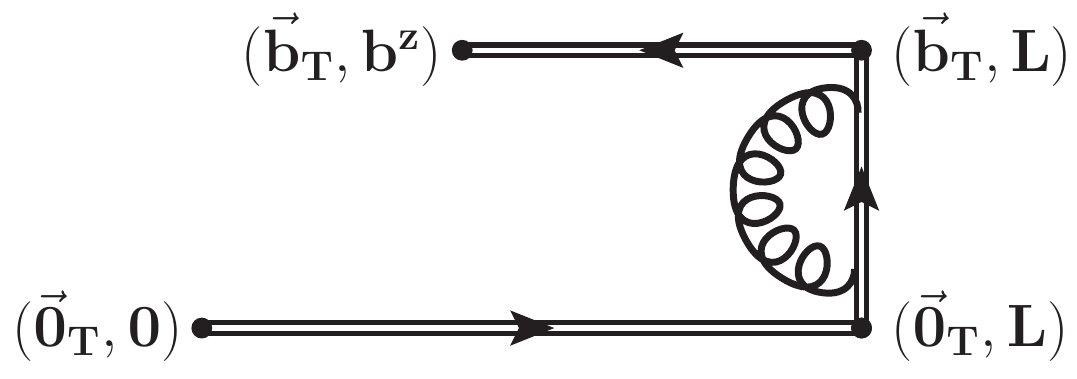}
  \caption{}
 \end{subfigure}
\\
 \begin{subfigure}{0.4\textwidth}
  \includegraphics[width=\textwidth]{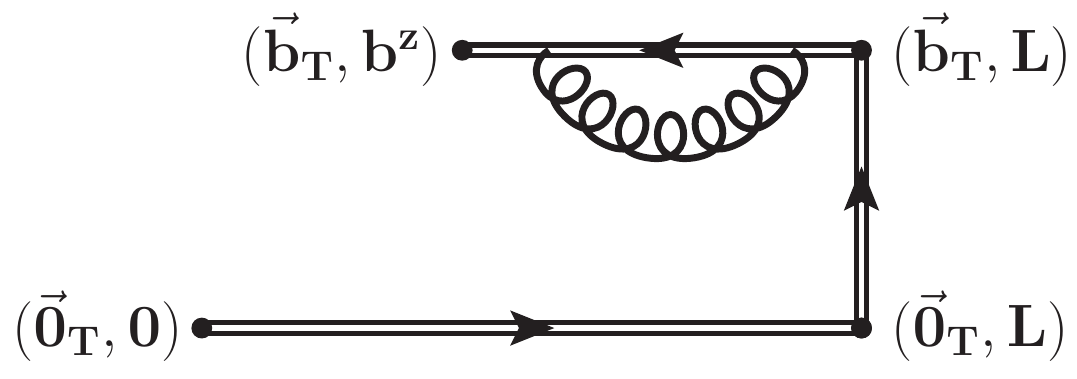}
  \caption{}
 \end{subfigure}
 \quad
 \begin{subfigure}{0.4\textwidth}
  \includegraphics[width=\textwidth]{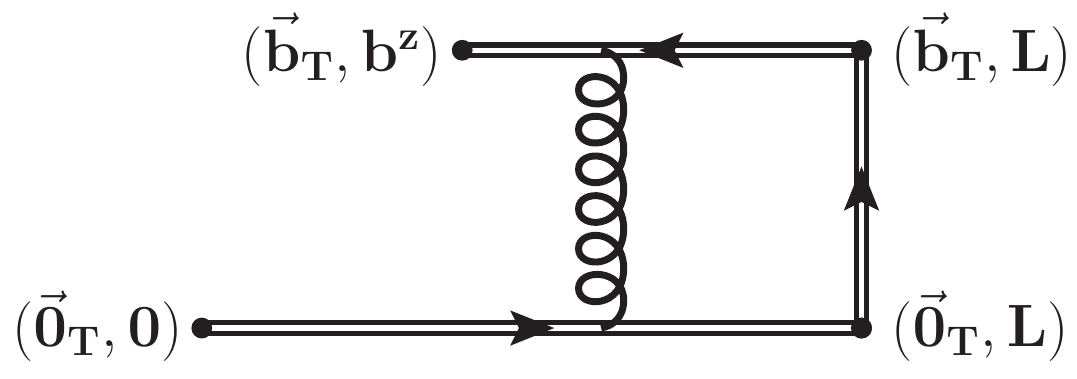}
  \caption{}
 \end{subfigure}
 \caption{Diagrams contributing to the Wilson line self energy correction to the quasi beam function.
 The coordinates illustrate the path in position space.
 Only diagrams (c) and (d) are sensitive to the vertex position $b^z$ and thus give a
 nontrivial contribution to the Fourier transform.}
 \label{fig:qtmd_tadpole}
\end{figure}

The general expression for the Wilson line self energy, \fig{TMD c},
in position space is given by
\begin{align} \label{eq:qtmd_c_1}
 \tilde q_n^{(c)}(b^\mu) &
 = \img g^2 C_F P^z e^{-\img b^z P^z}
  \times \frac{1}{2} \int_0^1 \df s \int_0^1 \df t \,
 [\gamma'(s) \cdot \gamma'(t)] \times \mu_0^{2\eps}
 \int\frac{\df^d k}{(2\pi)^d}  \frac{e^{\img k \cdot [\gamma(s) - \gamma(t)]}}{k^2 + \img 0}
\nn\\&
 = \frac{\as C_F}{2\pi} P^z e^{-\img b^z P^z} \, \mu_0^{2\eps} \pi^\eps \Gamma(1-\eps)
  \int_0^1 \df s \int_0^1 \df t \,
  \gamma'(s) \cdot \gamma'(t) \Bigl[-[\gamma(s) - \gamma(t)]^2\Bigr]^{-1+\eps}
\,,\end{align}
where a symmetry factor $1/2$ is included and $\gamma$ is the path of the Wilson line.
Since $\gamma$ can be split into three straight lines, see \fig{qbeam},
there are four distinct contributions in Feynman gauge, shown in \fig{qtmd_tadpole}.

\paragraph{Diagrams (a) and (b)} are independent of $b^z$, so the Fourier transform w.r.t.\ $b^z$ is trivial
and one can directly evaluate the line integrals
similar to the soft function calculation shown in \sec{qsoft_naive_nlo},
obtaining
\begin{align} \label{eq:qtmd_c_2}
 \tilde q_n^{(c,1)}(x, \bt, P^z) &
 = \frac{\as C_F}{2\pi} \delta(1-x) \left[ \frac{1}{\eps} + \ln\frac{L^2 \mu^2}{b_0^2} + 2 \right]
\,,\\
 \tilde q_n^{(c,2)}(x, \bt, P^z) &
 = \frac{\as C_F}{2\pi} \delta(1-x) \left[
   \frac{1}{\eps} + \Lb{} + 2 \right]
\,.\end{align}

\paragraph{Diagram (c)} depends on $b^z$, so care has to be taken with the Fourier transform.
The position space result is
\begin{align} \label{eq:qtmd_c_3}
 \tilde q_n^{(c,3)}(b^\mu) &
 = -\frac{\as C_F}{2\pi} P^z e^{-\img b^z P^z} \, (b^z-L)^{2\eps} \mu_0^{2\eps} \pi^\eps \Gamma(1-\eps)
  \frac{1}{\eps(2\eps-1)}
\nn\\&
 = \frac{\as C_F}{2\pi} P^z e^{-\img b^z P^z} \left[ \frac{1}{\eps} + \ln\frac{(b^z-L)^2 \mu^2}{b_0^2} + 2 \right]
  + \cO(\eps)
\,.\end{align}
The Fourier transform can be obtained using \cite{Izubuchi:2018srq}
\begin{align}
 \int\frac{\df b^z}{2\pi} e^{\img b^z (x-1) P^z} (b^z-L)^{2\eps}
 = \frac{4^\eps \Gamma(1/2+\eps)}{\sqrt{\pi} \Gamma(-\eps) (P^z)^{1+2\eps}}
   \frac{e^{- \img P^z L (1-x)}}{ |1-x|^{1+2\eps} }
\end{align}
as
\begin{equation} \label{eq:qtmd_c_4}
 \tilde q_n^{(c,3)}(x,\bt,P^z)
 = -\frac{\as C_F}{2\pi} \left[\frac{e^{-\img P^z L (1-x)}}{|1-x|} \right]\plusinf
   + \frac{\as C_F}{2\pi} \delta(1-x) \left[\frac{1}{\eps} + \ln\frac{\mu^2 L^2}{b_0^2} + 2 \right]
\,.\end{equation}
As shown in \app{pwr}, when convoluted with the TMDPDF the phase term will be power suppressed.
Note that if one were to expand \eq{qtmd_c_3} in $b^z \ll L$ prior to Fourier transforming,
one would instead obtain
\begin{align} \label{eq:qtmd_c_5}
 \tilde q_n^{(c,3)}(x,\bt,P^z)
 \stackrel{b^z \ll L}{=}
 &\int \frac{\df b^z}{2\pi} e^{\img b^z x P^z}
  \frac{\as C_F}{2\pi} P^z e^{-\img b^z P^z} \left[ \frac{1}{\eps} + \ln\frac{L^2 \mu^2}{b_0^2} + 2 \right]
\nn\\
 =~&\frac{\as C_F}{2\pi} \delta(1-x) \left[ \frac{1}{\eps} + \ln\frac{L^2 \mu^2}{b_0^2} + 2 \right]
\,,\end{align}
so dropping the phase in \eq{qtmd_c_4} is equivalent to performing a small-distance expansion
$b^z \ll L$ in position space.

\paragraph{Diagram (d)} gives rise to a factor $2$ to lift the symmetry factor.
Due to the finite separation $\bt$, it is UV finite, so we can let $\eps\to0$.
In position space, we obtain
\begin{align} \label{eq:qtmd_c_6}
 \tilde q_n^{(c,4)}(b^\mu) &
 = \frac{\as C_F}{2\pi} P^z e^{-\img b^z P^z}
   \biggl[ - 2\frac{b^z}{b_T} \arctan\frac{b^z}{b_T}
   + 2\frac{L}{b_T} \arctan\frac{L}{b_T}
   + 2 \frac{L - b^z}{b_T} \arctan\frac{L-b^z}{b_T}
   \nn\\&\hspace{3.2cm}
   - \ln\frac{[b_T^2 + L^2] [ b_T^2 + (L - b^z)^2]}{b_T^2 [b_T^2 + (b^z)^2]}
   \biggr]
\,.\end{align}
Here, it is quite difficult to take the Fourier transform while keeping the exact $b^z$ dependence.
It is easier to use the first line of \eq{qtmd_c_1} and leave the $k^z$ integration until the end,
\begin{align} \label{eq:qtmd_c_7}
 &\tilde q_n^{(c,4)}(x, \bt, P^z)
 = \img g^2 C_F P^z \mu_0^{2\eps} \int\frac{\df b^z}{2\pi} e^{\img (x-1)P^z b^z} L (L-b^z)
   \nn\\*&\hspace{3cm}\times
   \int_0^1 \df s \int_0^1 \df t \int\frac{\df^4 k}{(2\pi)^4}
   \frac{e^{-\img \kt \cdot \bt - \img k^z [L(1-s-t) + t b^z]}}{k^2 + \img 0}
\nn\\&
 = \frac{\as C_F}{2\pi} \frac{P^z}{b_T} \int\frac{\df b^z}{2\pi} e^{\img (x-1)P^z b^z}
   \int\df k^z \, e^{-b_T |k^z|} \frac{(1 - e^{\img k^z L})(e^{-\img b^z k^z} - e^{-\img k^z L})}{k_z^2}
\nn\\&
 = \frac{\as C_F}{2\pi} \frac{P^z}{b_T}
   \int\df k^z \, e^{-b_T |k^z|} \frac{1 - e^{\img k^z L}}{k_z^2}
   \left[ \delta[(x-1)P^z - k^z] - \delta[(x-1)P^z] e^{-\img k^z L} \right]
\,.\end{align}
While the second line is manifestly finite as $k^z\to0$, the last line has an apparent singularity,
which is as usual treated as a plus distribution,
\begin{align} \label{eq:qtmd_c_8}
 \tilde q_n^{(c,4)}(x, \bt, P^z) &
 = \Bigl[ \tilde q_n^{(c,4)}(x \ne 1, \bt, P^z) \Bigr]\plusinf
  + \delta(1-x) \int\df x'\,\tilde q_n^{(c,4)}(x', \bt, P^z)
\nn\\&
 = \frac{\as C_F}{2\pi} \left[ \frac{e^{-b_T P^z |1-x|}}{b_T P^z}
   \frac{1 - e^{- \img P^z L (1-x)}}{(1-x)^2} \right]\plusinf
   \nn\\&
 + \frac{\as C_F}{2\pi} \delta(1-x) \left[- 2 \ln\frac{b_T^2 + L^2}{b_T^2}
   + 4 \frac{L}{b_T} \arctan\frac{L}{b_T} \right]
\,.\end{align}
For comparison, the $b^z \ll L$ limit of \eq{qtmd_c_6} is given by
\begin{align} \label{eq:qtmd_c_9}
 \tilde q_n^{(c,4)}(b^\mu) &
 \stackrel{b^z \ll L}{=} \frac{\as C_F}{2\pi} P^z e^{-\img b^z P^z}
   \biggl[ - 2\frac{b^z}{b_T} \arctan\frac{b^z}{b_T}
   + 4 \frac{L}{b_T} \arctan\frac{L}{b_T}
   - \ln\frac{(b_T^2 + L^2)^2}{b_T^2 [b_T^2 + (b^z)^2]}
   \biggr]
\,.\end{align}
This still has a nontrivial $b^z$ dependence, so in order to obtain the $\delta(1-x)$
term in \eq{qtmd_c_8} we further need to expand in $b^z \ll b_T$, giving
\begin{align} \label{eq:qtmd_c_10}
 \tilde q_n^{(c,4)}(b^\mu) &
 \stackrel{b^z \ll L, b_T}{=}
 \frac{\as C_F}{2\pi} P^z e^{-\img b^z P^z}
   \biggl[ 4 \frac{L}{b_T} \arctan\frac{L}{b_T} - 2 \ln\frac{b_T^2 + L^2}{b_T^2} \biggr]
\nn\\*&
 = \frac{\as C_F}{2\pi} \delta(1-x) \biggl[ 4 \frac{L}{b_T} \arctan\frac{L}{b_T} - 2\ln\frac{b_T^2 + L^2}{b_T^2} \biggr]
\,.\end{align}
Hence dropping the phase term in \eq{qtmd_c_8} corresponds to a small-distance expansion
$b^z \ll L$ and $b^z \ll b_T$.

\paragraph{Combined result.}
Combining Eqs.\ \eqref{eq:qtmd_c_2}, \eqref{eq:qtmd_c_4} and \eqref{eq:qtmd_c_8},
we obtain the full exact tadpole diagram as
\begin{align} \label{eq:qtmd_c}
 \tilde q_n^{(c)}(x, \bt, P^z) &= \frac{\as C_F}{2\pi} \delta(1-x) \biggl[
   \frac{3}{\eps} + 3 \Lb{}  + 2 + \frac{2\pi L}{b_T} \biggr]
 + \Delta\tilde q_n^{(c)}(x, \bt, P^z)
\,,\end{align}
where the terms suppressed for $b^z \ll b_T \ll L$ are
\begin{align} \label{eq:qtmd_c_suppressed}
 \Delta\tilde q_n^{(c)}(x, \bt, P^z) &=
 -\frac{\as C_F}{2\pi} \left[\frac{e^{-\img P^z L (1-x)}}{|1-x|} \right]\plusinf
 + \frac{\as C_F}{2\pi} \left[ \frac{e^{-b_T P^z |1-x|}}{b_T P^z}
   \frac{1 - e^{- \img P^z L (1-x)}}{(1-x)^2} \right]\plusinf
 \nn\\&\quad
 + \frac{\as C_F}{2\pi} \delta(1-x) \biggl[- 2 \ln\frac{b_T^2 + L^2}{L^2}
   + 4 \frac{L}{b_T} \biggl(\arctan\frac{L}{b_T} - \frac{\pi}{2} \biggr) + 4\biggr]
\,.\end{align}

\subsection{Power-suppressed contributions to the matching kernel}
\label{app:pwr}

The unphysical contributions \eqss{qtmd_a_suppressed}{qtmd_b_suppressed}{qtmd_c_suppressed} contain fastly oscillating phases $\sim e^{-\img P^z L (1-x)}$,
and one may thus expect that these give vanishing contributions as $L P^z \to \infty$.
However, these phases vanish as $x\to1$, and furthermore can be associated with divergences in $1/(1-x)$,
so they can contribute nontrivially to the quasi beam function.
Thus, in order to neglect them, one has to show that they do not contribute to the matched TMDPDF in the limit $L P^z \to \infty$.

The unphysical terms cannot yield a multiplicative matching with the TMDPDF, as the support of quasi-TMD and TMD do not match. Given our discussion in \sec{schematic_matching}, this provides a direct indication for the fact that they are power suppressed, and we will show that this is consistent. Assuming that their contributions satisfy a convolution structure as in \eq{matching_schematic}, then the unphysical part of the matching is given by
\begin{align} \label{eq:delta_matching_1}
 \Delta \tilde f_i^\TMD(x, \bt, \mu, P^z) &
 =  \int_{-1}^1 \frac{\df y}{|y|} \, \biggl[\Delta \tilde q\Bigl(\frac{x}{y}, \bt, P^z\Bigr)\biggr]\plusinf f_i^\TMD(y, \bt, \mu, \zeta)
\,,\nn\\
 \Delta \tilde q(x, \bt, P^z) &= \Delta\tilde q_n^{(a)}(x, \bt, P^z) + \Delta\tilde q_n^{(b)}(x, \bt, P^z) + \Delta\tilde q_n^{(c)}(x, \bt, P^z)
\,.\end{align}
In the following, we suppress all arguments except $x$ and $y$ as well as all superscripts and subscripts and the flavor indices for brevity.
We also extend the integral in \eq{delta_matching_1} to infinity by implicitly assuming that $f^\TMD(y) = 0$ for $|y|\ge1$.
Changing the integration variable in \eq{delta_matching_1} from $y$ to $x/y$, we obtain
\begin{align} \label{eq:delta_matching_2}
 \Delta \tilde f(x) &
 = \int_{-\infty}^\infty \frac{\df y}{|y|} \, \bigl[\Delta \tilde q(y)\bigr]\plusinf  \, f\Bigl(\frac{x}{y}\Bigr)
\,.\end{align}
Next note that $\Delta\tilde q(y)$ is entirely given as a plus distribution, so we can rewrite \eq{delta_matching_2} as
\begin{align} \label{eq:delta_matching_3}
 \Delta \tilde f(x) &
 = \int_{-\infty}^\infty \df y \, \bigl[\Delta \tilde q(y)\bigr]\plusinf  \, \biggl[ \frac{1}{|y|} f\Bigl(\frac{x}{y}\Bigr) - f(x) \biggr]
   + f(x) \int_{-\infty}^\infty \df y \, \bigl[\Delta \tilde q(y)\bigr]\plusinf
\nn\\&
 = \int_{-\infty}^\infty \df y \, \Delta \tilde q(y) \, \biggl[ \frac{1}{|y|} f\Bigl(\frac{x}{y}\Bigr) - f(x) \biggr]
\,,\end{align}
where we can drop the plus prescription from now on, as the term in square bracket cancels an overall $1/(1-y)$ divergence.

Let us now consider the case where $\Delta\tilde q$ is given by an exponential phase factor, a regular function $r(y)$ and potentially a divergence as $y\to1$,
\begin{align}
 \bigl[\Delta \tilde q(y)\bigr]\plusinf \sim \biggl[ e^{-\img P^z L (1-y)} \frac{r(y)}{1-y} \biggr]\plusinf
\,.\end{align}
For this case, \eq{delta_matching_3} becomes
\begin{align} \label{eq:delta_matching_4}
  \Delta \tilde f(x) &
 \sim e^{-\img P^z L} \int_{-\infty}^\infty \df y \, e^{\img P^z L y} \, r(y) \, \frac{|y|^{-1} f(x/y) - f(x)}{1-y}
\nn\\&
 \equiv e^{-\img P^z L} \int_{-\infty}^\infty \df y \, e^{\img P^z L y} \, h(y)
\,.\end{align}
The function $h(y)$ is by construction regular as $z\to1$ due to the subtraction term.

Without giving an exact proof for all functions $h(y)$ appearing at one loop,
we can give strong arguments that \eq{delta_matching_4} vanishes as $P^z L \to \infty$ for all relevant functions $h(y)$.
The strongest criterion to prove this behavior is for functions $h$ that satisfy $\int \df y \, |h(y)| < \infty$, in which case the Riemann-Lebesgue lemma applies.
Typically, this is too restrictive, as for example $h(x) = \ln|x|$ is not integrable, but its Fourier transform is known to vanish as $1/(P^z L)$.
A less restrictive and more intuitive argument is that as long as $h(y)$ is sufficiently smooth such that $P^z L \gg |h'(y)|$, then the rapid oscillation of the Fourier kernel suppresses the integral.
This holds for all $(1-y) P^z L \gtrsim 1$, and thus the contribution of the integral not suppressed by the exponential phase must be of order $1/(P^z L)$.

All other functions appearing in \eqss{qtmd_a_suppressed}{qtmd_b_suppressed}{qtmd_c_suppressed} have an explicit suppression in $b_T P^z \to \infty$.
While they may contain divergences in $1/(1-y)$, they can be regulated the same way as the above,
after which the result is clearly suppressed by positive powers of $1/(b_T P^z)$ relative to the physical terms in the matching calculation.

Since all terms in \eqss{qtmd_a_suppressed}{qtmd_b_suppressed}{qtmd_c_suppressed} give power-suppressed contributions to the matching, they can be dropped in the final result for the unsubtracted beam function.
Upon adding only the physical contributions from \eqss{qtmd_a}{qtmd_b}{qtmd_c}, one then obtains \eq{qbeam_nlo}.

\bibliographystyle{JHEP}
\bibliography{../literature}

\end{document}